\documentclass[aps,prb,twocolumn,superscriptaddress,floatfix,letterpaper]{revtex4-1}
\usepackage[inline]{enumitem}
\usepackage{tikz}
\usepackage[export]{adjustbox}
\usepackage{float}

\def\arraystretch{1.25} 

\newcommand{\nn}{\nonumber}

\newcommand{\ZZ}{\mathbb{Z}}
\newcommand{\calB}{\mathcal{B}}
\newcommand{\calC}{\mathcal{C}}
\newcommand{\calF}{\mathcal{F}}
\newcommand{\calL}{\mathcal{L}}
\newcommand{\calM}{\mathcal{M}}

\newcommand{\calP}{\mathcal{P}}
\renewcommand{\tilde}{\widetilde}
\renewcommand{\bar}{\overline}

\renewcommand{\hat}[1]{\widehat{#1}}
\newcommand{\bchi}{\bar{\chi}}

\newcommand{\Is}{\text{Is}}

\newcolumntype{L}{>{$}l<{$}}
\newcolumntype{R}{>{$}r<{$}}
\newcolumntype{C}{>{$}c<{$}}

\renewcommand{\nn}{\nonumber}

\begin{document}

\begin{titlepage}

\title{Metallic states beyond Tomonaga-Luttinger liquids in one dimension}

\author{Wenjie Ji}
\affiliation{Department of Physics, Massachusetts Institute of
Technology, Cambridge, Massachusetts 02139, USA}
\author{Xiao-Gang Wen}
\affiliation{Department of Physics, Massachusetts Institute of
Technology, Cambridge, Massachusetts 02139, USA}

\begin{abstract} 

In this paper, we propose some new strongly correlated gapless states (or
critical states) of spin-1/2 electrons in 1+1-dimensions, such as doped
anti-ferromagnetic spin-1/2 Ising chain.  We find doped anti-ferromagnetic
Ising chain to be a different metallic phase from the doped ferromagnetic Ising
chain, despite the two have identical symmetry.  The  doped anti-ferromagnetic
Ising chain has a finite energy gap for all charge-1 fermionic excitations even
without pairing caused by attractive interactions, resembling the pseudo-gap phase of
underdoped high Tc superconductors.  Applying a transverse field to the
ferromagnetic and anti-ferromagnetic metallic phases can restore the $Z_2$
symmetry, which gives rise to two distinct critical points despite that the two
transitions have exactly the same symmetry breaking pattern.  We also propose
new chiral metallic states.  All those new gapless states are strongly
correlated in the sense that they do not belong to the usual Tomonaga-Luttinger
phase of fermions, {\it i.e.} they cannot be smoothly deformed into the
non-interacting fermion systems of the same symmetry.  Our non-perturbative
results are obtained by noticing that gapless quantum systems have emergent
\emph{categorical symmetries} (\emph{i.e.} non-invertible gravitational
anomalies), which are described by multi-component partition functions that
are modular covariant. This allows us to calculate the scaling dimensions and
quantum numbers of all the low energy operators for those strongly correlated
gapless states.  This demonstrates an application of emergent categorical
symmetries in determining low energy properties of strongly correlated gapless
states, which are hard to obtain otherwise.

\end{abstract}


\maketitle

\end{titlepage}

{\small \setcounter{tocdepth}{1} \tableofcontents }

\section{Introduction} 

The simplest 1d metallic states are Fermi liquids of non-interacting electrons,
whose low energy properties are described by non-interacting fermionic
quasiparticles.  In the low energy limit, Fermi liquids are described by
several decoupled sectors and each sector contains a few modes.  In this paper,
we will try to develop a general understanding of gapless states by viewing the
gapless states as formed by several decouples sectors, and using the notion of
\emph{categorical symmetry} \cite{JW191209391} (\ie modular covariance of
non-invertible gravitational anomaly \cite{JW190513279}). 

Readers who are just interested in 1d strongly interacting metallic states can
directly go to Section \ref{result}. The section \ref{gen} contains some
general discussions. 

If a strongly interacting metallic state is stable against all symmetry
preserving perturbations, then it will represent a stable phase of quantum
matter.  However, most strongly interacting metallic states are not stable
against certain symmetry preserving perturbations.  Those metallic states will
correspond to critical states (or multi-critical points) that describe
continuous phase transitions between different phases of quantum matter.  Thus
the constructions discussed in this paper can be viewed as a systematic way to
discover 1d gapless quantum phases, as well as 1d (multi-)critical points.  (In
this paper, we will use 1d to refer to 1-dimensional space and 1+1D to refer to
1+1-dimensional space-time.)

\section{A general picture for gapless quantum states} 
\label{gen}

After the development of last 30 years, we start to have a comprehensive
understanding of all gapped quantum states in 1-dimensional, 2-dimensional, and
3-dimensional spaces, in terms of spontaneous symmetry breaking
\cite{L3726,LL58}, group cohomology \cite{CGL1314,CGL1204}, and braided fusion
(higher) category
\cite{KW9327,FNS0428,LW0510,RSW0777,W150605768,BBC1440,LW150704673,LW160205946,LW170404221,LW180108530,ZW180809394}.
In fact, we have classified (or proposed to classify) all 1d
\cite{CGW1107,SPC1139,FK1103,PBT1225}, 2d
\cite{BBC1440,LW150704673,LW160205946,BW161207792}, and 3d
\cite{LW170404221,LW180108530,ZW180809394} gapped liquid\cite{ZW1490,SM1403}
states of boson/fermion systems with any finite on-site symmetry.  The
classification is achieved via the realization that gapped quantum phases are
described by symmetry breaking orders, topological orders
\cite{W9039,WN9077,KW9327}, and/or symmetry protected trivial (SPT) orders
\cite{GW0931,CLW1141}.

Such a systematic understating of topological orders \cite{W9039,WN9077,KW9327}
and SPT orders \cite{GW0931} (including topological insulators and
superconductors
\cite{KM0501,BZ0602,KM0502,MB0706,R0922,FKM0703,QHZ0824,SMF9945,RG0067,R0664,QHR0901,SF0904})
leads to a deeper understanding of gauge and gravitational anomalies
\cite{RML1204,W1313,KW1458,K1467}, in terms of the existence of lattice
realization in the same dimension (which is called the existence of a UV
completion).  This resulted in a generalization of anomalies to include
non-invertible
anomalies.\cite{KW1458,FV14095723,M14107442,KZ150201690,JW190513279} Those
generalized anomalies (inlcuding perturbative and global gauge/gravity
anomalies) are classified in terms of topological orders and SPT orders in one
higher dimension \cite{W1313,KW1458}.  Such an understanding of anomalies also
lead to a solution to the long-standing chiral fermion problem
\cite{W1301,YX14124784}.

In comparison, there is a lack of comprehensive understanding of gapless
quantum states of matter, despite that we know many examples of them, such as
superfluid, anti-ferromagnets, nodal $d$-wave superconductors, graphene, Weyl
semi-metals, \etc.  But in 1d, thanks to
Belavin-Polyakov-Zamolodchikov, we do have a good understanding
of gapless quantum states with linear velocities via conformal field theories
(CFT) \cite{BPZ8433,Ght9108028,MS8977}.  In particular, we can use the modular
invariant partition function, which is parametrized by a complex number $\tau$ describing the shape of the spacetime torus,\begin{align} \label{Minv}
Z(\tau)=Z(\tau+1)=Z(-1/\tau) \end{align} to systematically study 1d gapless
states.

In this paper, we will try to develop a systematic point of view of gapless
quantum matter based on gauge/gravity anomaly, hoping this may lead to a more
general understanding gapless states in higher dimensions.

\begin{figure}[t]
\begin{center}
\includegraphics[scale=0.6]{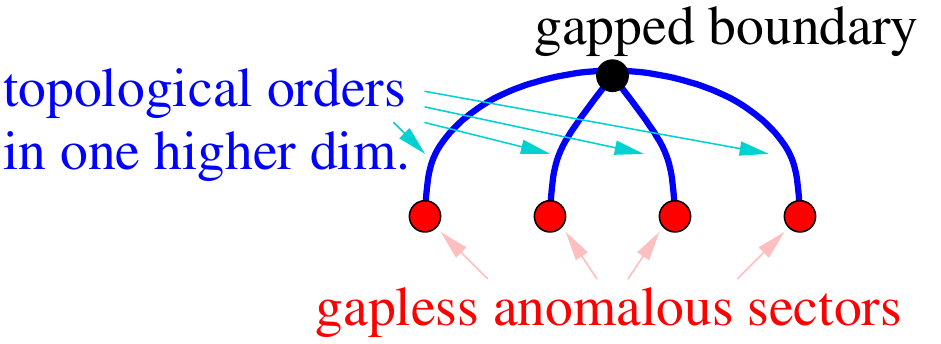} 
\end{center}
%1
\caption{(Color online) A general picture of a gapless quantum states, which
is formed by decoupled anomalous gapless sectors restricted to the symmetric
sub-Hilbert spaces (the red dots).  The emergent symmetry and emergent
anomalies are described by non-invertible gravitational anomalies (\ie the
topological orders in one higher dimension). Thus, the anomalous sectors are
boundary of corresponding topological orders in one higher dimension.  } \label{asect}
\end{figure}

First, the low energy part of a gapless state may becomes several decoupled
sectors, where the interactions between different sectors flow to zero in the
infrared limit under renormalization group flow. Consequently, in the low
energy limit, there are often emergent symmetries.  For example, the state charged under original
UV symmetry (the lattice symmetry) becomes several copies, one copy for each
decoupled sector.  Since each decoupled low energy sector is not a full system,
each sector by itself is often anomalous.  Thus there are also emergent
anomalies (\ie the low energy effective theory is anomalous).

Recently, it is pointed out that, when restricted to the symmetric sub-Hilbert
space, a symmetry can be fully characterized\cite{JW191209391} by a
non-invertible gravitational
anomaly.\cite{KW1458,FV14095723,M14107442,KZ150201690,JW190513279} So we can
treat the emergent symmetries and emergent anomalies in a unified way by
restricting to symmetric sub-Hilbert space.  In this case, we only have an
emergent non-invertible gravitational anomaly.  To stress this close connection
between non-invertible gravitational anomaly and symmetry, we refer to
non-invertible gravitational anomaly as categorical symmetry.\cite{JW190513279}
This point of view is very general. Not only emergent 0-symmetries (\ie the
usual global symmetries) can be viewed as emergent non-invertible gravitational
anomalies, emergent higher symmetries and even more general emergent higher
algebraic symmetries can also be viewed as emergent non-invertible
gravitational anomalies (\ie be viewed as emergent categorical
symmetry).\cite{KZ200308898,KZ200514178}

For example, 1+1D gapless state with on-site symmetry $G$ in the original
lattice system, also has a dual algebraic symmetry denoted by $\t G$
\cite{JW190513279}.  The total symmetry is the categorical symmetry denoted by
$G\vee \t G$ \cite{JW190513279}.  Note that a categorical symmetry is
nothing but a generalized gravitational anomaly.  Also note that a generalized
gravitational anomaly is nothing but a topological order in one higher
dimension \cite{W1313,KW1458,FV14095723,M14107442,KZ150201690}.  The
topological order in one higher dimension that describes the categorical
symmetry $G\vee \t G$ is the topological order described by $G$ gauge theory.
The 1+1D gapless state corresponds to minimal gapless boundary of the 2+1D  $G$
gauge theory \cite{JW190513279}, that have no condensation of gauge charge nor
gauge flux.

To have more information describing a gapless state, we want to decompose the
gapless state into smallest decoupled sectors. This allows us to see the
maximal emergent symmetry and emergent anomalies.  In other words, this allows
us to obtain the maximal categorical symmetry \cite{JW191209391}.  It is
possible that the maximal categorical symmetry fully characterizes the gapless
state.  This may be a way to systematically understand strongly correlated
gapless states.

Since each decoupled sector has a generalized graviational anomaly, it can be viewed as a boundary of topological order in one higher dimension
(see Fig. \ref{asect}).  For example, the right-moving sector of a 1+1D gapless
state has a perturbatively gravitational anomaly characterized by its central
charge $c_R$.  Similarly,  the left-moving sector also has a gravitational
anomaly characterized by its central charge $c_L$.  The right-moving sector is
a boundary of a 2+1D chiral topological order.  The left-moving sector is also
a boundary of a 2+1D chiral topological order.  The two chiral topological
orders allow us to describe the 1+1D gapless state.

For a system with generalized gravitational anomaly, its partition function
have multiple components. This multi-component  partition function transforms
covariantly under mapping-class-group transformations of the
spacetime\cite{JW190513279,CZ190312334}.  So the multi-component  partition
function forms a representation of the mapping class group.  Such a
representation turns out to be the
representation that describes the topological order in one higher dimension.
Since
\begin{align}
&\ \ \ \
\text{topological order in one higher dimension}
\nonumber\\
&=
\text{non-invertible gravitational anomaly}
\nonumber\\
&=
\text{categorical symmetry} ,
\end{align}
we see that the categorical symmetry determines the representation of the
mapping class group formed by the multi-component  partition function, which in
turn determines the dynamical properties (such as scaling dimensions) of the
1+1D gapless state.  This is how emergent maximal categorical symmetry
systematical describes strongly correlated gapless state.

In this paper, we will use this line of thinking, \ie use multi-component
partition functions and their modular covariance, to study strongly correlated
metals. This approach is beyond perturbation.

\section{Summary of results}
\label{result}

\subsection{Ising strongly correlated metal with ferromagnetic and
anti-ferromagnetic correlation }

In section \ref{IsingT}, we consider a  spin-$\frac12$ electron chain close to
one electron per site with strong on-site repulsive interaction and (anti-)ferromagnetic
Ising spin interaction. The model has $Z_2$ spin-flip symmetry: $S^z \to -S^z$
and the $U(1)$ electron conservation symmetry, as well as the translation
symmetry: $U(1)\times Z_2^s\times \Z$.   

We note that for the insulating Ising chain, there are two $Z_2$ symmetry
breaking phases, one for ferromagnetic and the other for anti-ferromagnetic
Ising interactions, and the anti-ferromagnetic phase breaks the translation
symmetry. There are also two $Z_2$ symmetry breaking phases in the metallic
states for ferromagnetic and anti-ferromagnetic Ising interactions. However,
the anti-ferromagnetic metallic phase does not break the translation symmetry.
We will show that despite the two metallic phases have the same symmetry, they
are two distinct phases separated by phase transitions, if we do not explicitly
break the symmetry.  In particular, the fermionic charge-1 excitation is
gapless in the ferromagnetic metallic phase, and is gapped in the
anti-ferromagnetic metallic phase.  Thus the ferromagnetic and
anti-ferromagnetic metallic phases provide examples of \textbf{symmetry
protected gapless phases} \cite{FO150307292,YO180506885}.  

The $Z_2$ spin-flip symmetry breaking in the two metallic phases can be
restored if we add a strong transverse magnetic field.  We find that the
critical theories of the transition points are different for the ferromagnetic
and anti-ferromagnetic cases. The ferromagnetic critical point is described by
a CFT
\begin{align}
\label{u1Is}
u1 \oplus \text{Is} \oplus
\overline{u1} \oplus \overline{\text{Is}},
\end{align}
while the anti-ferromagnetic critical point is described by a different CFT.
We see that even the same symmetry breaking pattern can have distinct critical
theories \cite{BS180807465}.

\subsection{Spin-rotation symmetric strongly correlated metal}

In this paper, we also construct some 1d chiral gapless states.  One way to do
so is to start with a 2d FQH stripe.  On one side of the stripe, we have a
gapless edge state (the bottom part of Fig. \ref{asect}), and on the other side
of stripe (the top part of Fig. \ref{asect}) we have a fully gapped edge
(assuming the FQH state support gapped edges).  This way, we can obtain a
strongly interacting gapless state.  In Appendix \ref{TLL}, we show that, if we
start with an Abelian FQH state and consider only $U(1)$ symmetry of electron
number conservation, the above construction actually will always give us a
Tomonaga-Luttinger (TL) liquids, not a new gapless phase.  Thus in this paper,
we consider electron systems with more than just the $U(1)$ symmetry.  As an
application, in Section \ref{chiral}, we start with 2d integer quantum Hall
stripe with $SO(3)$ spin rotation symmetry, and obtain a chiral metallic state
of spin-1/2 charge-1 electrons, where the right-moving and left-moving gapless
fermions carry different spins.  This chiral metallic state is beyond the TL
liquids of spin-1/2 electrons.  

Furthermore, we consider an electron system with $U(1)$ charge, $SU(2)$ spin,
and $\Z$ lattice translation symmetries.  The lattice fermions carry charge-1
and spin-1/2.  Such an electron system can realize a chiral metallic phase (see
Section \ref{u1su2Is}).  In this chiral metallic state, the low energy
excitations are described by a CFT
\begin{align}
\label{u1su2}
{su2}_2\oplus {u1} \oplus \text{Is} \oplus
\bar{{su2}}_1\oplus \bar{{su2}}_1 \oplus \bar{{u1}},
\end{align}
Note that the right movers and left movers are described by different CFTs (\ie
different chiral algebras), and those different sectors may have diffrent
velocities.  We see that the single lattice $SU(2)$ spin rotation symmetry is
enlarged to $SU(2)\times SU(2)\times SU(2) $ symmetry at low energies.  The
single lattice $U(1)$ charge conservation symmetry is enlarged to $U(1) \times
U(1)$ symmetry at low energies.  In the clean limit, the chiral metallic state
has a quantized two-terminal thermal conductance $\ka = c\frac{\pi}{6}
\frac{k_B^2 T}{\hbar}$, where $c=\frac{3}{2}+1+\frac12=3$ is the total central
charge for right movers (or left movers).  Since the spin $S_z$ is conserved,
we can treat it as a conserved charge where each electron carries $\pm \hbar/2$
$S_z$-charge.  The corresponding two-terminal $S_z$-conductance is also
quantized: 
\begin{align}
\label{nus}
\si_{S_z}= \nu_s\frac{(\hbar/2)^2}{h}=\nu_s \frac{\hbar}{8\pi},
\end{align}
with $\nu_s=4$.  For TL liquids of spin-1/2 electrons,
$c$ and $\nu_s$ are always integers, and they are always the same
\begin{align}
c=\nu_s.  
\end{align}
For the chiral metallic state \eq{u1su2}, $c=3$ and $\nu_s=4$.  Thus the
constructed chiral metallic state \eq{u1su2} is beyond TL liquids.  Note that
the central charges of some sectors are fractional.  Thus chiral metallic
state is a chiral ``non-Abelian'' metallic state.

\section{Ising phase transitions in metallic state of spin-$\frac12$ electron chain}
\label{IsingT}

In this section, we consider a  spin-$\frac12$ electron chain with
ferromagnetic or anti-ferromagnetic $S^z$-spin interactions.  The system has a
symmetry $U(1)\times Z_2^s\times \Z$.  We show that the $Z_2^s$ symmetry
breaking transitions for the two cases are described by different CFTs in the
metallic state, despite that the two transitions cause the identical symmetry
change, \ie reduce the symmetry group of the ground state from $U(1)\times
Z_2^s\times \Z$ to $U(1)\times \Z$.

\subsection{The model}

Let us first consider a spin-$\frac12$ chain with Ising
interaction
\begin{align}
\label{Ising}
 H = -J \sum_i \si^z_i \si^z_{i+1} - B \sum_i \si^x_i ,
\end{align}
where $B$ is the external magnetic field.  We then add some doping to obtain a
metallic state of spin-$\frac12$ electron chain. In this paper, we will mainly
consider the case when Fermi energy of the dropped electrons is much less than
$|J|,|B|$. In this case, the system is in $Z_2$ symmetry breaking phase when
$B=0$, with $\si^z=\pm 1$ (\ie all the electron either have $\si^z=+ 1$ or
$\si^z= -1$).  In the $Z_2$ symmetry breaking phase, the charge degree of
freedom remain gapless. The phase is described by $U(1)$ CFT with central
charge $c=\bar c=1$. The $Z_2$ symmetry breaking state has central charge
$c=\bar c=1$.  In the large $B$ limit, the system is in a $Z_2$ symmetric phase
where all the electrons have $\si^x=+1$.  The $Z_2$ symmetric state has central
charge $c=\bar c=1$.  We would like to consider the critical point of the $Z_2$
symmetry breaking transition.  The critical point of pure $Z_2$ symmetry
breaking has central charge $c=\bar c=\frac 12$. Plus the contribution from the
conserved $U(1)$ charge fluctuations, the critical point is expected to have a
total central charge $c=\bar c=\frac 32$.

The symmetry of a fermion system is described by a pair of groups:
$(Z_2^f,G_f)$. Here $G_f$ is the full symmetry group and $Z_2^f$ is generated
by fermion number parity, which is a central subgroup of $G_f$.  For our
spin-1/2 chain $G_f = U(1)\times Z_2^s$ and $Z_2^f$ is the subgroup of the
$U(1)$.

In general, to fully describe a critical theory with a global symmetry $G$, we
can consider the partition function twisted under the symmetry. More
specifically, a twisted partition function defined on a Euclidean spacetime
torus parametrized by a complex number $\tau$, is indexed by a pair of elements
$g,h$ of $G_f$,
\begin{align}
 Z_{g,h}(\tau) ,\ \ \ \ \ gh=hg,\ \ \ g,h \in G_f. 
\end{align}
It records all low energy excitations $\phi$ that satisfy twisted boundary conditions along spacial and temporal directions, $\phi (x+L,-\ii t)=g\phi (x,-\ii t), \phi (x,-\ii t +T)=h\phi (x,-\ii t)$, where $-\ii t$ denote the imaginary time.

If the symmetry $G_f$ is non-anomalous, the partition functions twisted under the symmetry satisfy the following relations:
\begin{align}
\label{ZpropG}
 Z_{h^{-1},g}(-1/\tau) &=  Z_{g,h}(\tau),
\nonumber\\
 Z_{g,hg}(\tau+1) &=  Z_{g,h}(\tau),
\nonumber\\
 Z_{ugu^{-1},uhu^{-1}}(\tau) &=  Z_{g,h}(\tau).
\end{align}
For example, for a fermionic system with only fermion-number-parity symmetry,
$G_f=Z_2^f$, the partition function depends on the boundary conditions along
temporal and spacial direction. Put it in plain words, we consider the
4-component partition function indexed by $g,h \in \{P,A\}$: $Z_{AP}(\tau)$,
$Z_{AA}(\tau)$, $Z_{PA}(\tau)$ and  $Z_{PP}(\tau)$, where $P$ and $A$
represents the periodic and anti-periodic boundary condition of a local
fermion.

\subsection{Partition functions}

Thus, for a CFT of a fermionic system, there are at least  four sectors of
partition functions defined as
\begin{align}
\begin{split}
 Z_{PE^f}(\tau) &=\Tr_{E} \ee^{-\Im(\tau) H_P -\ii \Re(\tau) K_P},\\
 Z_{PO^f}(\tau) &=\Tr_{O} \ee^{-\Im(\tau) H_P -\ii \Re(\tau) K_P},\\
 Z_{AE^f}(\tau) &=\Tr_{E} \ee^{-\Im(\tau) H_A -\ii \Re(\tau) K_A},\\
 Z_{AO^f}(\tau) &=\Tr_{O} \ee^{-\Im(\tau) H_A -\ii \Re(\tau) K_A}.
 \end{split}
\end{align}
where $\Tr_E$ is the trace over the states with even (total) numbers of
fermions and $\Tr_O$ is the trace over the states with odd numbers
of fermions.  $H_P$ ($H_A$) is the Hamiltonian for a system where fermion
fields satisfy an (anti-)periodic boundary condition in $x$ direction.
Similarly, $K_P$ ($K_A$) is the total momentum operator of the systems where
fermion fields satisfy a (anti-)periodic boundary condition in $x$ direction.

Alternatively, we may define the torus partition functions for fermion systems
through the space-time path integral, which also include four types,
$Z_{PP}(\tau)$, $Z_{PA}(\tau)$, $Z_{AP}(\tau)$, and $Z_{AA}(\tau)$.  Here the
first and second subscription $P$ or $A$ corresponds the periodic or
anti-periodic boundary condition for fermions in $x$ and $t$ direction,
respectively. The two sets of partition functions are related
\begin{align}
Z_{PE^f}=&\frac{1}{2}(Z_{PP}+Z_{PA}),\; Z_{PO^f}=-\frac{1}{2}(Z_{PP}-Z_{PA}),
\nonumber\\
Z_{AE^f}=&\frac{1}{2}(Z_{AP}+Z_{AA}),\; Z_{AO^f}=-\frac{1}{2}(Z_{AP}-Z_{AA}).
\end{align}
Each partition function can be expanded as 
\begin{align}
 Z(\tau)=q^{-\frac{c}{24}} (q^*)^{-\frac{\bar c}{24}}\sum_{(h,\bar h)} N_{h,\bar h} q^{h} (q^*)^{\bar h},
 \label{Zexpansion}
\end{align}
where $c,\ \bar c$ are the central charge for right and left movers,
\begin{align}
 q = \ee^{ -\ii \tau \frac{2\pi}{L}},
\end{align}
where $L$ is the size of the 1d system.  The summation $\sum_{(h,\bar h)}$ is
over a set of pairs $(h,\bar h)$, which gives rise to the spectrum of scaling
dimensions of local operators.  In particular, the expansion coefficients
$N_{h,\bar h}$ must be positive integers for each of $Z_{PE^f}(\tau)$,
$Z_{PO^f}(\tau)$, $Z_{AE^f}(\tau)$, and $Z_{AO^f}(\tau)$.

Unlike CFTs from bosonic lattice systems that have a modular invariant
partition function \eqn{Minv}, for a CFT realizable by a fermionic lattice
model, the above four types of partition functions transform covariantly under
modular transformations. More explicitly, under $S:\tau\rightarrow
-\frac{1}{\tau}$,
\begin{align}
\label{McovS}
\begin{split}
Z_{PP}\left(-\frac{1}{\tau}\right)=&Z_{PP}(\tau),\quad Z_{AA}\left(-\frac{1}{\tau}\right)=Z_{AA}(\tau),\\
Z_{AP}\left(-\frac{1}{\tau}\right)=&Z_{PA}(\tau),\quad Z_{PA}\left(-\frac{1}{\tau}\right)=Z_{AP}(\tau),
\end{split}
\end{align}
and under $T:\tau\rightarrow \tau+1$,
\begin{align}
\label{McovT}
\begin{split}
Z_{PP}(\tau+1)= &Z_{PP}(\tau),\quad Z_{AA}(\tau+1)= Z_{AP}(\tau),\\
Z_{AP}(\tau+1)= &Z_{AA}(\tau),\quad Z_{PA}(\tau+1)= Z_{PA}(\tau).
\end{split}
\end{align}

In the basis $(Z_{AE^f},Z_{PO^f},Z_{PE^f},Z_{AO^f})$, the
partition function transforms as\cite{JW190513279}
\begin{align}
 \label{Mcov}
Z_I(\tau+1) &= T_{IJ}^{Z_2^f}Z_J(\tau),\ \ \ \
Z_I(-1/\tau) = S_{IJ}^{Z_2^f}Z_J(\tau),
\end{align}
where $I,J  = AE^f,PO^f,PE^f,AO^f$ and
\begin{align}
\label{Z2STmat}
  T^{Z_2^f}&=
\begin{pmatrix}
    1&0&0&0\\
    0&1&0&0\\
    0&0&1&0\\
    0&0&0&-1
  \end{pmatrix} ,
&
  S^{Z_2^f}&=\frac12 \begin{pmatrix}
    1&1&1&1\\
    1&1&-1&-1\\
    1&-1&1&-1\\
    1&-1&-1&1
  \end{pmatrix}.
\end{align}

As a warm-up example, we consider the 1d charge-1 spinless non-interacting
fermions. The  4-component partition functions for a charge-1 spinless fermion
satisfy \eqn{Mcov} and are given by the characters of $u1_4$ CFT for right
movers near $k_F$, and by the characters of $\bar{u1}_4$ CFT for left movers
near $-k_F$ (see Appendix \ref{Mtrn}).  As a result, the 4-component partition
functions for a charge-1 spinless non-interacting fermion can be constructed
from the characters of $u1_4\oplus \bar{u1}_4$ CFT:
\begin{align}
\begin{split}
 Z_{AE^f} &= 
|\chi^{u1_4}_0 |^2 + |\chi^{u1_4}_2|^2,
  \\ 
Z_{PO^f} &= 
 \chi^{u1_4}_1 \bar \chi^{u1_4}_{-1} + \chi^{u1_4}_{-1} \bar \chi^{u1_4}_1,   \\ 
Z_{PE^f} &= 
|\chi^{u1_4}_1 |^2 + |\chi^{u1_4}_{-1}|^2,   \\ 
Z_{AO^f} &= 
 \chi^{u1_4}_0 \bar \chi^{u1_4}_2 + \chi^{u1_4}_2 \bar \chi^{u1_4}_0. 
\end{split}
\end{align}
Here, the primary field corresponding to $\chi_2^{u1_4}$
($\bar{\chi}_{2}^{u1_4}$) is the charge-1 right(left)-moving fermion.  We also
note that the right (left) movers near $k_F$ ($-k_F$) can be viewed as the edge
state for integer quantum Hall state with  filling fraction $\nu=1$ ($\nu=-1$).

To find modular covariant partition functions for the Ising critical point in
the spin-1/2 electron system, we use the characters of $u1_4$ CFT,
$\chi^{u1_4}_m$, and the characters of Ising CFT, $\chi^\text{Is}_h$, to
construct the 4-component partition functions that satisfy \eqn{Mcov} (details
shown in Appendix \ref{Mtrn}):
\begin{align}
\chi^{u1_M}_m(\tau), \quad& 0 \leq m < M=4,  
\nonumber\\
\chi^{\text{Is}}_h(\tau), \quad& h= 0, \frac12, \frac 1{16} 
\end{align}
Here the $u1$ CFT describes the gapless $U(1)$ charge fluctuations.  Also the
Ising CFT describes the gapless spin fluctuations at the Ising transition
point.  \Eqn{Mcov} can have many solutions. For example, the following
4-component partition functions represent a solution satisfy \eqn{Mcov}
\begin{align}
\label{Z2fafZ}
 Z_{AE} &= 
\big(|\chi^{u1_4}_0 |^2 + |\chi^{u1_4}_2|^2 \big) 
\big(|\chi^\text{Is}_0|^2+|\chi^\text{Is}_\frac{1}{2}|^2 \big)
\nonumber\\ &\ \ \ \
+\big(|\chi^{u1_4}_1 |^2 + |\chi^{u1_4}_{-1}|^2 \big) |\chi^\text{Is}_\frac{1}{16}|^2, 
 \nonumber \\ 
Z_{PO} &= 
\big( \chi^{u1_4}_0 \bar \chi^{u1_4}_2 + \chi^{u1_4}_2 \bar \chi^{u1_4}_0 \big)
 \big(\chi^\text{Is}_0 \bar \chi^\text{Is}_\frac{1}{2} + \chi^\text{Is}_\frac12 \bar \chi^\text{Is}_0\big)
\nonumber\\ &\ \ \ \
+
\big( \chi^{u1_4}_1 \bar \chi^{u1_4}_{-1} + \chi^{u1_4}_{-1} \bar \chi^{u1_4}_1 \big) |\chi^\text{Is}_\frac{1}{16}|^2,
 \nonumber \\ 
Z_{PE} &= 
\big(|\chi^{u1_4}_0 |^2 + |\chi^{u1_4}_2|^2 \big) 
|\chi^\text{Is}_\frac{1}{16}|^2 
\\ &\ \ \ \
+\big(|\chi^{u1_4}_1 |^2 + |\chi^{u1_4}_{-1}|^2 \big) 
\big(|\chi^\text{Is}_0|^2+|\chi^\text{Is}_\frac{1}{2}|^2 \big),
 \nonumber \\ 
Z_{AO} &= 
\big( \chi^{u1_4}_0 \bar \chi^{u1_4}_2 + \chi^{u1_4}_2 \bar \chi^{u1_4}_0 \big)
|\chi^\text{Is}_\frac{1}{16}|^2
\nonumber \\ &\ \ \ \
+
\big( \chi^{u1_4}_1 \bar \chi^{u1_4}_{-1} + \chi^{u1_4}_{-1} \bar \chi^{u1_4}_1 \big) 
 \big(\chi^\text{Is}_0 \bar \chi^\text{Is}_\frac{1}{2} + \chi^\text{Is}_\frac12 \bar \chi^\text{Is}_0\big).
\nonumber
\end{align}

In the above 4-component partition functions, we have considered the symmetry
twist and the quantum number of $Z_2^f$. To obtain more information, let us also consider the partition functions for the spin symmetry twist $Z_2^s$:
$Z_{PP}(\tau)$, $Z_{PA}(\tau)$, $Z_{AP}(\tau)$, and $Z_{AA}(\tau)$ which also
satisfy \eqn{McovT} and \eqn{McovS}.  We introduce $Z_{PE^s}(\tau)$,
$Z_{PO^s}(\tau)$, $Z_{AE^s}(\tau)$, and $Z_{AO^s}(\tau)$ in a similarly but
slightly different way
\begin{align}
Z_{PE^s}=&\frac{1}{2}(Z_{PP}+Z_{PA}),\; Z_{PO^s}=\frac{1}{2}(Z_{PP}-Z_{PA}),
\nonumber\\
Z_{AE^s}=&\frac{1}{2}(Z_{AP}+Z_{AA}),\; Z_{AO^s}=\frac{1}{2}(Z_{AP}-Z_{AA}),
\end{align}
where $Z_{PE^s}$ is the partition function in the $Z_2^s$ even sector, and $Z_{PO^s}$
is the partition function in the $Z_2^s$ odd sector.  Similarly, $Z_{AE^s}$ is
the partition function in the $Z_2^s$ even sector, and $Z_{AO^s}$ is the
partition function in the $Z_2^s$ odd sector, but now there is a $Z_2^s$
symmetry twist in the spatial direction.
In the basis $(Z_{PE^s},Z_{PO^s},Z_{AE^s},Z_{AO^s})$, the
partition function transforms as
\begin{align}
 \label{McovS}
Z_I(\tau+1) &= T_{IJ}^{Z_2^s}Z_J(\tau),\ \ \ \
Z_I(-1/\tau) = S_{IJ}^{Z_2^s}Z_J(\tau),
\end{align}
where $I,J  = PE^s,PO^s,AE^s,AO^s$ and
\begin{align}
\label{Z2sSTmat}
  T^{Z_2^s}&=
\begin{pmatrix}
    1&0&0&0\\
    0&1&0&0\\
    0&0&1&0\\
    0&0&0&-1
  \end{pmatrix} ,
&
  S^{Z_2^s}&=\frac12 \begin{pmatrix}
    1&1&1&1\\
    1&1&-1&-1\\
    1&-1&1&-1\\
    1&-1&-1&1
  \end{pmatrix},
\end{align}
which is the same as \eqn{Z2STmat}.

For example, the 4-component partition functions for the critical point of an
1d Ising model \eq{Ising} satisfy \eqn{McovS} and are given by the characters
of Ising CFT (see Appendix \ref{Mtrn}):
\begin{align}
\label{IsingZ}
\begin{split}
 Z_{PE^s} &= |\chi^\text{Is}_0|^2+|\chi^\text{Is}_\frac{1}{2}|^2, \\ 
Z_{PO^s} &= 
|\chi^\text{Is}_\frac{1}{16}|^2, \\ 
Z_{AE^s} &= 
|\chi^\text{Is}_\frac{1}{16}|^2, \\ 
Z_{AO^s} &= 
\chi^\text{Is}_0 \bar \chi^\text{Is}_\frac{1}{2} + \chi^\text{Is}_\frac12 \bar \chi^\text{Is}_0.
\end{split}
\end{align}

Now we like to include symmetry twists and the quantum numbers for both $Z_2^f$
and $Z_2^s$, which gives us the 16-component partition functions
$Z_{II'}(\tau)$, where $I  = AE^f,PO^f,PE^f,AO^f$.  and $I'  =
PE^s,PO^s,AE^s,AO^s$.  $Z_{II'}(\tau)$ satisfy the modular covariant condition
(see \Ref{JW190513279})
\begin{align}
 \label{Mcov1}
Z_{II'}(\tau+1) &= T_{II';JJ'}^{Z_2^f\times Z_2^s}Z_{JJ'}(\tau),
\nonumber\\
Z_{II'}(-1/\tau) &= S_{II',JJ'}^{Z_2^f\times Z_2^s}Z_{JJ'}(\tau),
\end{align}
where
\begin{align}
 T^{Z_2^f\times Z_2^s} &= T^{Z_2^f} \otimes T^{Z_2^s},
\nonumber\\
 S^{Z_2^f\times Z_2^s} &= S^{Z_2^f} \otimes S^{Z_2^s}.
\end{align}
\Eqn{Mcov1} has many solutions. The list of 36 solutions are given by Appendix
\ref{sols}.  But which one of the partition functions describe the Ising
transition of spin-1/2 electrons?

If the electron spins have a ferromagnetic interaction (\ie $J<0$ in
\eqn{Ising}), then we can view the doped holes as spinless fermions.  Thus, in
this case, we can view the Ising transition point as decoupled critical point
of Ising chain and the metalic state of spinless fermions.  Therefore, the
ferromagnetic Ising transition point of spin-1/2 electrons is described by the
following 16-component partition functions (see \eqn{sol1}):
\begin{align}
\label{sol1A}
 Z_{AE^f,PE^s} &= 
\big(|\chi^{u1_4}_0 |^2 + |\chi^{u1_4}_2|^2 \big) \big( |\chi^\text{Is}_0|^2+|\chi^\text{Is}_\frac{1}{2}|^2 \big),
  \\ 
Z_{PO^f,PE^s} &= 
\big( \chi^{u1_4}_1 \bar \chi^{u1_4}_{-1} + \chi^{u1_4}_{-1} \bar \chi^{u1_4}_1 \big) 
\big( |\chi^\text{Is}_0|^2+|\chi^\text{Is}_\frac{1}{2}|^2 \big),
 \nonumber \\ 
Z_{PE^f,PE^s} &= 
\big(|\chi^{u1_4}_1 |^2 + |\chi^{u1_4}_{-1}|^2 \big) 
\big( |\chi^\text{Is}_0|^2+|\chi^\text{Is}_\frac{1}{2}|^2 \big) ,
 \nonumber \\ 
Z_{AO^f,PE^s} &= 
\big( \chi^{u1_4}_0 \bar \chi^{u1_4}_2 + \chi^{u1_4}_2 \bar \chi^{u1_4}_0 \big) 
\big( |\chi^\text{Is}_0|^2+|\chi^\text{Is}_\frac{1}{2}|^2 \big)
\nonumber 
\end{align}
\begin{align}
\label{sol1B}
 Z_{AE^f,PO^s} &= 
\big(|\chi^{u1_4}_0 |^2 + |\chi^{u1_4}_2|^2 \big) 
|\chi^\text{Is}_\frac{1}{16}|^2,
  \\ 
Z_{PO^f,PO^s} &= 
\big( \chi^{u1_4}_1 \bar \chi^{u1_4}_{-1} + \chi^{u1_4}_{-1} \bar \chi^{u1_4}_1 \big) 
|\chi^\text{Is}_\frac{1}{16}|^2,
 \nonumber \\ 
Z_{PE^f,PO^s} &= 
\big(|\chi^{u1_4}_1 |^2 + |\chi^{u1_4}_{-1}|^2 \big) 
|\chi^\text{Is}_\frac{1}{16}|^2,
 \nonumber \\ 
Z_{AO^f,PO^s} &= 
\big( \chi^{u1_4}_0 \bar \chi^{u1_4}_2 + \chi^{u1_4}_2 \bar \chi^{u1_4}_0 \big) 
|\chi^\text{Is}_\frac{1}{16}|^2,
\nonumber 
\end{align}
\begin{align}
\label{sol1C}
 Z_{AE^f,AE^s} &= 
\big(|\chi^{u1_4}_0 |^2 + |\chi^{u1_4}_2|^2 \big) 
|\chi^\text{Is}_\frac{1}{16}|^2,
  \\ 
Z_{PO^f,AE^s} &= 
\big( \chi^{u1_4}_1 \bar \chi^{u1_4}_{-1} + \chi^{u1_4}_{-1} \bar \chi^{u1_4}_1 \big) 
|\chi^\text{Is}_\frac{1}{16}|^2,
 \nonumber \\ 
Z_{PE^f,AE^s} &= 
\big(|\chi^{u1_4}_1 |^2 + |\chi^{u1_4}_{-1}|^2 \big) 
|\chi^\text{Is}_\frac{1}{16}|^2,
 \nonumber \\ 
Z_{AO^f,AE^s} &= 
\big( \chi^{u1_4}_0 \bar \chi^{u1_4}_2 + \chi^{u1_4}_2 \bar \chi^{u1_4}_0 \big) 
|\chi^\text{Is}_\frac{1}{16}|^2,
\nonumber 
\end{align}
\begin{align}
\label{sol1D}
Z_{AE^f,AO^s} &= 
\big(|\chi^{u1_4}_0 |^2 + |\chi^{u1_4}_2|^2 \big) 
\big(\chi^\text{Is}_0 \bar \chi^\text{Is}_\frac{1}{2} + \chi^\text{Is}_\frac12 \bar \chi^\text{Is}_0\big),
  \\ 
Z_{PO^f,AO^s} &= 
\big( \chi^{u1_4}_1 \bar \chi^{u1_4}_{-1} + \chi^{u1_4}_{-1} \bar \chi^{u1_4}_1 \big) 
\big(\chi^\text{Is}_0 \bar \chi^\text{Is}_\frac{1}{2} + \chi^\text{Is}_\frac12 \bar \chi^\text{Is}_0\big),
 \nonumber \\ 
Z_{PE^f,AO^s} &= 
\big(|\chi^{u1_4}_1 |^2 + |\chi^{u1_4}_{-1}|^2 \big) 
\big(\chi^\text{Is}_0 \bar \chi^\text{Is}_\frac{1}{2} + \chi^\text{Is}_\frac12 \bar \chi^\text{Is}_0\big),
 \nonumber \\ 
Z_{AO^f,AO^s} &= 
\big( \chi^{u1_4}_0 \bar \chi^{u1_4}_2 + \chi^{u1_4}_2 \bar \chi^{u1_4}_0 \big) 
\big(\chi^\text{Is}_0 \bar \chi^\text{Is}_\frac{1}{2} + \chi^\text{Is}_\frac12 \bar \chi^\text{Is}_0\big).
\nonumber 
\end{align}
The above 16-component partition function is the multi-component partition
function mentioned in Section \ref{result}, which is a reflection of the
non-invertible gravitational anomaly if we restrict to the symmetric
sub-Hilbert space of the $Z_2^f\times Z_2^s$ symmetry.  The modular covariance
of the above multi-component partition function can help us to determine many
properties of the strongly correlated gapless state.  We remark that the above
16-component partition function only describe part of emergent non-invertible
gravitational anomaly (\ie part of emergent categorical symmetry), which is not
the maximal categorical symmetry.

We also note that the 16-component partition function reduces to the following
4-component partition function if we only consider the $Z_2^f$ symmetry twist:
\begin{align}\label{ferro}
 Z_{AE^f} &=  Z_{AE^f,PE^s} + Z_{AE^f,PO^s}
\nonumber\\
&=
\big(|\chi^{u1_4}_0 |^2 + |\chi^{u1_4}_2|^2 \big) Z_\text{Is} ,
 \nonumber \\ 
Z_{PO^f} &=  Z_{PO^f,PE^s} + Z_{PO^f,PO^s}
\nonumber\\
&=
\big( \chi^{u1_4}_1 \bar \chi^{u1_4}_{-1} + \chi^{u1_4}_{-1} \bar \chi^{u1_4}_1 \big) Z_\text{Is},
 \nonumber \\ 
Z_{PE^f} &=  Z_{PE^f,PE^s} + Z_{PE^f,PO^s}
\nonumber\\
&=
\big(|\chi^{u1_4}_1 |^2 + |\chi^{u1_4}_{-1}|^2 \big) Z_\text{Is} ,
 \nonumber \\ 
Z_{AO^f} &=  Z_{AO^f,PE^s} + Z_{AO^f,PO^s}
\nonumber\\
&=
\big( \chi^{u1_4}_0 \bar \chi^{u1_4}_2 + \chi^{u1_4}_2 \bar \chi^{u1_4}_0 \big) Z_\text{Is}, 
\end{align}
where
\begin{align}
 Z_\text{Is} = |\chi^\text{Is}_0|^2+|\chi^\text{Is}_\frac{1}{2}|^2 + |\chi^\text{Is}_\frac{1}{16}|^2.
\end{align}

When the electron spins have an anti-ferromagnetic interaction (\ie $J>0$ in
\eqn{Ising}),  the Ising transition point will be described by a different CFT.
This is because when there are an odd number of electrons on the ring, the spins
carried by the electrons will behave like those in a spin chain with a $Z_2^s$ symmetry
twist.  In other words, a state with an odd number of fermions is like a Neel ordered Ising spin configuration with an
odd number of spins, thus satisfying anti-periodic boundary condition. 

It means that in the partition functions whose first label is $AO^f$ or $PO^f$
(\ie with odd number of electrons), the (untwisted) spin part of the
excitations (the sectors labeled by $PO^s$ and $PE^s$) is given by the $Z_2^s$
twisted sector of Ising CFT. In specific,if the second label is $PO^s$, the
spin part is described by Ising character $\chi^\text{Is}_0 \bar
\chi^\text{Is}_\frac{1}{2} + \chi^\text{Is}_\frac12 \bar \chi^\text{Is}_0$
(which is $Z_{AO^s}$ shown in (\ref{IsingZ}));  if the second label is $PE^s$,
it is described by Ising character $|\chi^\text{Is}_\frac{1}{16}|^2$ (which is
$Z_{AE^s}$ shown in (\ref{IsingZ})). Still in the partition functions whose
first label is $AO^f$ or $PO^f$, the $\ZZ_2$ twisted spin part is given by the
$Z_2^s$ untwisted sector of Ising CFT. In summary, the partition functions,
whose first labels are $AO^f/PO^f$, are as follows,
\begin{align}
Z^\text{Is}_{AO^f/PO^f,PE^s}=&\left|\chi^\text{Is}_{\frac{1}{16}}\right|^2, \nn\\
Z^\text{Is}_{AO^f/PO^f,PO^s}=&\chi^\text{Is}_0\bar{\chi}^\text{Is}_{\frac{1}{2}}+\chi^\text{Is}_\frac{1}{2}\bar{\chi}^\text{Is}_0,\nn\\
Z^\text{Is}_{AO^f/PO^f,AE^s}=&\left|\chi^\text{Is}_0\right|^2+\left|\chi^\text{Is}_\frac{1}{2}\right|^2,\nn\\
Z^\text{Is}_{AO^f/PO^f,AO^s}=&\left|\chi^\text{Is}_{\frac{1}{16}}\right|^2.
\end{align}

Also, in the partition functions with first label $AE^f$ or $PE^f$ (\ie with even
number of electrons), the (untwisted) spin part is given by untwisted sector of Ising CFT, the $\ZZ_2$ twisted spin part is given by $Z_2^s$ twisted sector of Ising CFT. 
\begin{align}
Z^\text{Is}_{AE^f/PE^f,PE^s}=&\left|\chi^\text{Is}_0\right|^2+\left|\chi^\text{Is}_\frac{1}{2}\right|^2,\nn\\
Z^\text{Is}_{AE^f/PE^f,PO^s}=&\left|\chi^\text{Is}_{\frac{1}{16}}\right|^2, \nn\\
Z^\text{Is}_{AE^f/PE^f,AE^s}=&\left|\chi^\text{Is}_{\frac{1}{16}}\right|^2, \nn\\
Z^\text{Is}_{AE^f/PE^f,AO^s}=&\chi^\text{Is}_0\bar{\chi}^\text{Is}_{\frac{1}{2}}+\chi^\text{Is}_\frac{1}{2}\bar{\chi}^\text{Is}_0.
\end{align}

Furthermore, since a fermion always carries an odd number of the $U(1)$ charge, the
partition functions labeled by $AO^f$ and $PO^f$ (\ie with odd number of
electrons) must be described by $u1$ character $\chi^{u1_4}_m \bar
\chi^{u1_4}_n$ with $m-n=2$ mod 4.  We find the partition functions
\eqn{sol13} satisfy the above conditions. Thus, the anti-ferromagnetic Ising
transition point of spin-1/2 electrons is described by the following
16-component partition functions (see \eqn{sol13}):
% G13 
\begin{align}
\label{sol13A}
 Z_{AE^f,PE^s} &= \big(|\chi^{u1_4}_0 |^2 + |\chi^{u1_4}_2|^2 \big)\big( |\chi^\text{Is}_0|^2+|\chi^\text{Is}_\frac{1}{2}|^2 \big),
 \nonumber \\ 
Z_{PO^f,PE^s} &= \big( \chi^{u1_4}_1 \bar \chi^{u1_4}_{-1} + \chi^{u1_4}_{-1} \bar \chi^{u1_4}_1 \big)|\chi^\text{Is}_\frac{1}{16}|^2, 
 \\ 
Z_{PE^f,PE^s} &= \big(|\chi^{u1_4}_1 |^2 + |\chi^{u1_4}_{-1}|^2 \big)\big( |\chi^\text{Is}_0|^2+|\chi^\text{Is}_\frac{1}{2}|^2 \big), 
 \nonumber \\ 
Z_{AO^f,PE^s} &= \big( \chi^{u1_4}_0 \bar \chi^{u1_4}_2 + \chi^{u1_4}_2 \bar \chi^{u1_4}_0 \big)|\chi^\text{Is}_\frac{1}{16}|^2,
 \nonumber 
\end{align}
\begin{align}
\label{sol13B}
Z_{AE^f,PO^s} &= \big(|\chi^{u1_4}_1 |^2 + |\chi^{u1_4}_{-1}|^2 \big)|\chi^\text{Is}_\frac{1}{16}|^2, 
 \\ 
Z_{PO^f,PO^s} &= \big( \chi^{u1_4}_0 \bar \chi^{u1_4}_2 + \chi^{u1_4}_2 \bar \chi^{u1_4}_0 \big)\big(\chi^\text{Is}_0 \bar \chi^\text{Is}_\frac{1}{2} + \chi^\text{Is}_\frac12 \bar \chi^\text{Is}_0\big), 
 \nonumber \\ 
Z_{PE^f,PO^s} &= \big(|\chi^{u1_4}_0 |^2 + |\chi^{u1_4}_2|^2 \big)|\chi^\text{Is}_\frac{1}{16}|^2, 
 \nonumber \\ 
Z_{AO^f,PO^s} &= \big( \chi^{u1_4}_1 \bar \chi^{u1_4}_{-1} + \chi^{u1_4}_{-1} \bar \chi^{u1_4}_1 \big)\big(\chi^\text{Is}_0 \bar \chi^\text{Is}_\frac{1}{2} + \chi^\text{Is}_\frac12 \bar \chi^\text{Is}_0\big), 
 \nonumber 
\end{align}
\begin{align}
\label{sol13C}
Z_{AE^f,AE^s} &= \big(|\chi^{u1_4}_0 |^2 + |\chi^{u1_4}_2|^2 \big)|\chi^\text{Is}_\frac{1}{16}|^2, 
\\ 
Z_{PO^f,AE^s} &= \big( \chi^{u1_4}_1 \bar \chi^{u1_4}_{-1} + \chi^{u1_4}_{-1} \bar \chi^{u1_4}_1 \big)\big( |\chi^\text{Is}_0|^2+|\chi^\text{Is}_\frac{1}{2}|^2 \big),
 \nonumber \\ 
Z_{PE^f,AE^s} &= \big(|\chi^{u1_4}_1 |^2 + |\chi^{u1_4}_{-1}|^2 \big)|\chi^\text{Is}_\frac{1}{16}|^2, 
 \nonumber \\ 
Z_{AO^f,AE^s} &= \big( \chi^{u1_4}_0 \bar \chi^{u1_4}_2 + \chi^{u1_4}_2 \bar \chi^{u1_4}_0 \big)\big( |\chi^\text{Is}_0|^2+|\chi^\text{Is}_\frac{1}{2}|^2 \big),
 \nonumber 
\end{align}
\begin{align}
\label{sol13D}
Z_{AE^f,AO^s} &= \big(|\chi^{u1_4}_1 |^2 + |\chi^{u1_4}_{-1}|^2 \big)\big(\chi^\text{Is}_0 \bar \chi^\text{Is}_\frac{1}{2} + \chi^\text{Is}_\frac12 \bar \chi^\text{Is}_0\big), 
 \nonumber \\ 
Z_{PO^f,AO^s} &= \big( \chi^{u1_4}_0 \bar \chi^{u1_4}_2 + \chi^{u1_4}_2 \bar \chi^{u1_4}_0 \big)|\chi^\text{Is}_\frac{1}{16}|^2, 
  \\ 
Z_{PE^f,AO^s} &= \big(|\chi^{u1_4}_0 |^2 + |\chi^{u1_4}_2|^2 \big)\big(\chi^\text{Is}_0 \bar \chi^\text{Is}_\frac{1}{2} + \chi^\text{Is}_\frac12 \bar \chi^\text{Is}_0\big), 
 \nonumber \\ 
Z_{AO^f,AO^s} &= \big( \chi^{u1_4}_1 \bar \chi^{u1_4}_{-1} + \chi^{u1_4}_{-1} \bar \chi^{u1_4}_1 \big)|\chi^\text{Is}_\frac{1}{16}|^2. 
\nonumber 
\end{align}
The above 16-component partition functions reduce to the 4-component partition
functions given in \eqn{Z2fafZ}, if we only consider the $Z_2^f$ symmetry twist.

\subsection{Scaling operator and their quantum numbers}

\begin{table}[tb]
 \def\arraystretch{1.5}
\centering
\begin{tabular}{|c|c|c|c|}
\hline
operators & $\si^x$ & $k$ & $h,\bar h$\\
\hline 
$ \psi\bar\psi $ & 1 & $0$ & $\frac12,\frac12$ \\
\hline 
$ \si\bar\si$ & $-1$ & $0$ & $\frac1{16},\frac1{16}$ \\
\hline 
\hline 
$ \psi$ & $-1$ & $0$ & $\frac12,0$ \\
\hline 
$ \bar \psi$ & $-1$ & $0$ & $0,\frac12$ \\
\hline 
$ \si\bar\si\psi\sim \si\bar\si\bar\psi $ & $1$ & $0$ & $\frac1{16},\frac1{16}$ \\
\hline 
\end{tabular}
\caption{Quantum numbers of local and non-local operators in critical point of
ferromagnetic Ising model \eqn{Ising}.  Here $\si^x$ is the $Z_2$ spin quantum
number, $k$ is the crystal momentum, and $(h,\bar h)$ are the right and left
scaling dimension.  $\psi, \si$ are the $\text{Is}$ CFT primary fields
associated with the Ising character
$\chi^\text{Is}_\frac12,\chi^\text{Is}_\frac1{16}$, which have scaling
dimensions $\frac12$ and $\frac1{16}$ respectively.  Similarly, $\bar \psi,
\bar \si$ are the $\overline{\text{Is}}$ CFT fields.
}
\label{IsFM}
\end{table}

\begin{table}[tb]
 \def\arraystretch{1.5}
\centering
\begin{tabular}{|c|c|c|c|}
\hline
operators & $\si^x$ & $k$ & $h,\bar h$\\
\hline 
$ \psi\bar\psi $ & 1 & $0$ & $\frac12,\frac12$ \\
\hline 
$ \si\bar\si$ & $-1$ & $\pm \frac{\pi}{a}$ & $\frac1{16},\frac1{16}$ \\
\hline 
\hline 
$ \psi$ & $-1$ & $\pm \frac{\pi}{a}$ & $\frac12,0$ \\
\hline 
$ \bar \psi$ & $-1$ & $\pm \frac{\pi}{a}$ & $0,\frac12$ \\
\hline 
$ \si\bar\si\psi\sim\si\bar\si\bar\psi $ & $1$ & 0 & $\frac1{16},\frac1{16}$ \\
\hline 
\end{tabular}
\caption{Quantum numbers of local and non-local operators in critical point of
anti-ferromagnetic Ising model \eqn{Ising}.  
}
\label{IsAF}
\end{table}

Let us first consider the scaling operators and their quantum numbers of the
critical point (\ref{IsingZ}) of the Ising model \eqn{Ising} without doping.  The
partition functions \eqn{IsingZ} tell us the $Z_2^s$ quantum numbers.  For
ferromagnetic spin coupling $(J<0$ in \eqn{Ising}), the low energy states all
carry crystal momentum near zero.  The states described by
$|\chi^\text{Is}_\frac12|^2$ in $Z_{PE^s}$ are created by local operator
$\psi\bar\psi$ from the ground state in $|\chi^\text{Is}_0|^2$.  Thus the
operator $\psi\bar\psi$ carries $Z_2^s$ quantum number $\si^x=1$.  The states
described by $|\chi^\text{Is}_\frac1{16}|^2$ in $Z_{PO^s}$ are created by local
operator $\si\bar\si$ from the ground state.  Thus the operator $\si\bar\si$
carries $Z_2^s$ quantum number $\si^x=-1$.  The states described by
$\chi^\text{Is}_\frac12\bar\chi^\text{Is}_0$ in $Z_{AO^s}$ are created by
non-local operator $\psi$ from the ground state.  Thus the non-local operator $\psi$
carries $Z_2^s$ quantum number $\si^x=1$.  Similarly,  the non-local operator
$\bar\psi$ also carries $Z_2^s$ quantum number $\si^x=1$.  The states described
by $|\chi^\text{Is}_\frac1{16}|^2$ in $Z_{AE^s}$ are created by non-local
operator $\si\bar\si \psi \sim \si\bar\si \bar\psi$ from the ground state.
Thus the operator $\si\bar\si \psi \sim \si\bar\si \bar\psi$ carries $Z_2^s$
quantum number $\si^x=1$.  The above results are summarized in Table
\ref{IsFM}.

However, for anti-ferromagnetic spin coupling $(J>0$ in \eqn{Ising}), the low
energy states carry crystal momentum near $k=\pm\frac{\pi}{a}$ if the $Z_2^s$
quantum number $\si^x=-1$ (and carry crystal momentum near zero if the $Z_2^s$
quantum number $\si^x=1$).  The scaling operators and their quantum numbers for
anti-ferromagnetic Ising critical point are summarized in Table \ref{IsAF}.

\begin{table}[tb]
\centering
 \def\arraystretch{1.5}
\begin{tabular}{|c|c|c|c|cc|}
\hline
operators & $\si^x$ & $q$ & $k$ & $h,\bar h$ & ($\th=0$)\\
\hline 
$ \ee^{\pm\ii (\vphi+\bar \vphi)} $ & 1 & $0$ & $\pm 2k_F$ & $\frac{(\ch\th-\sh\th)^2}{2},\frac{(\ch\th-\sh\th)^2}{2}$ & ($\frac12,\frac12$) \\
\hline 
$ \ee^{\pm\ii (\vphi-\bar \vphi)} $ & 1 & $\pm 2$ & $ 0$ & $\frac{(\ch\th+\sh\th)^2}{2},\frac{(\ch\th+\sh\th)^2}{2}$ & ($\frac12,\frac12$) \\
\hline 
$ \psi\bar\psi $ & 1 & $0$ & $0$ & $\frac12,\frac12$ & ($\frac12,\frac12$)\\
\hline 
$ \si\bar\si$ & $-1$ & $0$ & $0$ & $\frac1{16},\frac1{16}$ & ($\frac1{16},\frac1{16}$)\\
\hline 
\hline 
$ \ee^{\pm \ii \vphi} $ & $1$ & $\pm 1$ & $\pm k_F$ & $\frac{\ch^2\th}{2},\frac{\sh^2\th}{2}$ & ($\frac12,0$) \\
\hline 
$ \ee^{\pm \ii \bar \vphi} $ & $1$ & $\mp 1$ & $\pm k_F$ & $\frac{\sh^2\th}{2},\frac{\ch^2\th}{2}$ & ($0,\frac12$) \\
\hline 
$ \ee^{\pm \ii \vphi}\si\bar\si $ & $-1$ & $\pm 1$ & $\pm k_F$ & $\frac{\ch^2\th}{2}+\frac1{16},\frac{\sh^2\th}{2}+\frac1{16}$ & ($\frac7{16},\frac1{16}$) \\
\hline 
$ \ee^{\pm \ii \bar \vphi} \si\bar\si $ & $-1$ & $\mp 1$ & $\pm k_F$ & $\frac{\sh^2\th}{2}+\frac1{16},\frac{\ch^2\th}{2}+\frac1{16}$ & ($\frac1{16},\frac7{16}$) \\
\hline 
\end{tabular}
\caption{Quantum numbers of local gapless bosonic and fermionic operators in
ferromagnetic Ising transition point of the strongly interacting spin-1/2
electron system (the doped ferromagnetic Ising model).  Here $\si^x$ is the
$Z_2$ spin quantum number, $q$ is the $U(1)$ charge, $k$ is the crystal
momentum, and $(h,\bar h)$ are the right and left scaling dimensions (the
values in bracket are for $\th=0$, see \eqn{hhbar}).  $\varphi$ is the bosonic
field to describe ${u1_4}$ CFT, where $\varphi$ is normalized such that
$\ee^{\ii \vphi}$ has a scaling dimension $\frac12$.  $\psi, \si$ are the
$\text{Is}$ CFT fields with scaling dimension $\frac12$ and $\frac1{16}$
respectively.  Similarly, $\bar \varphi$ is the bosonic field to describe
$\overline{u1}_4$ CFT and $\bar \psi, \bar \si$ are the $\overline{\text{Is}}$
CFT fields.
}
\label{u1IsFM}
\end{table}

Now let us consider the scaling operators and their quantum numbers for the
spin-1/2 electrons at the Ising transition point.  The partition functions
\eqn{sol1A}-(\ref{sol1D}) and \eqn{sol13A}-(\ref{sol13D}) tell us the $Z_2^f$ and $Z_2^s$ quantum numbers.
In the following, we will discuss the $U(1)$ and momentum quantum numbers. 

Let us first consider the ferromagnetic Ising transition point described by
\eqn{sol1A}-(\ref{sol1D}).  The $u1_4$ character
$\chi^{u1_4}_m$ describes states with $U(1)$ charge $q=\frac m 2$ mod 2, and
momentum $k=\frac {k_F} 2$ mod $2k_F$.  Here $k_F = \pi n_F$, where $n_F$ is
the fermion number per site.  The $\overline{u1}_4$ character $\bar
\chi^{u1_4}_m$ describes states with $U(1)$ charge $q=-\frac m 2$ mod 2, and
momentum $k=\frac {k_F} 2$ mod $2k_F$.  For such $U(1)$ charge assignment, we
see that the states described by the partition function $Z_{AE^f,\cdots}$
($Z_{AO^f,\cdots}$) carry even (odd) $U(1)$ charges.  The states described by
the Ising character do not carry any $U(1)$ charge and momentum.

The states described by the partition function $Z_{AE^f,\cdots}$
($Z_{AO^f,\cdots}$) are created by local gapless bosonic (fermionic) operators from
the ground state in the sector $|\chi^{u1_4}_0 |^2 |\chi^\text{Is}_0|^2 $.  So
the above discussion gives us a list of scaling operators, as well as their
quantum numbers and scaling dimensions.  The results are summarized in Table
\ref{u1IsFM}.  For example (see \eqn{sol1A}), the bosonic operator
$\ee^{\pm\ii (\vphi \pm \bar \vphi)}$ create the states in $ |\chi^{u1_4}_2|^2
|\chi^\text{Is}_0|^2 $.  The local fermionic operator $\ee^{\pm\ii \vphi }$
create the states in $ \chi^{u1_4}_2 \bar \chi^{u1_4}_0 |\chi^\text{Is}_0|^2
$.

From Table \ref{u1IsFM}, we see that there is only one relevant operator that
carries a trivial quantum number, $\bar \psi\psi$, with total scaling dimension
$h+\bar h=1$.  This is the operator that drives the ferromagnetic Ising
transition.

\begin{table}[tb]
 \def\arraystretch{1.5}
\centering
\begin{tabular}{|c|c|c|c|c|}
\hline
operators & $\si^x$ & $q$ & $k$ & $h,\bar h$\\
\hline 
$ \ee^{\pm\ii (\vphi+\bar \vphi)} $ & 1 & $0$ & $\pm 2k_F$ &  $\frac{(\ch\th-\sh\th)^2}{2},\frac{(\ch\th-\sh\th)^2}{2}$ \\
\hline 
$ \ee^{\pm\ii (\vphi-\bar \vphi)} $ & 1 & $\pm 2$ & $ 0$ &  $\frac{(\ch\th+\sh\th)^2}{2},\frac{(\ch\th+\sh\th)^2}{2}$ \\
\hline 
$ \psi\bar\psi $ & 1 & $0$ & $0$ & $\frac12,\frac12$ \\
\hline 
$ \ee^{\pm\ii \frac{\vphi+\bar \vphi}{2}} \si\bar\si$ & $-1$ & $0$ & $\pm k_F$ & $\frac{2(\ch\th-\sh\th)^2+1}{16},\frac{2(\ch\th-\sh\th)^2+1}{16}$ \\
\hline 
\hline 
$ \ee^{\pm \ii \vphi} \si\bar\si\psi$ & $ 1$ & $\pm 1$ & $\pm k_F$ & $\frac{\ch^2\th}{2}+\frac1{16},\frac{\sh^2\th}{2}+\frac1{16}$  \\
\hline 
$ \ee^{\pm \ii \bar \vphi} \si\bar\si\psi$ & $ 1$ & $\mp 1$ & $\pm k_F$ & $\frac{\sh^2\th}{2}+\frac1{16},\frac{\ch^2\th}{2}+\frac1{16}$ \\
\hline 
$ \ee^{\pm \ii \frac{\vphi-\bar \vphi}{2} } \psi$ & $-1$ & $\pm 1$ & $\pm k_F$ & $\frac{(\ch\th+\sh\th)^2+4}{8},\frac{(\ch\th+\sh\th)^2}{8}$  \\
\hline 
$ \ee^{\pm \ii \frac{\vphi-\bar \vphi}{2} } \bar \psi$ & $-1$ & $\pm 1$ & $\pm k_F$ & $\frac{(\ch\th+\sh\th)^2}{8},\frac{(\ch\th+\sh\th)^2+4}{8}$ \\
\hline 
\end{tabular}
\caption{
Quantum numbers of local gapless bosonic and fermionic operators in anti-ferromagnetic Ising
transition point of the strongly interacting spin-1/2 electron system (the
doped anti-ferromagnetic Ising model).  
}
\label{u1IsAF}
\end{table}

Next, let us consider the anti-ferromagnetic Ising transition point described
by \eqn{sol13A}-(\ref{sol13D}).  The $u1_4$
character $\chi^{u1_4}_m$ still describes states with $U(1)$ charge $q=\frac m
2$ mod 2, and momentum $k=\frac {k_F} 2$ mod $2k_F$.  The $\overline{u1}_4$
character $\bar \chi^{u1_4}_m$ still describes states with $U(1)$ charge
$q=-\frac m 2$ mod 2, and momentum $k=\frac {k_F} 2$ mod $2k_F$.  For such
$U(1)$ charge assignment, again the states described by the partition function
$Z_{AE^f,\cdots}$ ($Z_{AO^f,\cdots}$) carry even (odd) $U(1)$ charges.  The
states described by the Ising character do not carry any $U(1)$ charge.  But
they can carry momentum $\pm k_F$ if $\si_x=-1$.  The results are summarized
in Table \ref{u1IsAF}.  For example (see \eqn{sol13A}), the local gapless bosonic
operator $\ee^{\pm\ii (\vphi \pm \bar \vphi)}$ creates the states in $
|\chi^{u1_4}_2|^2 |\chi^\text{Is}_0|^2 $.  The local fermionic operator
$\ee^{\pm\ii \vphi } \si\bar\si\psi \sim \ee^{\pm\ii \vphi }
\si\bar\si\bar\psi $ creates the states in $ \chi^{u1_4}_2 \bar \chi^{u1_4}_0
|\chi^\text{Is}_\frac1{16}|^2 $.

From Table \ref{u1IsAF}, we see that there is only one relevant operator that
carries trivial quantum numbers, $\bar \psi\psi$, with total scaling dimension
$h+\bar h=1$.  This is the operator that drives the anti-ferromegnetic Ising
transition.

\begin{figure}[tb]
\centering
\includegraphics[scale=0.8]{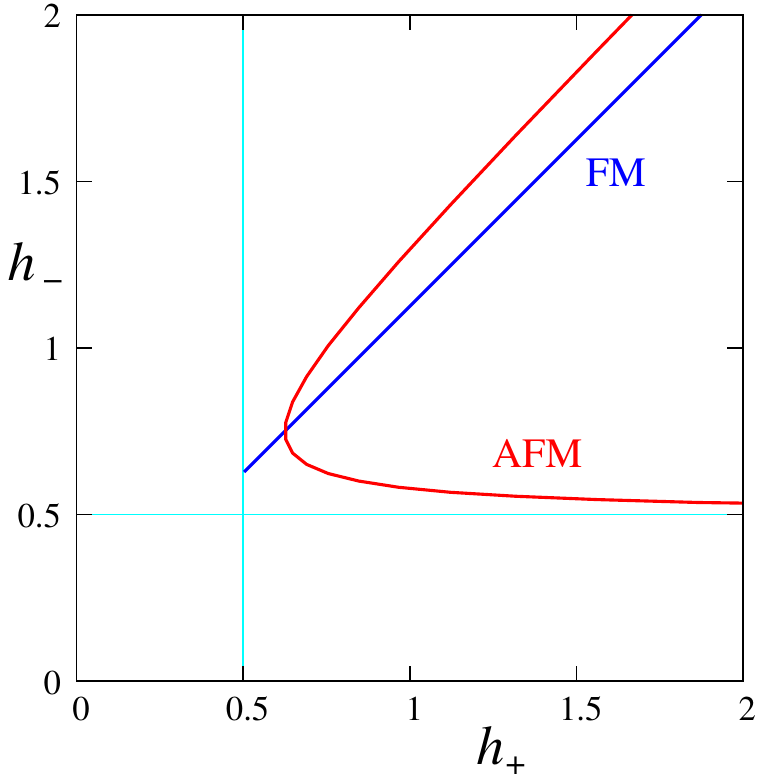}
%2
\caption{
The relations between the total scaling dimensions $h_+$ and $h_-$ of the
electron operators with $Z_2^s$ quantum number $\si^x=1$ and $\si^x=-1$, respectively, for the
ferromegnetic (FM) and anti-ferromegnetic (AFM) Ising transitions.
}\label{hpm}
\end{figure}

\subsection{Low energy effective theory}

Let us further compare the ferromagnetic and anti-ferromagnetic Ising
transition for the spin-$1/2$ electrons when there is interaction.  Both the
ferromagnetic and anti-ferromagnetic Ising transition points are described by
the same low energy effective field theory
\begin{align}
 \cL & =
 \frac1{4\pi} ( \prt_x \vphi \prt_t \vphi - \prt_x \vphi \prt_x \vphi - \prt_x \bar\vphi \prt_t \bar\vphi -  \prt_x \bar\vphi \prt_x \bar\vphi )
\nonumber\\
&
- \frac1{2\pi} V  \prt_x \vphi \prt_x \bar\vphi 
+ \psi (\prt_t - \prt_x) \psi + \bar \psi (\prt_t + \prt_x) \bar\psi.
\end{align}
However, the set of local operators are different for the two Ising transition
points.  For the ferromagnetic Ising transition point, the local operators are
given in Table \ref{u1IsFM}. While for the anti-ferromagnetic Ising transition
point, the local operators are given in Table \ref{u1IsAF}.  In the tables,
$\si(x)$ ($\bar\si(x)$) operator is the operator that creates the sign flip at
$x$ in the $\psi$ ($\bar\psi$) field.  

In the last section, we study the case with $V=0$. And the $U(1)$ charge
fluctuations are described by $u1_4\oplus \overline{u1}_4$ CFT.  Here we will
consider the effect of $V$ on the scaling dimensions $h,\bar h$.  Let us
introduce
\begin{align}
 \begin{pmatrix}
 \phi\\
\bar \phi\\
\end{pmatrix}
=&
\begin{pmatrix}
 \ch \th & \sh \th \\
 \sh \th & \ch \th \\
\end{pmatrix}
 \begin{pmatrix}
 \vphi\\
\bar \vphi\\
\end{pmatrix}
,
\nonumber\\
 \begin{pmatrix}
 \vphi\\
\bar \vphi\\
\end{pmatrix}
=&
\begin{pmatrix}
 \ch \th & -\sh \th \\
 -\sh \th & \ch \th \\
\end{pmatrix}
 \begin{pmatrix}
 \phi\\
\bar \phi\\
\end{pmatrix},
\end{align}
with $\th$ satisfying
\begin{align}
 V =\frac{2\ch\th\sh\th}{\ch^2\th+\sh^2\th}.
\end{align}
The Lagrangian for $\phi,\bar\phi$ is diagonal
\begin{align}
 \cL & =
 \frac1{4\pi} ( \prt_x \phi \prt_t \phi - v \prt_x \phi \prt_x \phi 
- \prt_x \bar\phi \prt_t \bar\phi -  v \prt_x \bar\phi \prt_x \bar\phi ).
\end{align}
Thus the scaling dimensions $h,\bar h$
for operator 
\begin{align}
\ee^{\ii (m \vphi +\bar m\,\bar \vphi)}
=
\ee^{\ii [m (\ch \th\, \phi -\sh \th\, \bar \phi) +\bar m (\ch \th\,\bar \phi -\sh \th \, \phi)]}
\end{align}
are given by
\begin{align}
\label{hhbar}
 h (\theta) &= \frac {(m\, \ch \th -\bar m\, \sh\th)^2 } 2 ,\ \ \
 \bar h (\theta) = \frac {(\bar m \,\ch \th - m\, \sh\th)^2 } 2 .
\end{align}

From Tables \ref{u1IsFM} and \ref{u1IsAF}, we see that the charge neutral
operator with $\si^x=-1$ has scaling dimensions $\frac{1}{16},\frac{1}{16}$ and
$\frac{2(\ch\th-\sh\th)^2+1}{16},\frac{2(\ch\th-\sh\th)^2+1}{16}$ for the
ferromagnetic and the anti-ferromagnetic critical points respectively.  The
scaling dimensions for the  anti-ferromagnetic critical points are always
larger then $\frac{1}{16},\frac{1}{16}$.  So the  the ferromagnetic and the
anti-ferromagnetic critical points are really distinct critical points, despite
they describe idential symmetry breaking pattern.

Next we compare the ferromagnetic  and anti-ferromagnetic Ising transition for
the spin-1/2 electrons by considering the total scaling dimension
$h_+(\th)=h+\bar h$ for the electron operator with $Z_2^s$ quantum number
$\si^x=1$ and $h_-(\th)=h+\bar h$ for the electron operator with $Z_2^s$
quantum number $\si^x=-1$.  As a function of interaction $\th$, $h_+(\th)$ and
$h_-(\th)$ have different relations for the ferromegnetic  and
anti-ferromegnetic Ising transitions, as shown in Fig. \ref{hpm}.  For example,
in the ferromagnetic transition, the $Z_2^s$ even and odd electron operator can
be $\ee^{\pm \ii \varphi}$ and $\ee^{\pm \ii \varphi}\sigma\bar\sigma$,
respectively. And in the antiferromagnetic transition, they can be $\ee^{\pm
\ii \varphi}\sigma\bar\sigma\psi$ and $\ee^{\pm \ii
\frac{\varphi-\bar\varphi}{2}}\psi$ respectively. From Fig. \ref{hpm}, we see
that in the ferromagnetic case, the scaling dimension of the $\Z_2^s$ odd
electron operator is always larger than that of $\Z_2^s$ one by $\frac{1}{8}$
that is independent of interaction. However, in the anti-ferromagnetic case,
the difference in the scaling dimension of $Z_2^s$ odd and $Z_2^s$ even
operator increases with the attractive interacting strength ($\theta >0$), and
decreases with the replusive interacting strength ($\theta<0$), comparing to
the non-interacting case $\theta=0$.

\subsection{Two metallic phases of spin-$\frac12$ electron chain with the same symmetry}

\begin{table}[tb]
 \def\arraystretch{1.6}
\centering
\begin{tabular}{|c|c|c|cc|}
\hline
operators & $q$ & $k$ & $h,\bar h$ & ($\th=0$)\\
\hline 
$ \ee^{\pm\ii (\vphi+\bar \vphi)} $ & $0$ & $\pm 2k_F$ & $\frac{(\ch\th-\sh\th)^2}{2},\frac{(\ch\th-\sh\th)^2}{2}$ & ($\frac12,\frac12$) \\
\hline 
$ \ee^{\pm\ii (\vphi-\bar \vphi)} $ & $\pm 2$ & $ 0$ & $\frac{(\ch\th+\sh\th)^2}{2},\frac{(\ch\th+\sh\th)^2}{2}$ & ($\frac12,\frac12$) \\
\hline 
\hline 
$ \ee^{\pm \ii \vphi} $ & $\pm 1$ & $\pm k_F$ & $\frac{\ch^2\th}{2},\frac{\sh^2\th}{2}$ & ($\frac12,0$) \\
\hline 
$ \ee^{\pm \ii \bar \vphi} $ & $\mp 1$ & $\pm k_F$ & $\frac{\sh^2\th}{2},\frac{\ch^2\th}{2}$ & ($0,\frac12$) \\
\hline 
\end{tabular}
\caption{Quantum numbers of local gapless bosonic and fermionic operators in metallic
phase of spin-1/2 electrons with strong ferromagnetic Ising interaction.  Here,
$q$ is the $U(1)$ charge, $k$ is the crystal momentum, and $(h,\bar h)$ are the
right and left scaling dimensions (the values in bracket are for $\th=0$, see
\eqn{hhbar}).  $\varphi$ is the bosonic field to describe ${u1_4}$ CFT.
Similarly, $\bar \varphi$ is the bosonic field to describe $\overline{u1}_4$
CFT.
}
\label{Fmetal}
\end{table}

\begin{table}[tb]
 \def\arraystretch{1.6}
\centering
\begin{tabular}{|c|c|c|cc|}
\hline
operators & $q$ & $k$ & $h,\bar h$ & ($\th=0$)\\
\hline 
$ \ee^{\pm\ii (\vphi+\bar \vphi)} $ & $0$ & $\pm 2k_F$ & $\frac{(\ch\th-\sh\th)^2}{2},\frac{(\ch\th-\sh\th)^2}{2}$ & ($\frac12,\frac12$) \\
\hline 
$ \ee^{\pm\ii (\vphi-\bar \vphi)} $ & $\pm 2$ & $ 0$ & $\frac{(\ch\th+\sh\th)^2}{2},\frac{(\ch\th+\sh\th)^2}{2}$ & ($\frac12,\frac12$) \\
\hline 
\end{tabular}
\caption{Quantum numbers of local gapless bosonic operators in metallic phase of
spin-1/2 electrons with strong anti-ferromagnetic Ising interaction.  Local
fermionic operators (\ie odd-charge operators) are all gapped.  
}
\label{AFmetal}
\end{table}

Let us consider the spin-$\frac12$ Ising chain \eqn{Ising} with $B>0$.
As we change $J$ from $0\to +\infty$, the Ising chain goes into a 
state that breaks the $Z_2^s$ spin-flip symmetry.
If we change $J$ from $0\to -\infty$, the Ising chain goes into a 
state that breaks both the $Z_2^s$ spin-flip and translation symmetries.

However, for a doped Ising chain which is a metallic state, both the $J \to
+\infty$ and the $J \to -\infty$ cases have the same symmetry: the $Z_2^s$
spin-flip symmetry is spontaneously broken while the translation symmetry is
not broken.  Despite the two large $|J|$  metallic phases have the same
symmetry, our previous discussions indicate that the transitions from the $J=0$
metallic phase to $J=\pm \infty$ metallic phases are described by two distinct
critical points.  Thus, even the transitions that have identical spontaneous
symmetry breaking pattern can be described by different critical points.

The two distinct critical points also suggest that $J=\pm \infty$ metallic
phases are two distinct metallic phases despite that they have the same
symmetry.  Thus, they are examples of \textbf{symmetry protected gapless
phases}, \ie distinct gapless phases with the same symmetry.  To understand
these two  distinct metallic phases, we consider modular covariant partition
functions with $U(1)\times Z_2^f$ symmetry.  We will consider the 16-component
partition functions with $Z_2^f\times Z_2^s$ symmetry twists.  Since $Z_2^s$
symmetry is spontaneously broken, the partition functions with non-trivial
$Z_2^s$ symmetry twist vanish.  Using the $u1_4$ CFT characters to construct
the  modular covariant partition functions, we identify the following two sets of
partition functions to describe the $J=\pm \infty$ metallic phases.

For the $J=+\infty$ metallic phase (ferromagnetic Ising interaction), we have
\begin{align}
\begin{split}
 Z_{AE^f,PE^s} &= 
|\chi^{u1_4}_0 |^2 + |\chi^{u1_4}_2|^2,  
  \\ 
Z_{PO^f,PE^s} &= 
 \chi^{u1_4}_1 \bar \chi^{u1_4}_{-1} + \chi^{u1_4}_{-1} \bar \chi^{u1_4}_1,   \\ 
Z_{PE^f,PE^s} &= 
|\chi^{u1_4}_1 |^2 + |\chi^{u1_4}_{-1}|^2,   \\ 
Z_{AO^f,PE^s} &= 
 \chi^{u1_4}_0 \bar \chi^{u1_4}_2 + \chi^{u1_4}_2 \bar \chi^{u1_4}_0, 
\end{split}
\end{align}
\begin{align}
\label{Fmetal}
\begin{split}
 Z_{AE^f,PO^s} &= 
|\chi^{u1_4}_0 |^2 + |\chi^{u1_4}_2|^2,  
  \\ 
Z_{PO^f,PO^s} &= 
 \chi^{u1_4}_1 \bar \chi^{u1_4}_{-1} + \chi^{u1_4}_{-1} \bar \chi^{u1_4}_1,  \\ 
Z_{PE^f,PO^s} &= 
|\chi^{u1_4}_1 |^2 + |\chi^{u1_4}_{-1}|^2,   \\ 
Z_{AO^f,PO^s} &= 
 \chi^{u1_4}_0 \bar \chi^{u1_4}_2 + \chi^{u1_4}_2 \bar \chi^{u1_4}_0,  
\end{split}
\end{align}
\begin{align}
\label{Fmetal1}
\begin{split}
Z_{AE^f,AE^s} &= 0,\\
Z_{PO^f,AE^s} &= 0,\\
Z_{PE^f,AE^s} &= 0,\\
Z_{AO^f,AE^s} &= 0,\\
\end{split}
\end{align}
\begin{align}
\begin{split}
Z_{AE^f,AO^s} &= 0,\\
Z_{PO^f,AO^s} &= 0,\\
Z_{PE^f,AO^s} &= 0,\\
Z_{AO^f,AO^s} &= 0.\\
\end{split}
\end{align}
The corresponding primary fields (\ie gapless operators) and their quantum numbers are listed in
Table \ref{Fmetal}.

For the $J=-\infty$ metallic phase (anti-ferromagnetic Ising interaction), we have
\begin{align}
\label{AFmetal}
\begin{split}
 Z_{AE^f,PE^s} &= 
|\chi^{u1_4}_0 |^2 + |\chi^{u1_4}_2|^2,  \\ 
Z_{PO^f,PE^s} &= 0,\\
Z_{PE^f,PE^s} &= 
|\chi^{u1_4}_1 |^2 + |\chi^{u1_4}_{-1}|^2,   \\ 
Z_{AO^f,PE^s} &= 0,
\end{split}
\end{align}
\begin{align}
\label{AFmetal1}
\begin{split}
 Z_{AE^f,PO^s} &= 
|\chi^{u1_4}_1 |^2 + |\chi^{u1_4}_{-1}|^2,   \\ 
Z_{PO^f,PO^s} &= 0,\\
Z_{PE^f,PO^s} &= 
|\chi^{u1_4}_0 |^2 + |\chi^{u1_4}_2|^2,  \\ 
Z_{AO^f,PO^s} &= 0,\\
\end{split}
\end{align}
\begin{align}
\begin{split}
Z_{AE^f,AE^s} &= 0,\\
Z_{PO^f,AE^s} &= 
 \chi^{u1_4}_1 \bar \chi^{u1_4}_{-1} + \chi^{u1_4}_{-1} \bar \chi^{u1_4}_1,   \\ 
Z_{PE^f,AE^s} &= 0,\\
Z_{AO^f,AE^s} &= 
 \chi^{u1_4}_0 \bar \chi^{u1_4}_2 + \chi^{u1_4}_2 \bar \chi^{u1_4}_0, 
\end{split}
\end{align}
\begin{align}
\begin{split}
Z_{AE^f,AO^s} &= 0,\\
Z_{PO^f,AO^s} &= 
 \chi^{u1_4}_0 \bar \chi^{u1_4}_2 + \chi^{u1_4}_2 \bar \chi^{u1_4}_0, \\ 
Z_{PE^f,AO^s} &= 0,\\
Z_{AO^f,AO^s} &= 
 \chi^{u1_4}_1 \bar \chi^{u1_4}_{-1} + \chi^{u1_4}_{-1} \bar \chi^{u1_4}_1 . 
\end{split}
\end{align}
The corresponding primary
fields (\ie gapless operators) and their quantum numbers are listed in Table
\ref{AFmetal}.
 
In particular, from the above partition function, we can read that in the
anti-ferromagnetic metallic phase, the single electron excitations are all
gapped.  For example, in \eqn{AFmetal}, we see that $Z_{AE^f,PE^s}\neq 0$ which
means a sector with even fermions and integer $S_z$ spins is gapless.  If add
an electron, we obtain a sector with odd fermions and half-integer $S_z$ spins
described by $Z_{AO^f,PO^s}$ in \eqn{AFmetal1}.  $Z_{AO^f,PO^s}=0$ means the
sector to have an energy gap.

This is in contrast to the ferromagnetic metallic phase.  $Z_{AE^f,PE^s}\neq 0$
in \eqn{Fmetal} and $Z_{AO^f,PO^s}\neq 0$ in \eqn{Fmetal1} implies that the
sectors differ by an electron are both gapless.  Thus the single electron
excitations are gapless.

The undertstand this result,
we note that the spins of electrons have a Neel-like $\up\down\up\down\cdots$ pattern.  As a
result, for even numbers of electrons, the partition function is non-zero only
when there is no $Z_2^s$ symmetry twist.  For odd numbers of electrons, the
partition function is non-zero only when there is a $Z_2^s$ symmetry twist.
Since the fermion number and the $Z_2^s$ symmetry twist are locked, the fermion
operators (\ie odd-charge operators) are all gapped.  We can also see that the
gapping of charge-1 fermions by noticing that applying a charge-1 fermion
operator to states in the sector $Z_{AE^f,PE^s}$ gives us states in the sector
$Z_{AO^f,PE^s}$ and $Z_{AO^f,PO^s}$, where $AE^f \to AO^f$ (adding a fermion)
and $PE^s \to PE^s,PO^s$ (the $Z_2^s$ symmetry twist cannot be changed).
Since $Z_{AO^f,PE^s} = Z_{AO^f,PO^s} =0$, meaning the two sectors are gapped,
thus the charge-1 fermionic excitations are all gapped.

\section{Chiral metallic phases of spin-$\frac12$ electrons} \label{chiral}

Following the ideas in \Ref{DW170604648}, we can also construct a strongly
interacting metallic phase of spin-1/2 electrons where the left-movers and
right-movers have very different behavior.  We will call such metallic phases
chiral metallic phases.

In the first example, the left-movers and right-movers have the same emergent symmetry at low energy. However, they carry different representations under the symmetries. In specific, one such chiral metallic phase has $SU(2)$-spin and $U(1)$-charge symmetries
with symmetry group $[SU(2)\times U(1)]/Z_2$.  At low energies, the chiral
metallic phase has $n$ left-moving and $n$ right-moving fermions, which are
non-interacting.  Those non-interacting fermions all carry charge-1.  But the
left-moving and right-moving fermions form different $SU(2)$ representations.
Let $S_z^R$ be the $n\times n$ hermitian matrix for the $S_z$-spin of the
right-moving fermions, and $S_z^L$ be the $n\times n$ hermitian matrix for the
left-moving fermions.  For the low energy fermions to be free from perturbative $SU(2)$ anomaly, the $SU(2)$ representations must satisfy
\begin{align}
\label{anomSU2}
 \Tr (S_z^R)^2 =\Tr (S_z^L)^2.
\end{align}
Then combining the results in \Ref{W1301,DW170604648}, we find that such a
chiral metallic phase is free of all $U(1)\times SU(2)$ anomalies, and can be
realized by interacting fermions on a 1d lattice.  

\Eqn{anomSU2} has solutions only when $n\geq 16$, if we require all
the fermions to have half-integer spins.  At $n=16$, we only have the following
two solutions:
\begin{center}
 \def\arraystretch{1.5}
\begin{tabular}{|r||c|c|c|c|}
\hline
 & \blue{right-movers} & \red{left-movers} & $\nu_s=4 \Tr (S_z^R)^2$ & $c$\\
\hline
\hline
spin & \blue{$\frac12,\frac12,\frac12,\frac92$}  & \red{$\frac72,\frac72$} & $336$ & $16$ \\
\hline
spin & \blue{$\frac12,\frac12,\frac12,\frac12,\frac12,\frac52$}  &  \red{$\frac32,\frac32,\frac32,\frac32$} &  $80$ & $16$ \\
\hline
\end{tabular}
\end{center}
All the above fermions carry charge-1.  The spin-1/2 fermions correspond to the
spin-1/2 electrons.  The fermions with higher spins can be viewed as bound
states of several spin-$\frac12$ electrons and spin-$\frac12$ holes.

The chiral metallic phases of charge-1 fermions carrying the above lists of
spins can be realized by interacting electrons on an 1d lattice, according to
the argument presented in \Ref{W1301,DW170604648}.  However, such chiral
metallic states cannot be smoothly deformed into the non-interacting spin-1/2
electron systems since $\nu_s$ (defined in \eqn{nus}) and the chiral central
charge $c$ are not equal.

\section{Chiral ``non-Abelian'' metallic phases}
\label{u1su2Is}

\subsection{The construction}
In this section, we are going to construct another stable chiral metallic phase
which is also ``non-Abelian''.  We start with a non-interacting 1d electron
system described by 
\begin{align}
\label{H0}
 \cH_0 = 
\psi_{\al a}^\dag(x) \ii v_0 \prt_x \psi_{\al a}(x) -
\bar \psi_{\al a}^\dag(x) \ii v_0 \prt_x \bar \psi_{\al a}(x), 
\end{align}
where $\alpha$ and $a$ are the spin $SU(2)$ and flavor $SU(2)$ labels, and $v_0$ is the Fermi velocity.
At low energy, the model has an emergent $[SU_s(2)\times SU_f(2) \times U(1)]_R
\times [SU_s(2)\times SU_f(2) \times U(1)]_L$ symmetry for right movers and
left movers.  The right movers of the above system are described by a CFT
\begin{align}
\label{su2u1su2u1}
su2^s_2\oplus su2^f_2\oplus u1^c,
\end{align}
where the excitations in $su2^s_2$ carry $SU_s(2)$ spin quantum numbers, the
excitations in $su2^f_2$ carry $SU_f(2)$ flavor quantum numbers, and the
excitations in ${u1}$ carry the $U(1)$ charges.  Similarly, the left movers of
the above system are described by a CFT
\begin{align}
\label{su2u1su2u1B}
\bar{su2^s_2}\oplus \bar{su2^f_2}\oplus \bar{u1^c}.
\end{align}

In the above, $suN_k$ denotes both the level-$k$ $su(N)$ Kac-Moody algebra and the CFT built from it. The CFT has central charge 
\begin{align}
c= \frac{k (N^2-1)}{k+N}.
\end{align}
Likewise $u1_M$ denotes the $U(1)$ current algebra, and the associated CFT,
whose central charge is $c=1$. For details, see Appendix \ref{appCFT}. 

\begin{figure}[tb]
\centering
\includegraphics[scale=0.5]{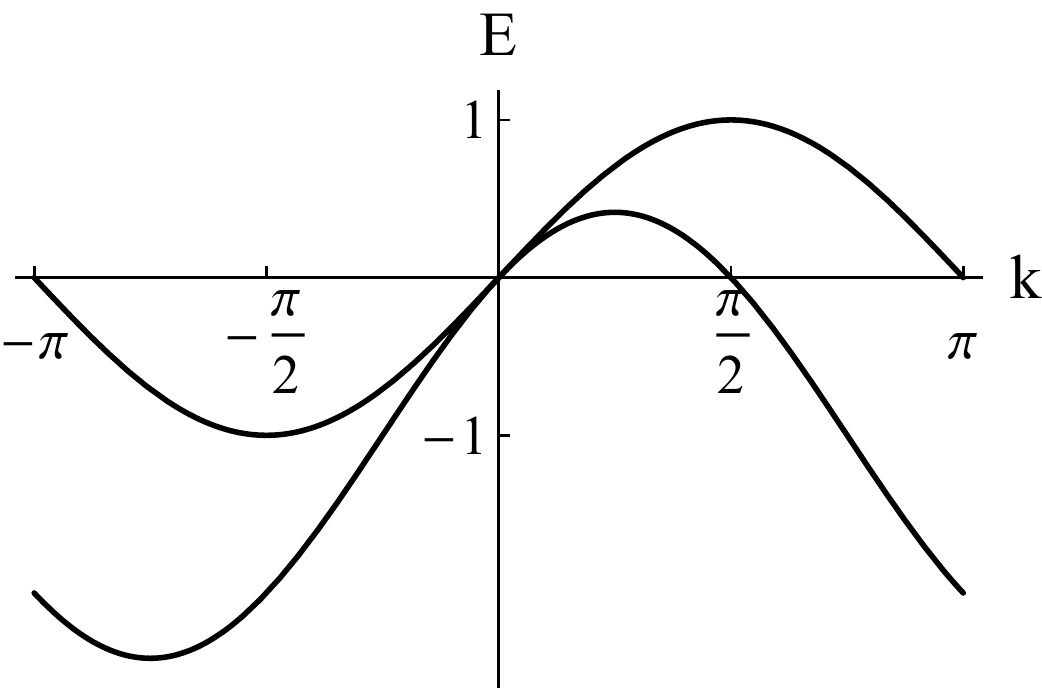}
%3
\caption{The band structure of free fermions to construct chiral metallic phase. The two bands are for two flavors. Each band is double degenerate corresponding to two spin-$\frac{1}{2}$. Velocites $v(k)$ at $k=0$ are the same, and likewise $v(\frac{\pi}{2})= v(\pi)$.}\label{bands}
\end{figure}

In \eqn{H0} the fermions also carry crystal momenta. In particular, $\psi_{\al
a}$ carry crystal momentum $k_F=0$, $\bar \psi_{\al 1}$ carry crystal momentum
$\bar{k}_{F1}=\pi$, and $\bar \psi_{\al 2}$ carry crystal momentum $\bar
k_{F2}=\pi/2$. Such free fermion model is easily realized, for example, with
band structure shown in Fig.\ref{bands}. In particular, the low energy fermion operator $\psi_{\alpha a,k-k_F}$ can be represented in terms of lattice fermion operator $c_{\alpha a,k}$ as follows. {For $k\sim k_F$
\begin{align}
 C_{\al a, k}= & \left(\sum_{\mu=0}^3  \ee^{\ii \mu(k-k_F)}\right) c_{\al a,k}
\sim \psi_{\al a,k-k_F},
\end{align}
and for $k\sim \bar k_{Fa}$,
\begin{align}
 \bar  C_{\al a, k}= &\left( \sum_{\mu=0}^3  \ee^{\ii \mu(k-\bar k_{Fa})}\right)c_{\al a,k}
\sim\bar \psi_{\al a,k-\bar k_{Fa}} .
\end{align}
And in real space,
\begin{align}
\psi_{\alpha a,i}=\sum_{\mu=0}^3 \ee^{-\ii \mu k_F a}c_{\alpha a,i+\mu},\ \bar \psi_{\alpha a,i}=\sum_{\mu=0}^3 \ee^{-\ii \mu \bar k_{Fa}}c_{\alpha a,i+\mu}.
\end{align}
$C_{\alpha a,k}$ reach the maximum at $k=0$ and vanish at $k=\frac{\pi}{2}, \pi$; $\bar C_{\alpha 1,k}$ reach the maximum at $k=\pi$ vanish at $k=0,\frac{\pi}{2}$; $\bar C_{\alpha 2,k}$ reach the maximum at $k=\frac{\pi}{2}$  vanish at $k=0,\pi$. }

To obtain the chiral metallic phase from the above free fermion model, we add
interactions that respect the spin $SU(2)$, charge $U(1)$, and translation
symmetry. We will add interactions in three steps and finally lead to the
Hamiltonian
\begin{align}
\cH=\cH_0+\delta \cH+\bar{\delta \cH}+\delta \cH',
\end{align}
It is crucial here that the interactions are different for left movers and
right movers. For the right movers, we add interaction
\begin{align}
 \del \cH= 
g_s \v J_s(x) \cdot \v J_s(x) 
+g_c J_c(x) J_c(x) ,
\end{align}
where
\begin{align}
\begin{split}
 J_c(x) &= \psi_{\al a}^\dag(x) \psi_{\al a}(x),\\
 \v J_s(x) &= \frac12 \psi_{\al a}^\dag(x) \v \si_{\al\bt}\psi_{\bt a}(x),
\end{split}
\end{align}
are the $U(1)$ charge and $SU(2)$ spin current (or density), and $\v \si$ are
the Pauli matrices.  As a current-current interaction, when the coupling
constant $g$'s are not too large, the above interaction term [with scaling
dimension $(h,\bar{h})=(2,0)$] is always \emph{exactly marginal}. It does not
open up any energy gap, but only modifies the velocities in the corresponding
sector.  With the interactions, the right movers have $SU(2)\times SU(2) \times
U(1)$ symmetry, and are described by a CFT
\begin{align}
\label{su2su2u1}
su2^f_2\oplus su2^s_2\oplus {u1^c}.
\end{align}
The three sectors, each containing the flavor, spin and charge degrees of
freedom respectively, can have separate velocities, while the excitations
{within} each sector have the same velocity.  

For left-movers we add interactions
\begin{align}
\bar{ \del \cH }&= 
\bar g_{s1} \bar{ \v J}_{s1} (x) \cdot \bar{\v J}_{s1}(x) 
+\bar g_{s2} \bar{ \v J}_{s2} (x) \cdot \bar{\v J}_{s2}(x) 
\nonumber\\
&\ \ \ \
+\bar g_{c1} \bar J_{c1}(x) \bar J_{c1}(x) 
+\bar g_{c2} \bar J_{c2}(x) \bar J_{c2}(x)  .
\end{align}
Here
\begin{align}
\begin{split}
 \bar J_{c1}(x) &= \bar \psi_{\al 1}^\dag(x) \bar \psi_{\al 1}(x),\\
 \bar J_{c2}(x) &= \bar \psi_{\al 2}^\dag(x) \bar \psi_{\al 2}(x),\\
 \bar{ \v J}_{s1}(x) &= \frac12 \bar \psi_{\al 1}^\dag(x) \v \si_{\al\bt}\bar \psi_{\bt 1}(x),\\
 \bar{ \v J}_{s2}(x) &= \frac12 \bar \psi_{\al 2}^\dag(x) \v \si_{\al\bt}\bar \psi_{\bt 2}(x).
 \end{split}
\end{align}
Such current-current interactions also do not open up gaps, but modify the
velocity in the corresponding sector.  With the interaction, the left movers
have the symmetry $SU(2)\times U(1) \times SU(2) \times U(1)$, and are described
by a CFT
\begin{align}
\label{su2u1su2u1}
\bar{su2^s_1}\oplus \bar{{u1}^{c1}}\oplus
\bar{su2^s_1}\oplus \bar{{u1}^{c2}}.
\end{align}

The chiral metallic phase that we have constructed so far (see \eqn{su2su2u1}
and \eqn{su2u1su2u1}) can be smoothly connected to TL
liquids (\ie interacting 1d Fermi liquids) \cite{T5044,L6354}, as we reduce
$g$'s to zero.  To construct a chiral chiral metallic phase that is not
connected to TL liquid, we add an additional interaction term
\begin{align}
\label{Hgap}
 \del \cH'
= g
(\psi^\dag_{\up 1}  \psi^\dag_{\down 1}
-\psi^\dag_{\down 1}  \psi^\dag_{\up 1})
(\bar \psi_{\up 1}  \bar \psi_{\down 1}-
\bar \psi_{\down 1}  \bar \psi_{\up 1}) + h.c.\,.
\end{align}
We note that the above operator carries a crystal momentum $k=0+0+\pi+\pi= 0$
mod $2\pi$.  Thus the term respects the translation symmetry.  Such a term is
not a current-current interaction and can induce energy gaps for some
excitations and drive the system into a new phase.

To understand the new phase, note that the above operator respects the spin
$SU(2)$, the {diagonal} charge $U(1)$, and the translation symmetry
(since the crystal momentum carried by the operator vanishes).  Such an
operator only causes interaction within the sector  $su2^f_2\oplus u1^c
\oplus \bar{{u1}^{c1}}$. 
If the interaction $g$ is strong enough, it will gap out the $\bar{{u1}^{c1}}$
and part of the $su2^f_2\oplus u1^c$ sector, which reduces  $su2^f_2\oplus u1^c
\oplus \bar{{u1}^{c1}}$ down to $\text{Is}\oplus u1^{cf}$, where $\text{Is}$
denotes Ising CFT.  In this way, we obtain a chiral metallic phase described by CFT
\begin{align}
su2_2\oplus u1^{cf} \oplus \text{Is} \oplus
\bar{su2}_1\oplus \bar{su2}_1 \oplus \bar{u1},
\end{align}
which is beyond TL liquids. 

\subsection{The gapping process}

We would like to show the gapping process of the interaction \eq{Hgap} more
explicitly. This is accomplished by using CFT and current algebras.
Furthermore, we can derive the physical properties such as local operators,
correlation function and partition functions, which will be done in the next
subsection. 

We start with $\cH_0$ in \eqn{H0} plus the interactions \eqn{su2su2u1} and
\eqn{su2u1su2u1}.
The resulting low energy theory has the following
emergent symmetry
\begin{align}
\begin{split}
\text{right movers:}&\quad U(1)\times SU_f(2)\times SU_s(2)\\
\text{left movers:}&\quad U_1(1)\times SU_s(2)\times U_2(1)\times SU_s(2)
\end{split}
\label{sym}
\end{align}
$\psi_{\alpha a}$ carry the $U(1)$ charge-1, and transform as doublets of both
the flavor and spin $SU(2)$. In contrast, $\bar{\psi}_{\al 1}$ carry the
charge-1 for $U_1(1)$, form a doublet of the first $SU_s(2)$; and
$\bar{\psi}_{\al 2}$ carry the charge-1 for $U_1(1)$, form a doublet of the
second $SU(2)_s$.
The low energy excitations are descirbed by the following current algebras:
\begin{align}
\begin{split}
\text{right movers:}&\quad {u1^c}\oplus su2^s_2\oplus su2^f_2,\\
\text{left movers:}&\quad \bar{{u1^1}}\oplus \bar{su2^1_1}\oplus \bar{{u1^2}}\oplus \bar{su2^2_1}.
\end{split}
\end{align}
The theory is free from gravitational anomaly, since the left central charge
$c=1+\frac{3}{2}+\frac{3}{2}=4$ is equal to the right central charge
$\bar{c}=1+1+1+1$. 

The local operators of our theory are powers of the fermion operators
$\psi_{\alpha a}, \bar{\psi}_{\alpha a}$. The fermion operators can be represented
in terms of the primary fields of the above CFT. In particular, they can be written in terms of simple free boson fields and free Majorana fermion
fields in $u1, su2_2$ and $su2_1$ CFTs  (see Appendix \ref{prime}):
\begin{align}
\label{ferm}
&\psi_{\alpha a}= \ee^{\ii\varphi_c/2}   
\sigma_s\ee^{\pm \ii \frac{\phi_s}{2}}\sigma_f\ee^{\pm \ii \frac{\phi_f}{2}} 
=\ee^{\ii\frac{\varphi_c}{2}}V^{su2^s_2}_{\frac12,\pm \frac12}V^{su2^f_2}_{\frac12,\pm \frac12},
\nonumber \\
 &\bar \psi_{\al 1}(\bar z)= \ee^{\ii \frac{\bar \varphi_1}{\sqrt 2}}
\ee^{\pm \ii \frac{\bar\phi_1}{\sqrt 2}}
=\ee^{\ii \frac{\bar \varphi_1}{\sqrt 2}}
\bar V^{su2^1_1}_{\frac12,\pm \frac12},
\\
 &\bar \psi_{\al 2}= \ee^{\ii  \frac{\bar \varphi_2}{\sqrt 2}}
\ee^{\pm \ii \frac{\bar\phi_2}{\sqrt 2}}
=\ee^{\ii \frac{\bar \varphi_1}{\sqrt 2}}
\bar V^{su2^2_1}_{\frac12,\pm \frac12}.\nonumber 
\end{align}
Here, for right movers,\\
\indent (1) $\varphi_c$ is the bosonic field to describe 
$ {u1^c}$,\\
\indent (2) $\eta_s, \si_s, \phi_s$ are the Ising CFT fields and 
the bosonic field to describe $su2^s_2$, and\\
\indent (3) $\eta_f, \si_f,
\phi_f$ are the Ising CFT fields and the bosonic field to describe 
$su2^f_2$.\\
  Similarly, for left movers,\\
\indent (1) $\bar\vphi_1$ is the 
bosonic field to describe $\bar{{u1^1}}$, \\
\indent (2) $\bar\vphi_2$ is the
bosonic field to describe $\bar{{u1^2}}$, \\
\indent (3) $\bar\phi_1$ is
the bosonic field to describe $\bar{su2^1_1}$, and \\
\indent (4)
$\bar\phi_2$ is the bosonic field to describe $\bar{su2^2_1}$.

We adopt the convention that the correlation function of all bosonic fields are
\begin{align}
\langle \phi (z_1)\phi (z_2)\rangle \sim -\ln (z_1-z_2),\label{bcorre}
\end{align}
where $z_i=\tau_i+\ii x_i$ is the complex cooridinate.  
The scaling dimensions of operators in (\ref{ferm}) all equal $\frac{1}{2}$, a necessary condition for chiral fermion operators. In fact, it fix the $u1$ parts of fermion operator representations in (\ref{ferm}).

Now the gapping term \eq{Hgap} can be rewritten as (via operator product expansion, see Appendix \ref{ope})
\begin{align}
\label{Hgap1}
\delta \cH' &\sim -g \cos \left(\varphi_c+\phi_f-\sqrt{2}\bar\varphi_1\right).
\end{align}
When $g>0$ is large, the sectors generated by $\varphi_c+\phi_f$ and
$\bar\varphi_1$ are fully gapped.  Other sectors are not affected.
Consequently, the gapless excitations in the new phase are described by the
following CFT
\begin{align}
\begin{split}
\text{right movers:}&\quad {u1^{cf}}\oplus su2^s_2\oplus \text{Is}^f,\\
\text{left movers:}&\quad \bar{su2^1_1}\oplus \bar{{u1^2}}\oplus 
\bar{su2^2_1},
\end{split}
\end{align}
where ${u1^{cf}}$ is $u1$ CFT represented by the field $\varphi_c-\phi_f$, the conjugate field of $\varphi_c+\phi_f$ that remains gapless. And
$\text{Is}^f = \frac{su2^f_2}{u1^f}$ is Ising CFT with primary fields $1, \sigma_f$ and $\eta_f$. Primary fields of above CFTs are summarised in Appendix \ref{prime}. 

The scaling dimension of $e^{\ii (\vphi_c+\phi_f-\sqrt{2}\bar{\varphi}_1)}$ is $\left(h,\bar{h}\right)=\left(1,1\right)$. We emphasize that, {even though the local interaction term gaps out the left-mode and right-mode in equal numbers, it selects an inequivalent combination of right-modes, comparing to the left-modes, since the operator is in the CFT
\begin{align}
 {u1^c_4}\oplus u1^{fz}_4\oplus  \bar{{u1^1_2}}\subset {u1^c_4}\oplus su2_2^{f}\oplus  \bar{{u1^1_2}}.
\end{align}
And the resulting phase is anomalous free and has lattice realization.
 }

\subsection{The local operators}

To compute the physical properties of the chiral metallic phase, we first
identify local operators in the above CFT.  In the chiral metallic phase, the
fermion operators $\psi_{\al 1}$ and $\bar \psi_{\al 1}$ are gapped (\ie their
imaginary-time correlations have exponential decay), since they contain either
$\ee^{\ii\frac{\vphi_c+\vphi_f}{2}}$ or $\ee^{\ii \frac{\bar \phi_1}{\sqrt
2}}$.  They are local operators but do not appear in the low energy CFT. All
other fermion operators, 
\begin{align}
 \label{fermop}
 \begin{split}
&\psi_{\alpha 2}= \ee^{\ii\frac{\varphi_c-\phi_f}{2}}   
\sigma_s\ee^{\pm \ii \phi_s/2}\sigma_f 
=
\ee^{\ii\frac{\varphi_c-\phi_f}{2}}  
V^{su2^s_2}_{\frac12,\pm \frac12} \si_f,
 \\
 &\bar \psi_{\al 2}
= \ee^{\ii  \frac{\bar \varphi_2}{\sqrt 2}}\ee^{\pm \ii \frac{\bar\phi_2}{\sqrt 2}}
= \ee^{\ii  \frac{\bar \varphi_2}{\sqrt 2}} \bar V^{su2^2_1}_{\frac12,\pm \frac12},
\end{split}
\end{align}
are still gapless, therefore are local operators in CFT. Operators generated from the OPEs of
$\psi_{\alpha 2}$'s and $\bar \psi_{\al 2}$ are also local
operators.

The above local operators are purely chiral with either only right movers or
left movers.  Another type of local operators containing both right movers and
left movers is 
\begin{align}
\label{fermop2}
\begin{split}
\psi_{\al 1}^\dag \bar  \psi_{\bt 1}
=&
\ee^{-\ii\frac{\varphi_c+\phi_f -\sqrt 2 \bar\phi_1}{2}}   
\sigma_s\ee^{\pm \ii \phi_s/2}\sigma_f 
\ee^{\pm \ii \frac{\bar\phi_1}{\sqrt 2}},
\\
\sim&  \sigma_s\ee^{\pm \ii \phi_s/2}\sigma_f 
\ee^{\pm \ii \frac{\bar\phi_1}{\sqrt 2}}=\si_f V^{su2^s_2}_{\frac12,\pm \frac12} \bar V^{su2^1_1}_{\frac12,\pm \frac12},
\end{split}
\end{align}
where we have used the knowledge that in the chiral metallic phase, the $\cos$ term in (\ref{Hgap1}) is frozen to the maximum value, \ie
$\ee^{-\ii\frac{\varphi_c+\phi_f -\sqrt 2 \bar\phi_1}{2}} \sim 1$.
Therefore $ \psi_{\al 1}^\dag \bar  \psi_{\bt 1}$'s are also low energy local operators in the chiral metallic phase.

\begin{table}[tb]
 \def\arraystretch{1.5}
\centering
\begin{tabular}{  |c|c|c|c|}
\hline
operators & spin & charge & $k$ \\
\hline 
\rule{0pt}{3.8ex}$\psi_{\al 2}
=\ee^{\ii\frac{\varphi_c-\phi_f}{2}}  
V^{su2^s_2}_{\frac12,\pm \frac12} \si_f
$ & $\frac12$ & -1 & 0 \\
\hline 
\rule{0pt}{3.8ex}$\bar \psi_{\al 2}
=\ee^{\ii  \frac{\bar \varphi_2}{\sqrt 2}} \bar V^{su2^2_1}_{\frac12,\pm \frac12}
$ & $\frac12$ & -1 & $-\frac{\pi}{2a}$ \\
\hline
\rule{0pt}{3.8ex}$\psi_{\al 1}^\dag \bar  \psi_{\bt 1}
=\si_f V^{su2^s_2}_{\frac12,\pm \frac12} \bar V^{su2^1_1}_{\frac12,\pm \frac12}
$ & 0, 1& 0 & $\frac{\pi}{a}$ \\
\hline
\end{tabular}
\caption{Quantum numbers of local operators, where $k$ is the crystal momentum.}
\label{quan}
\end{table}

\subsection{Partition functions}

To find modular covariant partition functions (see \eqn{Mcov}), we use the CFT characters for $u1_M$, $\text{Is}\cong
\frac{su2_2}{{u1}}$, $su2_1$, and $su2_2$ (details shown in Appendix \ref{Mtrn}):
\begin{align}
\begin{split}
\chi^{u1_M}_m, \quad& 0 \leq m < M,  \\
\chi^{\text{Is}}_\mu, \quad& \mu= 0, \eta, \sigma, \\
\chi^{su2_1}_\rho,\quad & \rho=0,\frac12, \\
\chi^{su2_2}_\nu, \quad & \nu=0,\frac12, 1.
\end{split}
\end{align}
The scaling dimension of the $U(1)$ part in $\psi_{\alpha 2},
\bar{\psi}_{\alpha 2}$ in (\ref{fermop}) are both $\frac{1}{4}$, thus
corresponds to $u1$ primary fields with $R=\sqrt{2}$. This determines the level of $u1_M$ CFT to be $M=2$ (See Eq \eq{u1Op}).

\begin{table*}[tb]
 \def\arraystretch{1.5}
\centering
\begin{tabular}{|c|c|c|c|c|}
\hline
operators & spin & charge & $k$ & $h,\bar h$\\
\hline 
$ \ee^{\pm \ii \frac{\vphi_c-\vphi_f}{2} \pm \ii \frac{ \bar \vphi_2}{\sqrt 2}
} V^{su2_2^s}_{1,l} \bar V^{su2_1^1}_{\frac12,\pm\frac12} \bar
V^{su2_1^2}_{\frac12,\pm\frac12} $ 
& 
{$0,1,2$} & $0,\pm 2$ & $\pi\pm \frac{\pi}{2}$ & $\frac34,\frac34$ \\
\hline 
\rule{0pt}{3.8ex}$ \ee^{\pm \ii \frac{\vphi_c-\vphi_f}{2} \pm \ii \frac{ \bar \vphi_2}{\sqrt 2} } 
\eta_f 
\bar V^{su2_1^1}_{\frac12,\pm\frac12} 
\bar V^{su2_1^2}_{\frac12,\pm\frac12} 
$ & $0,1$ & $0,\pm 2$ & $\pi\pm \frac{\pi}{2}$ & $\frac34,\frac34$ \\
\hline
\rule{0pt}{3.8ex}$ 
V^{su2_2^s}_{1,l} 
\eta_f 
$ & $1$ & $0$ & $0$ & $1,0$ \\
\hline
\rule{0pt}{3.8ex}$
V^{su2_2^s}_{\frac12,\pm\frac12} 
\si_f
\bar V^{su2_1^1}_{\frac12,\pm\frac12} 
$ & $0,1$ & $0$ & $\pi$ & $\frac14,\frac14$ \\
\hline 
\rule{0pt}{3.8ex}$\ee^{\pm \ii \frac{\vphi_c-\vphi_f}{2} \pm \ii \frac{ \bar \vphi_2}{\sqrt 2} } 
V^{su2_2^s}_{\frac12,\pm\frac12} 
\bar V^{su2_1^2}_{\frac12,\pm\frac12} 
$ & $0,1$ & $0,\pm 2$ & $\pm \frac{\pi}{2}$ & $\frac12,\frac12$ \\
\hline 
\end{tabular}
\caption{Quantum numbers of local gapless bosonic operators in the chiral metallic
phase $su2^s_2\times u1_2 \times \text{Is} \times \bar{su2^1_2}\times
\bar{su2^2_2}\times \bar{u1}_2$. Here $k$ is the crystal momentum.}
\label{quanCnabMS}
\end{table*}

We find the simplest solution of covariant partition functions that contains local operators (\ref{fermop}) and (\ref{fermop2}) is \\
\indent (1) Anti-periodic boundary condition with even number of fermions:
\begin{align}
\label{ZAE}
Z_{AE}=&\chi_0^{u1_2}(\chi_0^{su2_2}\chi_0^{\text{Is}}+\chi_1^{su2_2}\chi_\eta^{\text{Is}})\bchi_0^{u1_2}\bchi_0^{su2_1^1}\bchi_0^{su2_1^2}\nn\\
&+\chi_1^{u1_2}(\chi_1^{su2_2}\chi_0^{\text{Is}}+\chi_0^{su2_2}\chi_\eta^{\text{Is}})\bchi_1^{u1_2}\bchi_{1/2}^{su2_1^1}\bchi_{1/2}^{su2_1^2}\nn\\
&+\chi_0^{u1_2}\chi_{1/2}^{su2_2}\chi_\sigma^{\text{Is}}\bchi_0^{u1_2}\bchi_{1/2}^{su2_1^1}\bchi_0^{su2_1^2}\nn\\
&+\chi_1^{u1_2}\chi_{1/2}^{su2_2}\chi_\sigma^{\text{Is}}\bchi_1^{u1_2}\bchi_{0}^{su2_1^1}\bchi_{1/2}^{su2_1^2}.
\end{align}
The primary field corresponding to each term of characters in $Z_{AE}$ is bosonic with integral spin $h-\bar{h}\in \ZZ$. We list the scaling dimensions of all primary fields in Appendix \ref{nabMSscal}. 

\indent (2) Anti-periodic boundary condition with odd number of fermions:
\begin{align}
\label{ZAO}
Z_{AO}=&\chi_0^{u1_2}(\chi_0^{su2_2}\chi_0^{\text{Is}}+\chi_1^{su2_2}\chi_\eta^{\text{Is}})\bchi_1^{u1_2}\bchi_0^{su2_1^1}\bchi_{1/2}^{su2_1^2}\nn\\
&+\chi_1^{u1_2}(\chi_1^{su2_2}\chi_0^{\text{Is}}+\chi_0^{su2_2}\chi_\eta^{\text{Is}})\bchi_0^{u1_2}\bchi_{1/2}^{su2_1^1}\bchi_{0}^{su2_1^2}\nn\\
&+\chi_0^{u1_2}\chi_{1/2}^{su2_2}\chi_\sigma^{\text{Is}}\bchi_1^{u1_2}\bchi_{1/2}^{su2_1^1}\bchi_{1/2}^{su2_1^2}\nn\\
&+\chi_1^{u1_2}\chi_{1/2}^{su2_2}\chi_\sigma^{\text{Is}}\bchi_0^{u1_2}\bchi_{0}^{su2_1^1}\bchi_{0}^{su2_1^2}.
\end{align}

The primary field corresponding to each term of characters in $Z_{AE}$ is fermionic with half-integral spin $h-\bar{h}\in \ZZ+\frac{1}{2}$. We list the scaling dimensions of all primary fields in Appendix \ref{nabMSscal}. 

(3) Periodic boundary condition with even or odd number of fermions:
\begin{align}
\label{ZPE}
&Z_{PE}=Z_{PO} \nn\\
=&\frac{1}{2}(\chi_0^{u1_2}\chi_{1/2}^{su2_2}\chi_{\sigma}^{\text{Is}}\bchi_0^{su2_1^1}+\chi_1^{u1_2}\chi_0^{su2_2}\chi_0^{\text{Is}}\bchi_0^{su2_1^1})\nn\\
&\cdot (\bchi_1^{u1_2}\bchi_0^{su2_1^2}+\bchi_0^{u1_2}\bchi_{1/2}^{su2_1^2})\nn\\
&+\frac{1}{2}\left[\chi_0^{u1_2}(\chi_1^{su2_2}\chi_0^{\text{Is}}+\chi_0^{su2_2}\chi_\eta^{\text{Is}})\bchi_{1/2}^{su2_1^1}\right.\nn\\
&\left.+\chi_1^{u1_2}(\chi_{1}^{su2_2}\chi_{\eta}^{\text{Is}}\bchi_0^{su2_1^1}+\chi_{1/2}^{su2_2}\chi_{\sigma}^{\text{Is}}\bchi_{1/2}^{su2_1^1})\right]\nn\\
&\cdot (\bchi_1^{u1_2}\bchi_0^{su2_1^2}+\bchi_0^{u1_2}\bchi_{1/2}^{su2_1^2}),
\end{align}
where both terms in $Z_{PE} (Z_{PO})$ is $8$-fold degenerate, essentially contributed from $4$ Majorana zero modes for periodic boundary condition, as explained in Appendix \ref{spinhalfDirac}. 
The primary field corresponding to each term of characters in $Z_{PE}$ ($Z_{PO}$) is bosonic with integral spin $h-\bar{h}\in \ZZ$. We list the scaling dimensions of all primary fields in Appendix \ref{nabMSscal}. 

Here we have used the fact that $Z_{PP}=0$, since the chiral metallic phase
contains free fermions $\psi_{\al 2}$ and $\bar\psi_{\al 2}$. The zero modes
of free fermions in space-time path integral cause $Z_{PP}=0$.

There is a physical approach that leads to this solution of modular covariant partition functions. Consider the combination of a Heisenberg chain and a spin-$1/2$ Dirac fermion (refered as the HD hybrid system). The low energy theory is for the Heisenberg chain is $su2_1^1\oplus  \bar{su2_1^1}$ CFT. The low energy theory of free spin-1/2 Dirac fermion is $su2_1^2\oplus u1_2\oplus \bar{su2_1^2}\oplus \bar{u1_2}$. Its partition function $Z^{HD}_{AA}$ is thus the product of the partition functions of the two CFTs,
\begin{align}
Z^{HD}_{AA}=&
(\chi^{su2_1^1}_0\bchi^{su2_1^1}_0+\chi^{su2_1^1}_{1/2}\bchi^{su2_1^1}_{1/2})
(\chi^{su2_1^2}_0\chi^{u1_2}_0+\chi^{su2_1^2}_{1/2}\chi^{u1_2}_{1})\nn\\
&\cdot (\bchi^{su2_1^2}_0\bchi^{u1_2}_0+\bchi^{su2_1^2}_{1/2}\bchi^{u1_2}_{1}),
\end{align}
and can be reorganized as
\begin{align}
Z^{HD}_{AA}=&\left[(\chi^{su2_1^1}_0\chi^{su2_1^2}_0\chi^{u1_2}_0+\chi^{su2_1^1}_0\chi^{su2_1^2}_{1/2}\chi^{u1_2}_{1})\bchi^{su2_1^1}_0\right.\nn\\
&\left.+(\chi^{su2_1^1}_{1/2}\chi^{su2_1^2}_0\chi^{u1_2}_0+\chi^{su2_1^1}_{1/2}\chi^{su2_1^2}_{1/2}\chi^{u1_2}_{1})\bchi^{su2_1^1}_{1/2}\right]\nn\\
&\cdot(\bchi^{su2_1^2}_0\bchi^{u1_2}_0+\bchi^{su2_1^2}_{1/2}\bchi^{u1_2}_{1}).
\label{heigDiracZ}
\end{align}

The interesting fact is that $su2_1^1\times su2_1^2$ characters can all be represented precisely by $su2_2\times \text{Is}$ characters. More specifically,
\begin{align}\label{coset}
\chi_0^{su2_1^1}\chi_0^{su2_1^2}=&\chi_0^{su2_2}\chi^{\text{Is}}_0+\chi_1^{su2_2}\chi^{\text{Is}}_\eta,\nn\\
\chi_0^{su2_1^1}\chi_{1/2}^{su2_1^2}=&\chi_1^{su2_1^1}\chi_0^{su2_1^2}=\chi_{1/2}^{su2_2}\chi^{\text{Is}}_\sigma,\\
\chi_{1/2}^{su2_1^1}\chi_{1/2}^{su2_1^2}=&\chi_0^{su2_2}\chi^{\text{Is}}_\eta+\chi_1^{su2_2}\chi^{\text{Is}}_0.\nn
\end{align}

We find that after replacing $\chi^{su2_1^1}_\mu\chi^{su2_1^2}_\nu$ in the partition function (\ref{heigDiracZ}) with these identities above, and rewriting in the basis with fixed fermion number parity, we reach partition functions (\ref{ZAE}), (\ref{ZAO}) and (\ref{ZPE}).

The partition function \eq{ZAE} provides us a list of local gapless bosonic
operators and their scaling dimensions \eq{hAE} in the chiral metallic phase. The
result is summarized in Table \ref{quanCnabMS}.  

The crystal momenta of those local gapless bosonic operators is also an important quantum number. Note that
all right-movers carry zero crystal momentum.  For left-movers, the spin-1/2
operators in the $\bar{su^1_1}$ sector carry crystal momentum $\pi$.  In the
$\bar{u1_2}\oplus\bar{su^2_1}$ sector, the operator $\psi_{\al 2}\sim \ee^{\pm \ii \frac{ \bar
\vphi_2}{\sqrt 2}} \bar V^{su2_1^2}_{\frac12,\pm\frac12}$ carries crystal momentum $\pm \pi/2$. From these results,
 we obtain the crystal momenta of the local gapless bosonic operators in Table
\ref{quanCnabMS}.  

From
$\chi_0^{u1_2}\chi_0^{su2_2}\chi_0^{\text{Is}}\bchi_0^{u1_2}\bchi_0^{su2_1^1}\bchi_0^{su2_1^2}$
term in  $Z_{AE}$, we see that there is no discrete symmetry breaking in the
chiral metallic phase.  If there is, say, a $Z_2$  symmetry breaking,
$2\chi_0^{u1_2}\chi_0^{su2_2}\chi_0^{\text{Is}}\bchi_0^{u1_2}\bchi_0^{su2_1^1}\bchi_0^{su2_1^2}$
will appear in  $Z_{AE}$.  

We see that all the local gapless bosonic operators carry non-trivial quantum numbers.
Therefore, the chiral metallic phase is a stable phase.

\subsection{Phase transition from Tomonaga-Luttinger liquid to chiral metallic phase}
This procedure signals that there can be a direct phase transition between the HD hybrid system, whose low energy is described by Tomonaga-Luttinger liquid theory, and the chiral metallic phase, whose low energy physics is described by non-Abelian CFTs. The HD phase has $4$ emergent $SU(2)$ symmetries. The chiral metallic phase has $3$ emergent $SU(2)$ symmetries. 

Indeed, the interaction operator (\ref{Hgap1}) is a marginal operator with $h=\bar{h}=1$. It is a tempting indication that the zero-spin marginal perturbation can drive a transition between two stable gapless (under symmetry) phases.

\section{Examples of strongly interacting gapless metallic states in higher dimensions}

\begin{figure}[t]
\begin{center}
\includegraphics[scale=0.5]{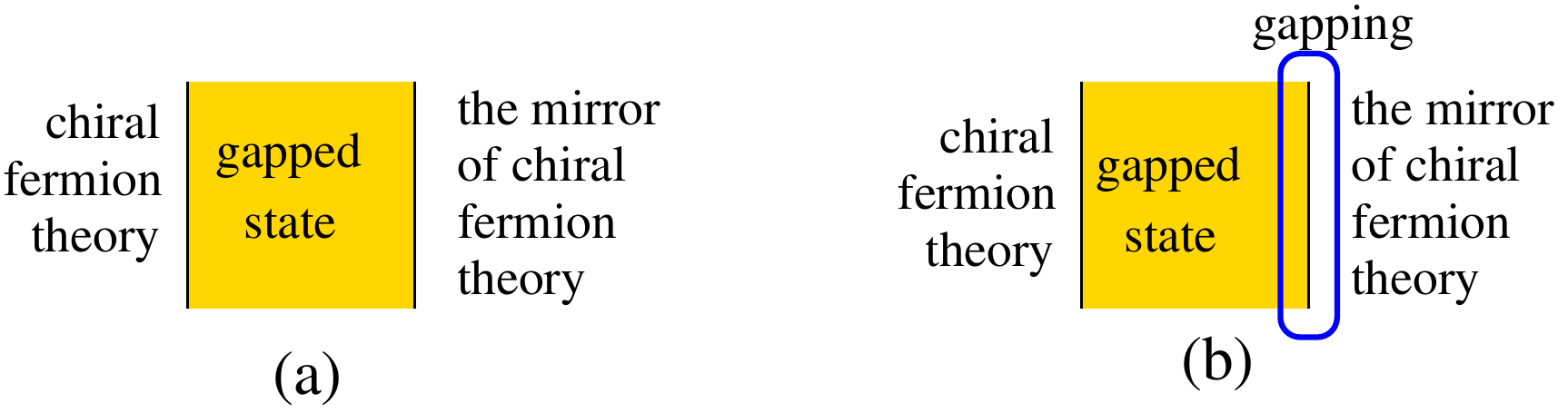} \end{center}
%4
\caption{ (Color online) 
(a) Chiral fermions and mirror of  chiral fermions can appear on the boundary
of a 4+1d slab of gapped state.  (b) Some times (such as the $SO(10)$ case) the
boundary mirror chiral fermions can be gapped by interaction, which leads to a
solution of chiral fermion problem.
}
\label{chferm}
\end{figure}

The fact that the emergent symmetry at low energies can be anomalous plays a
key role in the solution of chiral fermion problem \cite{W1301,YX14124784}.
For example, in the lattice realization of $SO(10)$ chiral fermions, we start
with a 4+1d slab, which can be viewed as a 3+1d system from far away.  We
design the gapped fermion state with $SO(10)$ on-site symmetry in the 4+1d bulk
properly, such that its surface is described by 16 massless Weyl fermions,
forming a 16-dimensional spinor representation of the $SO(10)$.  On the 4+1d
slab, one 3+1d surface give rise to 16 chiral Weyl fermions and the other 3+1d
surface give rise to 16 mirror chiral Weyl fermions (see Fig. \ref{chferm}a).
Each sector of the Weyl fermions has an emergent symmetry $U(16)$. Such an
emergent symmetry $U(16)$ is anomalous for each sector \cite{YX14124784}. In
\Ref{W1301} the sufficient conditions are given in order for a sector (such as
the 16 mirror chiral Weyl fermions) to be gappable via interactions without
breaking the lattice and on-site symmetry (see Fig. \ref{chferm}b). Applying to
the $SO(10)$ case, we find that the 16 mirror chiral Weyl fermions can be
gapped without breaking the $SO(10)$ on-site symmetry, and hence solved the
chiral fermion problem for the case of $SO(10)$ grant unification.  We like to
stress that the gaping of 16 chiral Weyl fermions is very special, in the sense
that there is no fermion mass term that can achieve such a gapping process
without breaking the $SO(10)$ symmetry.  It appears that the anomaly of the
emergent $U(16)$ symmetry protect the 16 chiral Weyl fermions to be gapless
against any small perturbations that respect the $SO(10)$ symmetry. 

In the above example, each sector of 16 massless Weyl fermions is free of all
anomalies.  It was also pointed out \cite{W1301} that even when each sector is
anomalous, it is also possible that an anomalous sector can be have
topologically ordered gapped phase \cite{VS1306}. This offers a more general
way to solve the chiral fermion problem.  In general, for a gapless system, its
low energy effective theory for the gapless modes can be anomalous. Even such
an anomalous low energy effective theory can sometimes be realized by a
well-defined lattice model in the same dimension, since the anomaly can be
canceled by a gapped (anomalous) topological sector.  

One such example is the 2d gapless theory of one single Weyl fermion with
$U(1)$ (fermion number $N_F$) and time reversal ($T$) symmetry.  The time
reversal transformation satisfies $T^2=(-)^{N_F}$.  Such a single-Weyl-fermion
theory has a parity anomaly (time reversal is a space-time parity
transformation).  It was believed (incorrectly) that there was no 2d lattice
theory with on-site $U(1)$ and time reversal symmetries that can produce low
energy effective theory of a single-Weyl-fermion.  Indeed, there are no
\emph{non-interacting} lattice theories with on-site $U(1)$ and time reversal
symmetries that can produce low energy effective theory of a
single-Weyl-fermion.  However, if we include interaction, then there are
\emph{interacting} lattice theory with on-site $U(1)$ and time reversal
symmetries that can produce low energy effective theory of a
single-Weyl-fermion without breaking those symmetries.  One way to construct
such an interacting 2d lattice model is to start with a slab of 3d lattice
model, which can be viewed as a 2d lattice model from far away.  On the 3d slab
we have the topological insulator with $U(1)$ charge and $T^2=(-)^{N_F}$ time
reversal symmetries.  The fermions do not interact near one surface of the slab
which give rise to the low energy effective theory of a single-Weyl-fermion.
Near the other surface of the slab, fermions interact strongly, which give rise
to a gapped non-Abelian topologically ordered state and do not contribute to
low energy modes (see Fig. \ref{chferm}b).

This research was partially supported by NSF DMS-1664412.  This work was also
partially supported by the Simons Collaboration on Ultra-Quantum Matter, which
is a grant from the Simons Foundation (651440).

\appendix \allowdisplaybreaks

\section{Tomonaga-Luttinger liquids as Abelian gapless phases} \label{TLL}

We define the Tomonaga-Luttinger liquid as the liquid, containing only excitations with
integral (or bosonic) statistics. It can always be written as 
\begin{align}
K_F=\begin{pmatrix}
1 & 0 \\ 0 & -1
\end{pmatrix}.
\end{align}
Thus the LL has the property that $\Gamma=\Gamma_0$. For simplicity, we assume $N_L=N_R=1$, there are only one left and one right mode. The essence of the proof does not depend on $N_L(=N_R)$, and can be generalized to $N_L>1$. The task is to prove that once all excitations in the Lagrangian subgroup are condensed, the partition function of the low energy theory is the same as that of LLs, \ie the $u1_1\oplus \bar{u1}_1$ CFT. 

We consider one kind of Abelian state, constructed from double-layered FQH
stripe, and gap sectors along one edge totally (the top part of Fig.
\ref{asect}), and sectors from the other edge remain gapless (the bottom part
of Fig. \ref{asect}).  The claim is that Abelian states realized by such
construction, are always LLs, whose low energy theory is $u1_1\oplus
\bar{u1}_1$ CFT. 

The edge theory of Abelian FQH state, is described by a symmetric integer matrix $K$. Quasiparticles, created by operator $\ee^{\ii l^T \phi}$ are labeled by an integer vector $l$. Given two quasiparticles $l,m \in\ZZ^2$, the self-statistics of $l$ quasiparticle and the mutual statistics of two quasiparticle are
\begin{align}
\frac{\theta_{l}}{\pi}=l^TK^{-1} l,\quad \frac{\theta_{lm}}{2\pi}=l^TK^{-1} m.
\end{align}
In particular, a local excitation is {that can be created by local operators, \ie bosonic or fermionic operators}. One set of  local excitations is $\Gamma_0=K\ZZ^2$. We see that basis vectors are columns of $K=(k_1,k_2)$. It follows that $K$ matrix encodes the statistics of these local operators, 
\begin{align}
\frac{\theta_{ij}}{2\pi}=k_i^T K^{-1} k_j=K_{ij},
\end{align}
which is integral. Another set of local operators are
\begin{align}
\ee^{\ii \v l^T (\phi +\bar \phi)}, \quad \v l\in \ZZ^2, \label{anyonpair}
\end{align}
where $\bar{\phi}$ are fields of the other edge, described by $-K$. We see that the statistics of 
\begin{align}
\frac{\theta_{\v l\v m}}{2\pi}=\v l^TK^{-1}\v m-\v l^TK^{-1}\v m=0,
\end{align}
and 
\begin{align}
\frac{\theta_{\v l m}}{2\pi}=\v l^TK^{-1} m \in \ZZ^2,
\end{align}
since $m\in K\ZZ^2$.

The gappable condition for a single edge is there are a set of quasiparticles $m\in \ZZ^2$ that forms a ``Lagrangian subgroup'' $\calM$\cite{levin2013protected}. One way to fully gap the edge is to add perturbation
\begin{align}
\delta \calL=\sum_{m\in \calM} g_n \cos (m^T\bar\phi).
\end{align}

We see that when $g_n>0$ are sufficiently large, the quasiparticles labeled by $m$ are condensed, \ie $\ee^{\ii m^T\bar \phi}\sim 1$. The question is what is the gapless theory for the other edge that remains gapless. 

When all $m$ are condensed, the local excitations in (\ref{anyonpair}) become
\begin{align}
\ee^{\ii \v m^T (\phi +\bar \phi)}\rightarrow \ee^{\ii m^T \phi},
\end{align}
Now the lattice of local opetors is
\begin{align}
\Gamma=\oplus_{m\in \calM} (m+K\ZZ^2).
\end{align}
This is still a two dimensional integeral lattice, and can thus be represented by $\Gamma=U\ZZ^2$, where $U$ is a integral matrix. Levin proved that now $P=U^T K^{-1}U$ is a symmetric integral matrix with vanishing signature, $\det P=\pm 1$. In fact, $P$ is the effective $K$ now 
\begin{align}
P=U^T K^{-1}U.
\end{align}
 Next by another linear superposition, 
\begin{align}
(W^T)^{-1}=UW_0,\quad W^TK W=\eta,
\end{align}
this means that when the null vectors become local operators as well. Meanwhile, we tune the interactions at the upper edge appropriately, so that $\tilde{V}_{ij}=v\delta_{ij}$, 
\begin{align}
V=(W^T)^{-1}\tilde{V}W^{-1}.
\end{align}

Let us illustrate the proof with two examples. First, we consider 
\begin{align}
K=\begin{pmatrix}
m & 0 \\ 0 &  -m
\end{pmatrix}.
\end{align}

Pick a null vector $\v l= (1,1)^T$, and $k_1$ to form the new basis, 
\begin{align}
U=\begin{pmatrix}
1 & m \\
1 & 0
\end{pmatrix}
\end{align}
has mutual statistics
\begin{align}
P= \begin{pmatrix}
0 & 1 \\
1 & m
\end{pmatrix}.
\end{align}
By a second basis transformation $W_0$ we find the basis . 
\begin{align}
W_0=\begin{pmatrix}
\frac{1-m}{2} & \frac{1+m}{2}\\ 1 & -1
\end{pmatrix}, \quad UW_0=\frac{1}{2}\begin{pmatrix}
1+m & 1-m \\ 1-m & 1+m
\end{pmatrix}.
\end{align}
With this basis, the statistics is
\begin{align}
K_{eff}=K_F.
\end{align}
The interaction is tuned to
\begin{align}
V=\frac{v}{2}\begin{pmatrix}
1+m^2 & 1-m^2 \\ 1-m^2 & 1+m^2
\end{pmatrix}.
\end{align}
In this basis, all vectors are mutually trivial. 
\begin{align}
\ee^{\ii u^T\phi}=\ee^{\ii \gamma^T W^{-1}\phi}=\ee^{\ii \gamma^T\tilde{\phi}}
\end{align}
where $\gamma=W^Tu$
it contributes 
\begin{align}
Z(\tau)=\sum_{\gamma\in \tilde{\Gamma}}|\eta (q)|^{-2} q^{\frac{1}{2}\gamma_1^2} (q^*)^{\frac{1}{2}\gamma_2^2} 
\end{align}
to $Z(\tau)$. 

$\gamma$ is in the lattice
\begin{align}
\tilde{\Gamma}=W^T U\ZZ^2=W_0^{-1}U^{-1}U\ZZ^2=W_0^{-1}\ZZ^2.
\end{align}
Since $W_0$ is an integer matrix with $\det W_0=\pm 1$, so is $W_0^{-1}$. Then from the theorem of lattice theory, 
\begin{align}
\tilde{\Gamma}\equiv \ZZ^2.
\end{align}
Therefore
\begin{align}
Z(\tau)=\sum_{\gamma\in \ZZ^2}|\eta (q)|^{-2} q^{\frac{1}{2}\gamma_1^2} (q^*)^{\frac{1}{2}\gamma_2^2} =\chi_0^{u1_1}\bar \chi_0^{u1_1},
\end{align}
and is the same as the partition function of $u1_1$ CFT. 
\begin{align}
Z(\tau)=\sum_{m\in \calM} \sum_{\gamma\in\Gamma_m}|\eta (q)|^{-2} q^{\frac{1}{2}\gamma_1^2} (q^*)^{\frac{1}{2}\gamma_2^2}.
\end{align}

We point out that for the velocity matrix 
\begin{align}
V=\begin{pmatrix}
v & 0 \\ 0 & v
\end{pmatrix},
\end{align}
the matrix to make $K\rightarrow \eta$ and $V$ diagonal is 
\begin{align}
W'=\frac{1}{\sqrt{m}}\eta,
\end{align}
and in this case, the partition function is
\begin{align}
Z(\tau)=\sum_{m\in P\ZZ} Z_{m}(\tau).
\end{align}
It is modular invariant, since
\begin{align}
Z(-\frac{1}{\tau})=&\sum_{m,n\in P\ZZ} S_{mn}Z_{n}(\tau),\nn\\
S_{mn}=&\frac{1}{|\det P|}\ee^{\ii 2\pi m^T K_{eff}^{-1}n}.
\end{align}
The low energy theory is $u1_M\times \bar{u1}_M$ CFT. And 
\begin{align}
(W')^{-1}W=\frac{1}{2\sqrt{m}}\begin{pmatrix}
1+m & -1+m \\ -1+m & 1+m
\end{pmatrix},
\end{align}
is a boosting matrix in $SO(1,1)$. 

Second, we consider a general case with non-trivial Lagrangian subgroup, 
\begin{align}
K=\begin{pmatrix}
-1 & 1 \\ 1 & 3
\end{pmatrix}.
\end{align}
We can choose a Lagrange subgroup $\calM=\{ (0,0)^T, (1,1)^T\}$. And we find
\begin{align}
U=\begin{pmatrix}
1 & 1 \\ 1 & -1
\end{pmatrix},\  W_0=\begin{pmatrix}
1 & 0 \\ -1 & 1
\end{pmatrix},\ W=\begin{pmatrix}
\frac{1}{2} & 1 \\ \frac{1}{2} & 0
\end{pmatrix}.
\end{align}

We conclude that from the double-layered FQH and gapping one edge, the gapless phase obtained is still a Tomonaga-Luttinger liquid. 

\section{Conformal field theory extended with current algebras}
\label{appCFT}

The theory of conformal field theory with extended symmetry is well-known. In this section, we summarize some defining knowledge to introduce our convention. We refer the readers to Francesco's textbook \cite{francesco2012conformal} for further details.

\subsection{Current and primary fields}
\label{prime}

The ${u1}$, ${su2}_1$, and ${su2}_2$ CFT's are not only invariant under conformal symmetry, but also invariant under current algebras. Current algebras are generated by \emph{currents}, chiral primary fields with scaling dimension $1$, and denoted as $J^a (z)$. The defining OPE of the level-$k$ current algebra $g_k$ is
\begin{align}
J^a (z) J^b (w)\sim \frac{k\delta_{ab}}{(z-w)^2}+\sum_c \ii f_{abc}\frac{J^c(w)}{(z-w)},
\end{align}
where $f_{abc}$ is the structure constant of the corresponding Lie algebra $g$. In particular, for $u1_M$ current algebra, 
\begin{align}
&J^0(z)J^0(w)\sim \frac{M}{(z-w)^2},
\end{align}
for $su2_k$ current algebra, in the spin basis, whose $su2$ generators satisfying
\begin{align}
[J^+, J^-]=2J^0, \quad [J^0, J^\pm]=\pm J^\pm,
\end{align}
 the OPEs are
\begin{align}
&J^0(z)J^0(w)\sim \frac{k/2}{(z-w)^2},\quad J^0(z) J^\pm (w)\sim \frac{\pm J^\pm (w)}{(z-w)},\nn\\
&J^+(z) J^-(w)\sim\frac{k}{(z-w)^2}+\frac{2J^0(w)}{(z-w)}.
\end{align}

Analogous to highest weight representations of Lie algebras, the
highest weight representations of the current algebras are labeled by primary
fields. The defining OPE of a primary field $V_\lambda(z)$ is
\begin{align}
J^a (z) V_\lambda (w)\sim \frac{t^a_\lambda V_\lambda (w)}{z-w},
\end{align}
where $t^a_\lambda$ is the representation matrix for $J^a$ of $g$ in the representation labeled by $\lambda$.

Current algebras can be represented in
terms of different quantum fields, as long as the different representations
produce the same correlation functions (so-called \emph{quantum equivalence}).
In particular, primary fields of the above current algebras can be expressed
in terms of the chiral compactified bosonic field $\phi$ and primary fields in $\text{Is}$ CFT. The representations of current fields and primary fields, and their scaling dimensions $h$ are listed in Table \ref{calgfields}. Table \ref{Isfields} lists the primary fields and the scaling dimensions $h$.
\begin{table}
\begin{tabular}{ |C | L  C |}
\hline
\rule{0pt}{3ex} \hbox{CFT} & \hbox{field} & h \\
\hline
\rule{0pt}{3.3ex} u1_M & J^0=\ii\sqrt{M}\partial \phi & 1\\
\rule{0pt}{3ex}& V_{k}=\ee^{\ii \frac{k}{\sqrt{M}}\phi}, k=0,\ldots, M-1  & \frac{k^2}{2M} \\
\hline
\rule{0pt}{3.3ex} {su2}_1 & J^0=\frac{\ii}{\sqrt{2}} \partial \phi  & 1 \\
\rule{0pt}{3ex} & J^\pm= \ee^{\pm \ii \sqrt{2} \phi}  & 1 \\
\rule{0pt}{3ex} & V_{\frac12,\pm \frac12}=\ee^{\pm \ii \frac{1}{\sqrt{2}}\phi}  &\frac{1}{4} \\
\hline
\rule{0pt}{3.3ex}  {su2}_2 & J^0=\ii \partial \phi  & 1 \\
\rule{0pt}{3ex} & J^\pm= \sqrt{2}\eta \ee^{\pm \ii \phi}  & 1 \\
\rule{0pt}{3ex} & V_{\frac12,\pm \frac12}=\sigma \ee^{\pm \ii \phi /2} &\frac{3}{16} \\
\rule{0pt}{3ex} & V_{1,\pm 1}=\ee^{\pm \ii \phi }  &\frac{1}{2} \\
\rule{0pt}{3ex} & V_{1,0}=\eta  & \frac{1}{2}\\
\hline
\end{tabular}
\caption{Fields of CFTs with current algebras. $J^i$ are current fields, and others are primary fields (except the identity field with scaling dimension $0$) of the current algebra. $a_0$ is the cut-off length scale.}\label{calgfields}
\begin{tabular}{ | C  C |}
\hline
\rule{0pt}{2.5ex}  \hbox{Primary field} & h \\
\hline
 1 & 0 \\
 \si & \frac{1}{16} \\
 \eta & \frac12 \\
\hline
\end{tabular}
\caption{Primary fields of Ising CFT}\label{Isfields}
\end{table}

\subsection{Operator product expansion}\label{ope}
The OPE of fermion operators is
\begin{align}
\eta (z)\eta (w)\sim \frac{1}{z-w}.
\label{fcorre}
\end{align}
The OPE of Ising primary fields are
\begin{align}
\sigma (z)\sigma (w)\sim & \frac{1}{(z-w)^{\frac{1}{8}}}+C(z-w)^{\frac{3}{8}}\eta (w),\nn\\
\eta (z)\sigma (w)\sim &\frac{1}{(z-w)^{\frac{1}{2}}}\mu (w),\\
\eta (z)\mu (w)\sim &\frac{1}{(z-w)^{\frac{1}{2}}}\sigma (w).\nn
\label{iscorre}
\end{align}
where $\mu$ denotes the disorder operator dual to the spin operator $\sigma$, and it has the same OPE and conformal dimensions as $\sigma$. All other OPEs can be derived from (\ref{bcorre}), (\ref{fcorre}) and (\ref{iscorre}). 

\subsection{Characters and modular transformations}
Each primary field corresponds to a highest weight-representation of the current algebra. The \emph{character} of a highest weight-representation encodes the degeneracy, or \emph{multiplicities} of states with same quantum numbers. 

\label{Mtrn}

\subsubsection{$u1_M$ character}

The $u1_M$ character $\chi^{u1_M}_m$ is given by 
\begin{align}
 \chi^{u1_M}_m(\tau) = q^{-\frac{1}{24}} \frac{\sum_{n=-\infty}^\infty 
q^{\frac{1}{2} (\frac{m}{R}+n R)^2}
}{\prod_{n=1}^\infty (1-q^n)},\ \ \ \ R^2=M,
\end{align}
which contains primary fields of conformal symmetry,
\begin{align}
\label{u1Op}
\ee^{\ii (\frac{m}{R}+n R) \phi}.
\end{align}
When $M=$ even,
under modular transformation, $\chi^{u1_M}_m$ transform as
\begin{align}
 \chi^{u1_M}_i(-\frac{1}{\tau}) =&
\sum_j S_{ij} \chi^{u1_M}_j(\tau),\ \ \
S_{ij}=\frac{1}{\sqrt{M}} \ee^{-\ii 2\pi \frac{ij}{M} },
\nonumber\\
 \chi^{u1_M}_i(\tau+1)=& \ee^{\ii 2\pi \left(\frac{1}{2} \frac{i^2}{M}-\frac{1}{24}\right)}\chi^{u1_M}_i(\tau).
\end{align} 

When $M=$ odd,
$\chi^{u1_M}_m$ corresponds to partition function of a fermionic system.

\subsubsection{$su2_k$ character}
The $\hat{su}(2)_k$ character $\chi_{j}^{su2_k}(\tau)$ is
\begin{align}
\chi_{j}^{su2_k}(q)=&\frac{q^{(2j+1)^2/4(k+2)}}{[\eta (q)]^3}\nn\\
&\cdot \sum_{n\in\ZZ} \left[2j+1+2n(k+2)\right] q^{n[2j+1+(k+2)n]}
\end{align}
where $j\in\calP=\left\{0,\frac{1}{2},\cdots,\frac{k}{2}\right\}$. The modular transformations are
\begin{align}
&\chi^{su2_k}_j(-1/\tau)=\sum_{l\in\calP}S_{jl}\chi^{su2_k}_l(\tau),\nn\\
&S_{jl}=\sqrt{\frac{2}{k+2}}\sin \left[ \frac{\pi (2j+1)(2l+1)}{k+2}\right]\\
&\chi^{su2_k}_j(\tau+1)=\sum_{l\in\calP}T_{jl}\chi^{su2_k}_l(\tau),\ T_{jl}=\delta_{jl}\ee^{\ii 2\pi \left(\frac{j(j+1)}{k+2}-\frac{c}{24}\right)}.\nn
\end{align}
and $c=\frac{3k}{k+2}$.

\subsubsection{Ising characters}

The Ising characters are
\begin{align}
\chi_{r,s}(q)=\eta^{-1}(q)\sum_{n\in \ZZ}\left(q^{(24n+4r-3s)^2/48}-q^{(24n+4r+3s)^2/48}\right),
\end{align}
where
\begin{align}
\chi_{\bf 1}\equiv \chi_{1,1},\quad \chi_\sigma\equiv \chi_{1,2},\quad \chi_\psi\equiv \chi_{2,1}.
\end{align}

In the basis $(\chi_{\bf 1}, \chi_\sigma,\chi_\eta)$, the $S$ matrix is
\begin{align}
S=\frac{1}{2}\begin{pmatrix}
1 & \sqrt{2} & 1 \\
 \sqrt{2} & 0 & -\sqrt{2} \\
 1 & -\sqrt{2} & 1
\end{pmatrix},
\end{align}
and the $T$ operation is
\begin{align}
T\chi_\mu=\ee^{\ii2\pi \left(h_\mu-\frac{1}{48}\right)}\chi_\mu,
\end{align}
where $h_{\bf 1}=0, h_\sigma=\frac{1}{16}$ and $h_\eta=\frac{1}{2}$.

\section{Exactly marginal operators}

Consider a perturbation $\delta S=\frac{1}{2\pi}\sum_i g_i\int d^2z \phi_i(z,\bar z)$, where $\phi_i (z,\bar{z})$ is the marginal Virasora primary field with weights $(h_i, \bar h_i)$. The correction of the correlations of $O(z,\bar z)$, a product of primary fields, are
\begin{align}
\frac{\delta }{\delta g_j} \langle O \rangle =\frac{1}{2\pi}\int d^2 w\langle \phi_j (w,\bar w) O\rangle.
\end{align}
In particular, by taking $O=\phi_i (z_1,\bar z_1)\phi_i (z_2,\bar z_2)$, one can show that to the first order in $\delta g_i$, the correction to the weights are
\begin{align}
\delta h_i=\delta \bar h_i =-\sum_j c_{iij} \delta g_j.
\end{align}
The necessary condition for a marginal operator to be exactly marginal, i.e., preserving conformal symmetry when $g_i$ is turned on continuously, is that $c_{iij}=0$, for any primary field $\phi_j$.

\section{Partition functions of free spin-$\frac{1}{2}$ Dirac fermions}\label{spinhalfDirac}

The spin-$\frac{1}{2}$ Dirac fermions can be considered as the representation of the $u1_2\oplus su2_1$ current algebra.
The partition functions are
\begin{align}
Z^\text{Dirac}_{AE}=&\chi^{u1_2}_0\chi^{su2_1}_0\bchi^{u1_2}_0\bchi^{su2_1}_0+\chi^{u1_2}_{1}\chi^{su2_1}_{1/2}\bchi^{u1_2}_{1}\bchi^{su2_1}_{1/2},\nn\\
=&\frac{1}{2}\left(\left|\frac{\theta_3(q)}{\eta(q)}\right|^4+\left|\frac{\theta_4(q)}{\eta(q)}\right|^4\right),\nn\\
=&\sum_{k=0,2,4}\binom{4}{k}(Z^\text{Maj}_{AE})^k(Z^\text{Maj}_{AO})^{4-k},\nn\\
Z^\text{Dirac}_{AO}=&\chi^{u1_2}_0\chi^{su2_1}_0\bchi^{u1_2}_{1}\bchi^{su2_1}_{1/2}-\chi^{u1_2}_{1}\chi^{su2_1}_{1/2}\bchi^{u1_2}_0\bchi^{su2_1}_0,\nn\\
=&\frac{1}{2}\left(\left|\frac{\theta_3(q)}{\eta(q)}\right|^4-\left|\frac{\theta_4(q)}{\eta(q)}\right|^4\right),\nn\\
=&\sum_{k=1,3}\binom{4}{k}(Z^\text{Maj}_{AE})^k(Z^\text{Maj}_{AO})^{4-k},\nn\\
Z^\text{Dirac}_{PE}=&Z^\text{Dirac}_{PO},\nn\\
=\frac{1}{2}(&\chi^{u1_2}_0\chi^{su2_1}_{1/2}+\chi^{u1_2}_{1}\chi^{su2_1}_{0})(\bchi^{u1_2}_0\bchi^{su2_1}_{1/2}+\bchi^{u1_2}_{1}\bchi^{su2_1}_{0}),\nn\\
=&\frac{1}{2}\left|\frac{\theta_2(q)}{\eta(q)}\right|^4=8(Z^\text{Maj}_{PE})^4,
\end{align}
where we have used Jacobi's theta functions $\theta_a(\tau)$ to track the various identities between characters. To understand the multiplicity in $Z_{PE}^\text{Dirac} (Z_{PO}^\text{Dirac})$, we compare it with
\begin{align}
Z_{PE}^\text{Maj}=&Z_{PO}^\text{Maj}=\frac{1}{2}\left|\frac{\theta_2(q)}{\eta(q)}\right|.
\end{align}
Since $\theta_2/\eta$ has $2$-fold degeneracy, there are no degeneracy in $Z_{PE}^\text{Maj}$, but a $8$-fold degeneracy in $Z_{PE}^\text{Dirac}$. Physically, in $n$ chains of Majorana fermion with periodic boundary condition, there is a ground state degeneracy of $2^{n}$ due to $n$ zero modes. For fixed fermion number parity, the degeneracy is
\begin{align}
\sum_{k=0,2,\cdots, k\leq n} \binom{n}{k}=\sum_{k=1,3,\cdots, k\leq n}\binom{n}{k}=2^{n-1}.
\end{align}

\section{Scaling dimensions of primary fields in chiral metallic phase}
\label{nabMSscal}

Here we list the scaling dimensions $(h,\bar{h})$ of primary fields corresponding to each term of characters $Z_{AE}$ (\ref{ZAE}),  are
\begin{enumerate}[label=(\arabic*)]
\item In $Z_{AE}$ (\ref{ZAE}):
\begin{align}
\label{hAE}
(0,0), \ \left(\frac{3}{4},\frac{3}{4}\right),\ \left(\frac{3}{4},\frac{3}{4}\right),\ (1,0)\ \left(\frac{1}{4},\frac{1}{4}\right),\ \left(\frac{1}{2},\frac{1}{2}\right).
\end{align}
\item In $Z_{AO}$ (\ref{ZAO}):
\begin{align}
\label{hAO}
\left(0,\frac{1}{2}\right), \ \left(\frac{3}{4},\frac{1}{4}\right),\ \left(\frac{3}{4},\frac{1}{4}\right),\ \left(1,\frac{1}{2}\right)\ \left(\frac{1}{4},\frac{3}{4}\right),\ \left(\frac{1}{2},0\right).
\end{align}
\item In $Z_{PE}$ or $Z_{PO}$  (\ref{ZPE}):
\begin{align}
&\quad \left(\frac{1}{4},\frac{1}{4}\right),\quad \left(\frac{1}{4},\frac{1}{4}\right), \quad \left(\frac{1}{2},\frac{1}{2}\right), \quad \left(\frac{1}{2},\frac{1}{2}\right), \nn\\
&\quad \left(\frac{1}{2},\frac{1}{2}\right), \quad \left(\frac{1}{2},\frac{1}{2}\right), \quad \left(\frac{5}{4},\frac{1}{4}\right),\quad \left(\frac{5}{4},\frac{1}{4}\right),\nn\\
&\quad \left(\frac{1}{4},\frac{1}{4}\right),\quad \left(\frac{1}{4},\frac{1}{4}\right), \quad \left(\frac{1}{2},\frac{1}{2}\right), \quad \left(\frac{1}{2},\frac{1}{2}\right).
\end{align}
\end{enumerate}

\section{Computation of modular invariant partition functions}\label{minvZ}

\subsection{Fusion algebra in chiral metallic phase}

Fields generated from OPEs of fermion operators can be summarised in terms of fusion algebras. 
The primary fields in the partion functions can be organized in the following fashion. We denote the vacuum and local operators as
\begin{align}
\nu_1 =& V^{u1_2}_0V^{su2_2}_{0}{\bf 1}_{\text{Is}} \bar V^{u1_2}_0\bar V^{su2_1^1}_{0}\bar V^{su2_1^2}_{0},\nn\\
\nu_2\equiv &\bar{\psi}_{\alpha 2}= V^{u1_2}_0V^{su2_2}_{0}{\bf 1}_{\text{Is}} \bar V^{u1_2}_1\bar V^{su2_1^1}_{0}\bar V^{su2_1^2}_{\frac{1}{2}},\nn\\
\mu_1\equiv &\psi_{\alpha 2}= V^{u1_2}_1V^{su2_2}_{\frac{1}{2}}\sigma \bar V^{u1_2}_0\bar V^{su2_1^1}_{0}\bar V^{su2_1^2}_{0},\nn\\
\mu_2\equiv &\psi_{\alpha 1}^\dagger\bar \psi_{\beta 1}= V^{u1_2}_0V^{su2_2}_{\frac{1}{2}}\sigma \bar V^{u1_2}_0\bar V^{su2_1^1}_{\frac{1}{2}}\bar V^{su2_1^2}_{0}.
\end{align}

All the primary fields generated from the fusion of them are
\begin{align}
\mu_3=\nu_2\times \mu_1=&V^{u1_2}_1V^{su2_2}_{\frac{1}{2}}\sigma \bar V^{u1_2}_1\bar V^{su2_1^1}_{0}\bar V^{su2_1^2}_{\frac12},\nn\\
\mu_4=\nu_2\times \mu_2=&V^{u1_2}_0V^{su2_2}_{\frac{1}{2}}\sigma \bar V^{u1_2}_1\bar V^{su2_1^1}_{\frac12}\bar V^{su2_1^2}_{\frac12}.\nn\\
\end{align}

The other fields are $\nu_{ij}, 1\leq i\leq 4, 0\leq j\leq 3$ defined as
\begin{align}
\begin{split}
\nu_{i0}=&\nu_{i},\\
\nu_{i1}=&V^{u1_2}_0V^{su2_2}_{1}{\bf 1}_{\text{Is}}\times \nu_{i0},\\
\nu_{i2}=&V^{u1_2}_0V^{su2_2}_{0}\eta\times \nu_{i0},\\
\nu_{i3}=&V^{u1_2}_0V^{su2_2}_{1}\eta\times \nu_{i0}.
\end{split}
\label{nuij}
\end{align}

$\nu_1,\nu_2$ has been given. And $\nu_3, \nu_4$ is determined by
\begin{align}
\mu_1\times\mu_2=\sum_{j=0}^3 \nu_{3j},\quad \mu_1\times\mu_4=\sum_{j=0}^3 \nu_{4j}.
\end{align}
The solution is
\begin{align}
\begin{split}
\nu_3=&V^{u1_2}_1V^{su2_2}_{0}{\bf 1}_{\text{Is}} \bar V^{u1_2}_0\bar V^{su2_1^1}_{\frac12}\bar V^{su2_1^2}_{0},\\
\nu_4=&V^{u1_2}_1V^{su2_2}_{0}{\bf 1}_{\text{Is}} \bar V^{u1_2}_1\bar V^{su2_1^1}_{\frac12}\bar V^{su2_1^2}_{\frac12}.
\end{split}
\end{align}
Note that all $\nu_{i0}\equiv \nu_i$ has the partial vacuum $V^{su2_2}_{0}{\bf 1}_{\text{Is}}$, thus the fusion in (\ref{nuij}) is Abelian and trivial. 

The primary fields defined above form a complete set such that the fusion algebra is closed. We denote this fusion algebra as
\begin{align}
\calC=\{ \mu_{l},\nu_{ij}|\ 1\leq l,i\leq 4, 0\leq j\leq 3\}.
\end{align}
\subsection{A receipt to look for modular invariant partition functions}
From the fusion algebra $\calC$, we can look for solutions of modular covariant $Z$. 
\begin{enumerate}
\item Assign non-negative integeral multiplicities for the characters $\chi_{c_j}$ for $c_j\in\calC$, and sum over them to get an initial partition function $z$. 
\item Generate a set of partition functions,
\begin{align}
z \quad Sz,\quad TSz,\quad STSz,\quad Tz,\quad STq
\end{align}
Since $S^2=1$,\footnote{We do not consider the more involved case that $S^2=C$ that $C\neq 1$.} $T^2=1, (TS)^3=1$, they are all the partition functions generated by modular transformations. 
Next one check that all the multiplicities in these vectors are non-negative integer, and the primary fields in these partition funcitons are either bosonic or fermionic fields. 
\item There are three basis vector of $S$ invariant partion funcitons,
\begin{align}
z_1=z+Sz,\; z_T=Tz+STz,\; z_{TS}=TSz+STSz.
\end{align}
Therefore all vectors $Z_{AA}=\sum_{i=1}^3 a_i z_i, \ a_i\in \ZZ, a_i\geq 0$ are invariant under $S$ transformation. 
\item The other sectors of partion function can be generated by $Z_{AP}=TZ_{AA}, Z_{PA}=SZ_{AP}$.
\item $Z_{PP}$ is to find $\sum_{i=1}^3 a_i z_i, \ a_i\in \ZZ, a_i\geq 0$ that is purely bosonic. 
\end{enumerate}

The only free choice in the receipt is in the first step, the general guide is to assign small integers to the multiplicites.

\subsubsection{Another modular covariant partition function}
There is another independent solution for modular covariant partition functions, as shown below.
\begin{align}
Z'_{AE}=\chi_0^{u1_2}\chi_0^{su2_2}\chi_0^{\text{Is}}\bchi_0^{u1_2}\bchi_0^{su2_1^1}\bchi_0^{su2_1^2}\nn\\
+\chi_1^{u1_2}\chi_1^{su2_2}\chi_0^{\text{Is}}\bchi_1^{u1_2}\bchi_{1/2}^{su2_1^1}\bchi_{1/2}^{su2_1^2}\nn\\
+\chi_1^{u1_2}\chi_0^{su2_2}\chi_\eta^{\text{Is}}\bchi_1^{u1_2}\bchi_{1/2}^{su2_1^1}\bchi_{1/2}^{su2_1^2}\nn\\
+\chi_0^{u1_2}\chi_1^{su2_2}\chi_\eta^{\text{Is}}\bchi_0^{u1_2}\bchi_{0}^{su2_1^1}\bchi_{0}^{su2_1^2}\nn\\
+\chi_1^{u1_2}\chi_0^{su2_2}\chi_0^{\text{Is}}\bchi_0^{u1_2}\bchi_{1/2}^{su2_1^1}\bchi_0^{su2_1^2}\nn\\
+\chi_0^{u1_2}\chi_1^{su2_2}\chi_0^{\text{Is}}\bchi_1^{u1_2}\bchi_{0}^{su2_1^1}\bchi_{1/2}^{su2_1^2}\nn\\
+\chi_0^{u1_2}\chi_0^{su2_2}\chi_\eta^{\text{Is}}\bchi_1^{u1_2}\bchi_{0}^{su2_1^1}\bchi_{1/2}^{su2_1^2}\nn\\
+\chi_1^{u1_2}\chi_1^{su2_2}\chi_\eta^{\text{Is}}\bchi_0^{u1_2}\bchi_{1/2}^{su2_1^1}\bchi_{0}^{su2_1^2}.
\end{align}
The scaling dimensions of primary field corresponding to each term are
\begin{align}
& (0,0), \quad \left(\frac{3}{4},\frac{3}{4}\right),\quad \left(\frac{3}{4},\frac{3}{4}\right),\quad (1,0)\nn\\
&\left(\frac{1}{4},\frac{1}{4}\right),\quad \left(\frac{1}{2},\frac{1}{2}\right),\quad \left(\frac{1}{2},\frac{1}{2}\right),\quad \left(\frac{5}{4},\frac{1}{4}\right).
\end{align}

\begin{align}
Z'_{AO}=&\chi_0^{u1_2}\chi_0^{su2_2}\chi_0^{\text{Is}}\bchi_1^{u1_2}\bchi_0^{su2_1^1}\bchi_{1/2}^{su2_1^2}\nn\\
&+\chi_1^{u1_2}\chi_1^{su2_2}\chi_0^{\text{Is}}\bchi_0^{u1_2}\bchi_{1/2}^{su2_1^1}\bchi_{0}^{su2_1^2}\nn\\
&+\chi_1^{u1_2}\chi_0^{su2_2}\chi_\eta^{\text{Is}}\bchi_0^{u1_2}\bchi_{1/2}^{su2_1^1}\bchi_{0}^{su2_1^2}\nn\\
&+\chi_0^{u1_2}\chi_1^{su2_2}\chi_\eta^{\text{Is}}\bchi_1^{u1_2}\bchi_{0}^{su2_1^1}\bchi_{1/2}^{su2_1^2}\nn\\
&+\chi_1^{u1_2}\chi_0^{su2_2}\chi_0^{\text{Is}}\bchi_{1}^{u1_2}\bchi_{1/2}^{su2_1^1}\bchi_{1/2}^{su2_1^2}\nn\\
&+\chi_0^{u1_2}\chi_1^{su2_2}\chi_0^{\text{Is}}\bchi_0^{u1_2}\bchi_{0}^{su2_1^1}\bchi_{0}^{su2_1^2}\nn\\
&+\chi_0^{u1_2}\chi_0^{su2_2}\chi_\eta^{\text{Is}}\bchi_0^{u1_2}\bchi_{0}^{su2_1^1}\bchi_{0}^{su2_1^2}\nn\\
&+\chi_1^{u1_2}\chi_1^{su2_2}\chi_\eta^{\text{Is}}\bchi_1^{u1_2}\bchi_{1/2}^{su2_1^1}\bchi_{1/2}^{su2_1^2}.
\end{align}
The scaling dimensions of primary field corresponding to each term are
\begin{align}
\begin{split}
& \left(0,\frac{1}{2}\right), \quad \left(\frac{3}{4},\frac{1}{4}\right),\quad \left(\frac{3}{4},\frac{1}{4}\right),\quad \left(1,\frac{1}{2}\right)\\
& \left(\frac{1}{4},\frac{3}{4}\right),\quad \left(\frac{1}{2},0\right),\quad \left(\frac{1}{2},0\right),\quad \left(\frac{5}{4},\frac{3}{4}\right)
\end{split}.
\end{align}
\begin{align}
Z'_{PE}=&Z'_{PO},\nn\\
=&\chi_0^{u1_2}\chi_{1/2}^{su2_2}\chi_{\sigma}^{\text{Is}}\bchi_1^{u1_2}\bchi_0^{su2_1^1}\bchi_0^{su2_1^2}\nn\\
&+\chi_0^{u1_2}\chi_{1/2}^{su2_2}\chi_{\sigma}^{\text{Is}}\bchi_0^{u1_2}\bchi_0^{su2_1^1}\bchi_{1/2}^{su2_1^2}\nn\\
&+\chi_1^{u1_2}\chi_{1/2}^{su2_2}\chi_{\sigma}^{\text{Is}}\bchi_1^{u1_2}\bchi_{1/2}^{su2_1^1}\bchi_0^{su2_1^2}\nn\\
&+\chi_1^{u1_2}\chi_{1/2}^{su2_2}\chi_{\sigma}^{\text{Is}}\bchi_0^{u1_2}\bchi_{1/2}^{su2_1^1}\bchi_{1/2}^{su2_1^2}.
\end{align}
The scaling dimensions of primary field corresponding to each term are
\begin{align}
\quad \left(\frac{1}{4},\frac{1}{4}\right),\quad \left(\frac{1}{4},\frac{1}{4}\right), \quad \left(\frac{1}{2},\frac{1}{2}\right), \quad \left(\frac{1}{2},\frac{1}{2}\right).
\end{align}

\section{Solutions of \eqn{Mcov1}}
\label{sols}

Using the characters of $u1_4$ CFT, $\chi^{u1_4}_m$, and the characters of
Ising CFT, $\chi^\text{Is}_h$, we can construct many solution of \eqn{Mcov1}.
The following is a list of 36 solutions.  (The list may not be complete.)

\begin{align}
\label{sol1}
 Z_{AE^f,PE^s} &= \big(|\chi^{u1_4}_0 |^2 + |\chi^{u1_4}_2|^2 \big)\big( |\chi^\text{Is}_0|^2+|\chi^\text{Is}_\frac{1}{2}|^2 \big) 
 \nonumber \\ 
Z_{PO^f,PE^s} &= \big( \chi^{u1_4}_1 \bar \chi^{u1_4}_{-1} + \chi^{u1_4}_{-1} \bar \chi^{u1_4}_1 \big)\big( |\chi^\text{Is}_0|^2+|\chi^\text{Is}_\frac{1}{2}|^2 \big) 
 \nonumber \\ 
Z_{PE^f,PE^s} &= \big(|\chi^{u1_4}_1 |^2 + |\chi^{u1_4}_{-1}|^2 \big)\big( |\chi^\text{Is}_0|^2+|\chi^\text{Is}_\frac{1}{2}|^2 \big) 
 \nonumber \\ 
Z_{AO^f,PE^s} &= \big( \chi^{u1_4}_0 \bar \chi^{u1_4}_2 + \chi^{u1_4}_2 \bar \chi^{u1_4}_0 \big)\big( |\chi^\text{Is}_0|^2+|\chi^\text{Is}_\frac{1}{2}|^2 \big) 
 \nonumber \\ 
Z_{AE^f,PO^s} &= \big(|\chi^{u1_4}_0 |^2 + |\chi^{u1_4}_2|^2 \big)|\chi^\text{Is}_\frac{1}{16}|^2 
 \nonumber \\ 
Z_{PO^f,PO^s} &= \big( \chi^{u1_4}_1 \bar \chi^{u1_4}_{-1} + \chi^{u1_4}_{-1} \bar \chi^{u1_4}_1 \big)|\chi^\text{Is}_\frac{1}{16}|^2 
 \nonumber \\ 
Z_{PE^f,PO^s} &= \big(|\chi^{u1_4}_1 |^2 + |\chi^{u1_4}_{-1}|^2 \big)|\chi^\text{Is}_\frac{1}{16}|^2 
 \nonumber \\ 
Z_{AO^f,PO^s} &= \big( \chi^{u1_4}_0 \bar \chi^{u1_4}_2 + \chi^{u1_4}_2 \bar \chi^{u1_4}_0 \big)|\chi^\text{Is}_\frac{1}{16}|^2 
 \nonumber \\ 
Z_{AE^f,AE^s} &= \big(|\chi^{u1_4}_0 |^2 + |\chi^{u1_4}_2|^2 \big)|\chi^\text{Is}_\frac{1}{16}|^2 
 \nonumber \\ 
Z_{PO^f,AE^s} &= \big( \chi^{u1_4}_1 \bar \chi^{u1_4}_{-1} + \chi^{u1_4}_{-1} \bar \chi^{u1_4}_1 \big)|\chi^\text{Is}_\frac{1}{16}|^2 
 \nonumber \\ 
Z_{PE^f,AE^s} &= \big(|\chi^{u1_4}_1 |^2 + |\chi^{u1_4}_{-1}|^2 \big)|\chi^\text{Is}_\frac{1}{16}|^2 
 \\ 
Z_{AO^f,AE^s} &= \big( \chi^{u1_4}_0 \bar \chi^{u1_4}_2 + \chi^{u1_4}_2 \bar \chi^{u1_4}_0 \big)|\chi^\text{Is}_\frac{1}{16}|^2 
 \nonumber \\ 
Z_{AE^f,AO^s} &= \big(|\chi^{u1_4}_0 |^2 + |\chi^{u1_4}_2|^2 \big)\big(\chi^\text{Is}_0 \bar \chi^\text{Is}_\frac{1}{2} + \chi^\text{Is}_\frac12 \bar \chi^\text{Is}_0\big) 
 \nonumber \\ 
Z_{PO^f,AO^s} &= \big( \chi^{u1_4}_1 \bar \chi^{u1_4}_{-1} + \chi^{u1_4}_{-1} \bar \chi^{u1_4}_1 \big)\big(\chi^\text{Is}_0 \bar \chi^\text{Is}_\frac{1}{2} + \chi^\text{Is}_\frac12 \bar \chi^\text{Is}_0\big) 
 \nonumber \\ 
Z_{PE^f,AO^s} &= \big(|\chi^{u1_4}_1 |^2 + |\chi^{u1_4}_{-1}|^2 \big)\big(\chi^\text{Is}_0 \bar \chi^\text{Is}_\frac{1}{2} + \chi^\text{Is}_\frac12 \bar \chi^\text{Is}_0\big) 
 \nonumber \\ 
Z_{AO^f,AO^s} &= \big( \chi^{u1_4}_0 \bar \chi^{u1_4}_2 + \chi^{u1_4}_2 \bar \chi^{u1_4}_0 \big)\big(\chi^\text{Is}_0 \bar \chi^\text{Is}_\frac{1}{2} + \chi^\text{Is}_\frac12 \bar \chi^\text{Is}_0\big) .
\nonumber 
\end{align}

\begin{align}
\label{sol2}
 Z_{AE^f,PE^s} &= \big(|\chi^{u1_4}_0 |^2 + |\chi^{u1_4}_2|^2 \big)\big( |\chi^\text{Is}_0|^2+|\chi^\text{Is}_\frac{1}{2}|^2 \big) 
 \nonumber \\ 
Z_{PO^f,PE^s} &= \big(|\chi^{u1_4}_1 |^2 + |\chi^{u1_4}_{-1}|^2 \big)\big( |\chi^\text{Is}_0|^2+|\chi^\text{Is}_\frac{1}{2}|^2 \big) 
 \nonumber \\ 
Z_{PE^f,PE^s} &= \big( \chi^{u1_4}_1 \bar \chi^{u1_4}_{-1} + \chi^{u1_4}_{-1} \bar \chi^{u1_4}_1 \big)\big( |\chi^\text{Is}_0|^2+|\chi^\text{Is}_\frac{1}{2}|^2 \big) 
 \nonumber \\ 
Z_{AO^f,PE^s} &= \big( \chi^{u1_4}_0 \bar \chi^{u1_4}_2 + \chi^{u1_4}_2 \bar \chi^{u1_4}_0 \big)\big( |\chi^\text{Is}_0|^2+|\chi^\text{Is}_\frac{1}{2}|^2 \big) 
 \nonumber \\ 
Z_{AE^f,PO^s} &= \big(|\chi^{u1_4}_0 |^2 + |\chi^{u1_4}_2|^2 \big)|\chi^\text{Is}_\frac{1}{16}|^2 
 \nonumber \\ 
Z_{PO^f,PO^s} &= \big(|\chi^{u1_4}_1 |^2 + |\chi^{u1_4}_{-1}|^2 \big)|\chi^\text{Is}_\frac{1}{16}|^2 
 \nonumber \\ 
Z_{PE^f,PO^s} &= \big( \chi^{u1_4}_1 \bar \chi^{u1_4}_{-1} + \chi^{u1_4}_{-1} \bar \chi^{u1_4}_1 \big)|\chi^\text{Is}_\frac{1}{16}|^2 
 \nonumber \\ 
Z_{AO^f,PO^s} &= \big( \chi^{u1_4}_0 \bar \chi^{u1_4}_2 + \chi^{u1_4}_2 \bar \chi^{u1_4}_0 \big)|\chi^\text{Is}_\frac{1}{16}|^2 
 \nonumber \\ 
Z_{AE^f,AE^s} &= \big(|\chi^{u1_4}_0 |^2 + |\chi^{u1_4}_2|^2 \big)|\chi^\text{Is}_\frac{1}{16}|^2 
 \nonumber \\ 
Z_{PO^f,AE^s} &= \big(|\chi^{u1_4}_1 |^2 + |\chi^{u1_4}_{-1}|^2 \big)|\chi^\text{Is}_\frac{1}{16}|^2 
  \\ 
Z_{PE^f,AE^s} &= \big( \chi^{u1_4}_1 \bar \chi^{u1_4}_{-1} + \chi^{u1_4}_{-1} \bar \chi^{u1_4}_1 \big)|\chi^\text{Is}_\frac{1}{16}|^2 
 \nonumber \\ 
Z_{AO^f,AE^s} &= \big( \chi^{u1_4}_0 \bar \chi^{u1_4}_2 + \chi^{u1_4}_2 \bar \chi^{u1_4}_0 \big)|\chi^\text{Is}_\frac{1}{16}|^2 
 \nonumber \\ 
Z_{AE^f,AO^s} &= \big(|\chi^{u1_4}_0 |^2 + |\chi^{u1_4}_2|^2 \big)\big(\chi^\text{Is}_0 \bar \chi^\text{Is}_\frac{1}{2} + \chi^\text{Is}_\frac12 \bar \chi^\text{Is}_0\big) 
 \nonumber \\ 
Z_{PO^f,AO^s} &= \big(|\chi^{u1_4}_1 |^2 + |\chi^{u1_4}_{-1}|^2 \big)\big(\chi^\text{Is}_0 \bar \chi^\text{Is}_\frac{1}{2} + \chi^\text{Is}_\frac12 \bar \chi^\text{Is}_0\big) 
 \nonumber \\ 
Z_{PE^f,AO^s} &= \big( \chi^{u1_4}_1 \bar \chi^{u1_4}_{-1} + \chi^{u1_4}_{-1} \bar \chi^{u1_4}_1 \big)\big(\chi^\text{Is}_0 \bar \chi^\text{Is}_\frac{1}{2} + \chi^\text{Is}_\frac12 \bar \chi^\text{Is}_0\big) 
 \nonumber \\ 
Z_{AO^f,AO^s} &= \big( \chi^{u1_4}_0 \bar \chi^{u1_4}_2 + \chi^{u1_4}_2 \bar \chi^{u1_4}_0 \big)\big(\chi^\text{Is}_0 \bar \chi^\text{Is}_\frac{1}{2} + \chi^\text{Is}_\frac12 \bar \chi^\text{Is}_0\big) .
\nonumber 
\end{align}

\begin{align}
\label{sol3}
 Z_{AE^f,PE^s} &= \big(|\chi^{u1_4}_0 |^2 + |\chi^{u1_4}_2|^2 \big)\big( |\chi^\text{Is}_0|^2+|\chi^\text{Is}_\frac{1}{2}|^2 \big) 
 \nonumber \\ 
Z_{PO^f,PE^s} &= \big(|\chi^{u1_4}_1 |^2 + |\chi^{u1_4}_{-1}|^2 \big)|\chi^\text{Is}_\frac{1}{16}|^2 
 \nonumber \\ 
Z_{PE^f,PE^s} &= \big( \chi^{u1_4}_1 \bar \chi^{u1_4}_{-1} + \chi^{u1_4}_{-1} \bar \chi^{u1_4}_1 \big)|\chi^\text{Is}_\frac{1}{16}|^2 
 \nonumber \\ 
Z_{AO^f,PE^s} &= \big( \chi^{u1_4}_0 \bar \chi^{u1_4}_2 + \chi^{u1_4}_2 \bar \chi^{u1_4}_0 \big)\big( |\chi^\text{Is}_0|^2+|\chi^\text{Is}_\frac{1}{2}|^2 \big) 
 \nonumber \\ 
Z_{AE^f,PO^s} &= \big(|\chi^{u1_4}_0 |^2 + |\chi^{u1_4}_2|^2 \big)|\chi^\text{Is}_\frac{1}{16}|^2 
 \nonumber \\ 
Z_{PO^f,PO^s} &= \big(|\chi^{u1_4}_1 |^2 + |\chi^{u1_4}_{-1}|^2 \big)\big( |\chi^\text{Is}_0|^2+|\chi^\text{Is}_\frac{1}{2}|^2 \big) 
 \nonumber \\ 
Z_{PE^f,PO^s} &= \big( \chi^{u1_4}_1 \bar \chi^{u1_4}_{-1} + \chi^{u1_4}_{-1} \bar \chi^{u1_4}_1 \big)\big( |\chi^\text{Is}_0|^2+|\chi^\text{Is}_\frac{1}{2}|^2 \big) 
 \nonumber \\ 
Z_{AO^f,PO^s} &= \big( \chi^{u1_4}_0 \bar \chi^{u1_4}_2 + \chi^{u1_4}_2 \bar \chi^{u1_4}_0 \big)|\chi^\text{Is}_\frac{1}{16}|^2 
 \nonumber \\ 
Z_{AE^f,AE^s} &= \big( \chi^{u1_4}_0 \bar \chi^{u1_4}_2 + \chi^{u1_4}_2 \bar \chi^{u1_4}_0 \big)\big(\chi^\text{Is}_0 \bar \chi^\text{Is}_\frac{1}{2} + \chi^\text{Is}_\frac12 \bar \chi^\text{Is}_0\big) 
 \nonumber \\ 
Z_{PO^f,AE^s} &= \big( \chi^{u1_4}_1 \bar \chi^{u1_4}_{-1} + \chi^{u1_4}_{-1} \bar \chi^{u1_4}_1 \big)|\chi^\text{Is}_\frac{1}{16}|^2 
 \nonumber \\ 
Z_{PE^f,AE^s} &= \big(|\chi^{u1_4}_1 |^2 + |\chi^{u1_4}_{-1}|^2 \big)|\chi^\text{Is}_\frac{1}{16}|^2 
 \nonumber \\ 
Z_{AO^f,AE^s} &= \big(|\chi^{u1_4}_0 |^2 + |\chi^{u1_4}_2|^2 \big)\big(\chi^\text{Is}_0 \bar \chi^\text{Is}_\frac{1}{2} + \chi^\text{Is}_\frac12 \bar \chi^\text{Is}_0\big) 
 \nonumber \\ 
Z_{AE^f,AO^s} &= \big( \chi^{u1_4}_0 \bar \chi^{u1_4}_2 + \chi^{u1_4}_2 \bar \chi^{u1_4}_0 \big)|\chi^\text{Is}_\frac{1}{16}|^2 
 \nonumber \\ 
Z_{PO^f,AO^s} &= \big( \chi^{u1_4}_1 \bar \chi^{u1_4}_{-1} + \chi^{u1_4}_{-1} \bar \chi^{u1_4}_1 \big)\big(\chi^\text{Is}_0 \bar \chi^\text{Is}_\frac{1}{2} + \chi^\text{Is}_\frac12 \bar \chi^\text{Is}_0\big) 
 \nonumber \\ 
Z_{PE^f,AO^s} &= \big(|\chi^{u1_4}_1 |^2 + |\chi^{u1_4}_{-1}|^2 \big)\big(\chi^\text{Is}_0 \bar \chi^\text{Is}_\frac{1}{2} + \chi^\text{Is}_\frac12 \bar \chi^\text{Is}_0\big) 
 \nonumber \\ 
Z_{AO^f,AO^s} &= \big(|\chi^{u1_4}_0 |^2 + |\chi^{u1_4}_2|^2 \big)|\chi^\text{Is}_\frac{1}{16}|^2.
\end{align}

\begin{align}
\label{sol4}
 Z_{AE^f,PE^s} &= \big(|\chi^{u1_4}_0 |^2 + |\chi^{u1_4}_2|^2 \big)\big( |\chi^\text{Is}_0|^2+|\chi^\text{Is}_\frac{1}{2}|^2 \big) 
 \nonumber \\ 
Z_{PO^f,PE^s} &= \big( \chi^{u1_4}_1 \bar \chi^{u1_4}_{-1} + \chi^{u1_4}_{-1} \bar \chi^{u1_4}_1 \big)|\chi^\text{Is}_\frac{1}{16}|^2 
 \nonumber \\ 
Z_{PE^f,PE^s} &= \big(|\chi^{u1_4}_1 |^2 + |\chi^{u1_4}_{-1}|^2 \big)|\chi^\text{Is}_\frac{1}{16}|^2 
 \nonumber \\ 
Z_{AO^f,PE^s} &= \big( \chi^{u1_4}_0 \bar \chi^{u1_4}_2 + \chi^{u1_4}_2 \bar \chi^{u1_4}_0 \big)\big( |\chi^\text{Is}_0|^2+|\chi^\text{Is}_\frac{1}{2}|^2 \big) 
 \nonumber \\ 
Z_{AE^f,PO^s} &= \big(|\chi^{u1_4}_0 |^2 + |\chi^{u1_4}_2|^2 \big)|\chi^\text{Is}_\frac{1}{16}|^2 
 \nonumber \\ 
Z_{PO^f,PO^s} &= \big( \chi^{u1_4}_1 \bar \chi^{u1_4}_{-1} + \chi^{u1_4}_{-1} \bar \chi^{u1_4}_1 \big)\big( |\chi^\text{Is}_0|^2+|\chi^\text{Is}_\frac{1}{2}|^2 \big) 
 \nonumber \\ 
Z_{PE^f,PO^s} &= \big(|\chi^{u1_4}_1 |^2 + |\chi^{u1_4}_{-1}|^2 \big)\big( |\chi^\text{Is}_0|^2+|\chi^\text{Is}_\frac{1}{2}|^2 \big) 
 \nonumber \\ 
Z_{AO^f,PO^s} &= \big( \chi^{u1_4}_0 \bar \chi^{u1_4}_2 + \chi^{u1_4}_2 \bar \chi^{u1_4}_0 \big)|\chi^\text{Is}_\frac{1}{16}|^2 
 \nonumber \\ 
Z_{AE^f,AE^s} &= \big( \chi^{u1_4}_0 \bar \chi^{u1_4}_2 + \chi^{u1_4}_2 \bar \chi^{u1_4}_0 \big)\big(\chi^\text{Is}_0 \bar \chi^\text{Is}_\frac{1}{2} + \chi^\text{Is}_\frac12 \bar \chi^\text{Is}_0\big) 
 \nonumber \\ 
Z_{PO^f,AE^s} &= \big(|\chi^{u1_4}_1 |^2 + |\chi^{u1_4}_{-1}|^2 \big)|\chi^\text{Is}_\frac{1}{16}|^2 
 \nonumber \\ 
Z_{PE^f,AE^s} &= \big( \chi^{u1_4}_1 \bar \chi^{u1_4}_{-1} + \chi^{u1_4}_{-1} \bar \chi^{u1_4}_1 \big)|\chi^\text{Is}_\frac{1}{16}|^2 
 \nonumber \\ 
Z_{AO^f,AE^s} &= \big(|\chi^{u1_4}_0 |^2 + |\chi^{u1_4}_2|^2 \big)\big(\chi^\text{Is}_0 \bar \chi^\text{Is}_\frac{1}{2} + \chi^\text{Is}_\frac12 \bar \chi^\text{Is}_0\big) 
 \nonumber \\ 
Z_{AE^f,AO^s} &= \big( \chi^{u1_4}_0 \bar \chi^{u1_4}_2 + \chi^{u1_4}_2 \bar \chi^{u1_4}_0 \big)|\chi^\text{Is}_\frac{1}{16}|^2 
 \nonumber \\ 
Z_{PO^f,AO^s} &= \big(|\chi^{u1_4}_1 |^2 + |\chi^{u1_4}_{-1}|^2 \big)\big(\chi^\text{Is}_0 \bar \chi^\text{Is}_\frac{1}{2} + \chi^\text{Is}_\frac12 \bar \chi^\text{Is}_0\big) 
 \nonumber \\ 
Z_{PE^f,AO^s} &= \big( \chi^{u1_4}_1 \bar \chi^{u1_4}_{-1} + \chi^{u1_4}_{-1} \bar \chi^{u1_4}_1 \big)\big(\chi^\text{Is}_0 \bar \chi^\text{Is}_\frac{1}{2} + \chi^\text{Is}_\frac12 \bar \chi^\text{Is}_0\big) 
 \nonumber \\ 
Z_{AO^f,AO^s} &= \big(|\chi^{u1_4}_0 |^2 + |\chi^{u1_4}_2|^2 \big)|\chi^\text{Is}_\frac{1}{16}|^2. 
\end{align}

\begin{align}
\label{sol5}
 Z_{AE^f,PE^s} &= \big(|\chi^{u1_4}_0 |^2 + |\chi^{u1_4}_2|^2 \big)\big( |\chi^\text{Is}_0|^2+|\chi^\text{Is}_\frac{1}{2}|^2 \big) 
 \nonumber \\ 
Z_{PO^f,PE^s} &= \big(|\chi^{u1_4}_1 |^2 + |\chi^{u1_4}_{-1}|^2 \big)|\chi^\text{Is}_\frac{1}{16}|^2 
 \nonumber \\ 
Z_{PE^f,PE^s} &= \big( \chi^{u1_4}_1 \bar \chi^{u1_4}_{-1} + \chi^{u1_4}_{-1} \bar \chi^{u1_4}_1 \big)|\chi^\text{Is}_\frac{1}{16}|^2 
 \nonumber \\ 
Z_{AO^f,PE^s} &= \big( \chi^{u1_4}_0 \bar \chi^{u1_4}_2 + \chi^{u1_4}_2 \bar \chi^{u1_4}_0 \big)\big( |\chi^\text{Is}_0|^2+|\chi^\text{Is}_\frac{1}{2}|^2 \big) 
 \nonumber \\ 
Z_{AE^f,PO^s} &= \big( \chi^{u1_4}_0 \bar \chi^{u1_4}_2 + \chi^{u1_4}_2 \bar \chi^{u1_4}_0 \big)\big(\chi^\text{Is}_0 \bar \chi^\text{Is}_\frac{1}{2} + \chi^\text{Is}_\frac12 \bar \chi^\text{Is}_0\big) 
 \nonumber \\ 
Z_{PO^f,PO^s} &= \big( \chi^{u1_4}_1 \bar \chi^{u1_4}_{-1} + \chi^{u1_4}_{-1} \bar \chi^{u1_4}_1 \big)|\chi^\text{Is}_\frac{1}{16}|^2 
 \nonumber \\ 
Z_{PE^f,PO^s} &= \big(|\chi^{u1_4}_1 |^2 + |\chi^{u1_4}_{-1}|^2 \big)|\chi^\text{Is}_\frac{1}{16}|^2 
 \nonumber \\ 
Z_{AO^f,PO^s} &= \big(|\chi^{u1_4}_0 |^2 + |\chi^{u1_4}_2|^2 \big)\big(\chi^\text{Is}_0 \bar \chi^\text{Is}_\frac{1}{2} + \chi^\text{Is}_\frac12 \bar \chi^\text{Is}_0\big) 
 \nonumber \\ 
Z_{AE^f,AE^s} &= \big(|\chi^{u1_4}_0 |^2 + |\chi^{u1_4}_2|^2 \big)|\chi^\text{Is}_\frac{1}{16}|^2 
 \nonumber \\ 
Z_{PO^f,AE^s} &= \big(|\chi^{u1_4}_1 |^2 + |\chi^{u1_4}_{-1}|^2 \big)\big( |\chi^\text{Is}_0|^2+|\chi^\text{Is}_\frac{1}{2}|^2 \big) 
 \nonumber \\ 
Z_{PE^f,AE^s} &= \big( \chi^{u1_4}_1 \bar \chi^{u1_4}_{-1} + \chi^{u1_4}_{-1} \bar \chi^{u1_4}_1 \big)\big( |\chi^\text{Is}_0|^2+|\chi^\text{Is}_\frac{1}{2}|^2 \big) 
 \nonumber \\ 
Z_{AO^f,AE^s} &= \big( \chi^{u1_4}_0 \bar \chi^{u1_4}_2 + \chi^{u1_4}_2 \bar \chi^{u1_4}_0 \big)|\chi^\text{Is}_\frac{1}{16}|^2 
 \nonumber \\ 
Z_{AE^f,AO^s} &= \big( \chi^{u1_4}_0 \bar \chi^{u1_4}_2 + \chi^{u1_4}_2 \bar \chi^{u1_4}_0 \big)|\chi^\text{Is}_\frac{1}{16}|^2 
 \nonumber \\ 
Z_{PO^f,AO^s} &= \big( \chi^{u1_4}_1 \bar \chi^{u1_4}_{-1} + \chi^{u1_4}_{-1} \bar \chi^{u1_4}_1 \big)\big(\chi^\text{Is}_0 \bar \chi^\text{Is}_\frac{1}{2} + \chi^\text{Is}_\frac12 \bar \chi^\text{Is}_0\big) 
 \nonumber \\ 
Z_{PE^f,AO^s} &= \big(|\chi^{u1_4}_1 |^2 + |\chi^{u1_4}_{-1}|^2 \big)\big(\chi^\text{Is}_0 \bar \chi^\text{Is}_\frac{1}{2} + \chi^\text{Is}_\frac12 \bar \chi^\text{Is}_0\big) 
 \nonumber \\ 
Z_{AO^f,AO^s} &= \big(|\chi^{u1_4}_0 |^2 + |\chi^{u1_4}_2|^2 \big)|\chi^\text{Is}_\frac{1}{16}|^2. 
\end{align}

\begin{align}
\label{sol6}
 Z_{AE^f,PE^s} &= \big(|\chi^{u1_4}_0 |^2 + |\chi^{u1_4}_2|^2 \big)\big( |\chi^\text{Is}_0|^2+|\chi^\text{Is}_\frac{1}{2}|^2 \big) 
 \nonumber \\ 
Z_{PO^f,PE^s} &= \big( \chi^{u1_4}_1 \bar \chi^{u1_4}_{-1} + \chi^{u1_4}_{-1} \bar \chi^{u1_4}_1 \big)|\chi^\text{Is}_\frac{1}{16}|^2 
 \nonumber \\ 
Z_{PE^f,PE^s} &= \big(|\chi^{u1_4}_1 |^2 + |\chi^{u1_4}_{-1}|^2 \big)|\chi^\text{Is}_\frac{1}{16}|^2 
 \nonumber \\ 
Z_{AO^f,PE^s} &= \big( \chi^{u1_4}_0 \bar \chi^{u1_4}_2 + \chi^{u1_4}_2 \bar \chi^{u1_4}_0 \big)\big( |\chi^\text{Is}_0|^2+|\chi^\text{Is}_\frac{1}{2}|^2 \big) 
 \nonumber \\ 
Z_{AE^f,PO^s} &= \big( \chi^{u1_4}_0 \bar \chi^{u1_4}_2 + \chi^{u1_4}_2 \bar \chi^{u1_4}_0 \big)\big(\chi^\text{Is}_0 \bar \chi^\text{Is}_\frac{1}{2} + \chi^\text{Is}_\frac12 \bar \chi^\text{Is}_0\big) 
 \nonumber \\ 
Z_{PO^f,PO^s} &= \big(|\chi^{u1_4}_1 |^2 + |\chi^{u1_4}_{-1}|^2 \big)|\chi^\text{Is}_\frac{1}{16}|^2 
 \nonumber \\ 
Z_{PE^f,PO^s} &= \big( \chi^{u1_4}_1 \bar \chi^{u1_4}_{-1} + \chi^{u1_4}_{-1} \bar \chi^{u1_4}_1 \big)|\chi^\text{Is}_\frac{1}{16}|^2 
 \nonumber \\ 
Z_{AO^f,PO^s} &= \big(|\chi^{u1_4}_0 |^2 + |\chi^{u1_4}_2|^2 \big)\big(\chi^\text{Is}_0 \bar \chi^\text{Is}_\frac{1}{2} + \chi^\text{Is}_\frac12 \bar \chi^\text{Is}_0\big) 
 \nonumber \\ 
Z_{AE^f,AE^s} &= \big(|\chi^{u1_4}_0 |^2 + |\chi^{u1_4}_2|^2 \big)|\chi^\text{Is}_\frac{1}{16}|^2 
 \nonumber \\ 
Z_{PO^f,AE^s} &= \big( \chi^{u1_4}_1 \bar \chi^{u1_4}_{-1} + \chi^{u1_4}_{-1} \bar \chi^{u1_4}_1 \big)\big( |\chi^\text{Is}_0|^2+|\chi^\text{Is}_\frac{1}{2}|^2 \big) 
 \nonumber \\ 
Z_{PE^f,AE^s} &= \big(|\chi^{u1_4}_1 |^2 + |\chi^{u1_4}_{-1}|^2 \big)\big( |\chi^\text{Is}_0|^2+|\chi^\text{Is}_\frac{1}{2}|^2 \big) 
 \nonumber \\ 
Z_{AO^f,AE^s} &= \big( \chi^{u1_4}_0 \bar \chi^{u1_4}_2 + \chi^{u1_4}_2 \bar \chi^{u1_4}_0 \big)|\chi^\text{Is}_\frac{1}{16}|^2 
 \nonumber \\ 
Z_{AE^f,AO^s} &= \big( \chi^{u1_4}_0 \bar \chi^{u1_4}_2 + \chi^{u1_4}_2 \bar \chi^{u1_4}_0 \big)|\chi^\text{Is}_\frac{1}{16}|^2 
 \nonumber \\ 
Z_{PO^f,AO^s} &= \big(|\chi^{u1_4}_1 |^2 + |\chi^{u1_4}_{-1}|^2 \big)\big(\chi^\text{Is}_0 \bar \chi^\text{Is}_\frac{1}{2} + \chi^\text{Is}_\frac12 \bar \chi^\text{Is}_0\big) 
 \nonumber \\ 
Z_{PE^f,AO^s} &= \big( \chi^{u1_4}_1 \bar \chi^{u1_4}_{-1} + \chi^{u1_4}_{-1} \bar \chi^{u1_4}_1 \big)\big(\chi^\text{Is}_0 \bar \chi^\text{Is}_\frac{1}{2} + \chi^\text{Is}_\frac12 \bar \chi^\text{Is}_0\big) 
 \nonumber \\ 
Z_{AO^f,AO^s} &= \big(|\chi^{u1_4}_0 |^2 + |\chi^{u1_4}_2|^2 \big)|\chi^\text{Is}_\frac{1}{16}|^2. 
\end{align}

\begin{align}
\label{sol7}
 Z_{AE^f,PE^s} &= \big(|\chi^{u1_4}_0 |^2 + |\chi^{u1_4}_2|^2 \big)\big( |\chi^\text{Is}_0|^2+|\chi^\text{Is}_\frac{1}{2}|^2 \big) 
 \nonumber \\ 
Z_{PO^f,PE^s} &= \big( \chi^{u1_4}_1 \bar \chi^{u1_4}_{-1} + \chi^{u1_4}_{-1} \bar \chi^{u1_4}_1 \big)|\chi^\text{Is}_\frac{1}{16}|^2 
 \nonumber \\ 
Z_{PE^f,PE^s} &= \big(|\chi^{u1_4}_1 |^2 + |\chi^{u1_4}_{-1}|^2 \big)\big( |\chi^\text{Is}_0|^2+|\chi^\text{Is}_\frac{1}{2}|^2 \big) 
 \nonumber \\ 
Z_{AO^f,PE^s} &= \big( \chi^{u1_4}_0 \bar \chi^{u1_4}_2 + \chi^{u1_4}_2 \bar \chi^{u1_4}_0 \big)|\chi^\text{Is}_\frac{1}{16}|^2 
 \nonumber \\ 
Z_{AE^f,PO^s} &= \big(|\chi^{u1_4}_0 |^2 + |\chi^{u1_4}_2|^2 \big)|\chi^\text{Is}_\frac{1}{16}|^2 
 \nonumber \\ 
Z_{PO^f,PO^s} &= \big( \chi^{u1_4}_1 \bar \chi^{u1_4}_{-1} + \chi^{u1_4}_{-1} \bar \chi^{u1_4}_1 \big)\big( |\chi^\text{Is}_0|^2+|\chi^\text{Is}_\frac{1}{2}|^2 \big) 
 \nonumber \\ 
Z_{PE^f,PO^s} &= \big(|\chi^{u1_4}_1 |^2 + |\chi^{u1_4}_{-1}|^2 \big)|\chi^\text{Is}_\frac{1}{16}|^2 
 \nonumber \\ 
Z_{AO^f,PO^s} &= \big( \chi^{u1_4}_0 \bar \chi^{u1_4}_2 + \chi^{u1_4}_2 \bar \chi^{u1_4}_0 \big)\big( |\chi^\text{Is}_0|^2+|\chi^\text{Is}_\frac{1}{2}|^2 \big) 
 \nonumber \\ 
Z_{AE^f,AE^s} &= \big(|\chi^{u1_4}_1 |^2 + |\chi^{u1_4}_{-1}|^2 \big)|\chi^\text{Is}_\frac{1}{16}|^2 
 \nonumber \\ 
Z_{PO^f,AE^s} &= \big( \chi^{u1_4}_0 \bar \chi^{u1_4}_2 + \chi^{u1_4}_2 \bar \chi^{u1_4}_0 \big)\big(\chi^\text{Is}_0 \bar \chi^\text{Is}_\frac{1}{2} + \chi^\text{Is}_\frac12 \bar \chi^\text{Is}_0\big) 
 \nonumber \\ 
Z_{PE^f,AE^s} &= \big(|\chi^{u1_4}_0 |^2 + |\chi^{u1_4}_2|^2 \big)|\chi^\text{Is}_\frac{1}{16}|^2 
 \nonumber \\ 
Z_{AO^f,AE^s} &= \big( \chi^{u1_4}_1 \bar \chi^{u1_4}_{-1} + \chi^{u1_4}_{-1} \bar \chi^{u1_4}_1 \big)\big(\chi^\text{Is}_0 \bar \chi^\text{Is}_\frac{1}{2} + \chi^\text{Is}_\frac12 \bar \chi^\text{Is}_0\big) 
 \nonumber \\ 
Z_{AE^f,AO^s} &= \big(|\chi^{u1_4}_1 |^2 + |\chi^{u1_4}_{-1}|^2 \big)\big(\chi^\text{Is}_0 \bar \chi^\text{Is}_\frac{1}{2} + \chi^\text{Is}_\frac12 \bar \chi^\text{Is}_0\big) 
 \nonumber \\ 
Z_{PO^f,AO^s} &= \big( \chi^{u1_4}_0 \bar \chi^{u1_4}_2 + \chi^{u1_4}_2 \bar \chi^{u1_4}_0 \big)|\chi^\text{Is}_\frac{1}{16}|^2 
 \nonumber \\ 
Z_{PE^f,AO^s} &= \big(|\chi^{u1_4}_0 |^2 + |\chi^{u1_4}_2|^2 \big)\big(\chi^\text{Is}_0 \bar \chi^\text{Is}_\frac{1}{2} + \chi^\text{Is}_\frac12 \bar \chi^\text{Is}_0\big) 
 \nonumber \\ 
Z_{AO^f,AO^s} &= \big( \chi^{u1_4}_1 \bar \chi^{u1_4}_{-1} + \chi^{u1_4}_{-1} \bar \chi^{u1_4}_1 \big)|\chi^\text{Is}_\frac{1}{16}|^2.
\end{align}

\begin{align}
\label{sol8}
 Z_{AE^f,PE^s} &= \big(|\chi^{u1_4}_0 |^2 + |\chi^{u1_4}_2|^2 \big)\big( |\chi^\text{Is}_0|^2+|\chi^\text{Is}_\frac{1}{2}|^2 \big) 
 \nonumber \\ 
Z_{PO^f,PE^s} &= \big(|\chi^{u1_4}_1 |^2 + |\chi^{u1_4}_{-1}|^2 \big)|\chi^\text{Is}_\frac{1}{16}|^2 
 \nonumber \\ 
Z_{PE^f,PE^s} &= \big( \chi^{u1_4}_1 \bar \chi^{u1_4}_{-1} + \chi^{u1_4}_{-1} \bar \chi^{u1_4}_1 \big)\big( |\chi^\text{Is}_0|^2+|\chi^\text{Is}_\frac{1}{2}|^2 \big) 
 \nonumber \\ 
Z_{AO^f,PE^s} &= \big( \chi^{u1_4}_0 \bar \chi^{u1_4}_2 + \chi^{u1_4}_2 \bar \chi^{u1_4}_0 \big)|\chi^\text{Is}_\frac{1}{16}|^2 
 \nonumber \\ 
Z_{AE^f,PO^s} &= \big(|\chi^{u1_4}_0 |^2 + |\chi^{u1_4}_2|^2 \big)|\chi^\text{Is}_\frac{1}{16}|^2 
 \nonumber \\ 
Z_{PO^f,PO^s} &= \big(|\chi^{u1_4}_1 |^2 + |\chi^{u1_4}_{-1}|^2 \big)\big( |\chi^\text{Is}_0|^2+|\chi^\text{Is}_\frac{1}{2}|^2 \big) 
 \nonumber \\ 
Z_{PE^f,PO^s} &= \big( \chi^{u1_4}_1 \bar \chi^{u1_4}_{-1} + \chi^{u1_4}_{-1} \bar \chi^{u1_4}_1 \big)|\chi^\text{Is}_\frac{1}{16}|^2 
 \nonumber \\ 
Z_{AO^f,PO^s} &= \big( \chi^{u1_4}_0 \bar \chi^{u1_4}_2 + \chi^{u1_4}_2 \bar \chi^{u1_4}_0 \big)\big( |\chi^\text{Is}_0|^2+|\chi^\text{Is}_\frac{1}{2}|^2 \big) 
 \nonumber \\ 
Z_{AE^f,AE^s} &= \big( \chi^{u1_4}_1 \bar \chi^{u1_4}_{-1} + \chi^{u1_4}_{-1} \bar \chi^{u1_4}_1 \big)|\chi^\text{Is}_\frac{1}{16}|^2 
 \nonumber \\ 
Z_{PO^f,AE^s} &= \big( \chi^{u1_4}_0 \bar \chi^{u1_4}_2 + \chi^{u1_4}_2 \bar \chi^{u1_4}_0 \big)\big(\chi^\text{Is}_0 \bar \chi^\text{Is}_\frac{1}{2} + \chi^\text{Is}_\frac12 \bar \chi^\text{Is}_0\big) 
 \nonumber \\ 
Z_{PE^f,AE^s} &= \big(|\chi^{u1_4}_0 |^2 + |\chi^{u1_4}_2|^2 \big)|\chi^\text{Is}_\frac{1}{16}|^2 
 \nonumber \\ 
Z_{AO^f,AE^s} &= \big(|\chi^{u1_4}_1 |^2 + |\chi^{u1_4}_{-1}|^2 \big)\big(\chi^\text{Is}_0 \bar \chi^\text{Is}_\frac{1}{2} + \chi^\text{Is}_\frac12 \bar \chi^\text{Is}_0\big) 
 \nonumber \\ 
Z_{AE^f,AO^s} &= \big( \chi^{u1_4}_1 \bar \chi^{u1_4}_{-1} + \chi^{u1_4}_{-1} \bar \chi^{u1_4}_1 \big)\big(\chi^\text{Is}_0 \bar \chi^\text{Is}_\frac{1}{2} + \chi^\text{Is}_\frac12 \bar \chi^\text{Is}_0\big) 
 \nonumber \\ 
Z_{PO^f,AO^s} &= \big( \chi^{u1_4}_0 \bar \chi^{u1_4}_2 + \chi^{u1_4}_2 \bar \chi^{u1_4}_0 \big)|\chi^\text{Is}_\frac{1}{16}|^2 
 \nonumber \\ 
Z_{PE^f,AO^s} &= \big(|\chi^{u1_4}_0 |^2 + |\chi^{u1_4}_2|^2 \big)\big(\chi^\text{Is}_0 \bar \chi^\text{Is}_\frac{1}{2} + \chi^\text{Is}_\frac12 \bar \chi^\text{Is}_0\big) 
 \nonumber \\ 
Z_{AO^f,AO^s} &= \big(|\chi^{u1_4}_1 |^2 + |\chi^{u1_4}_{-1}|^2 \big)|\chi^\text{Is}_\frac{1}{16}|^2. 
\end{align}

\begin{align}
\label{sol9}
 Z_{AE^f,PE^s} &= \big(|\chi^{u1_4}_0 |^2 + |\chi^{u1_4}_2|^2 \big)\big( |\chi^\text{Is}_0|^2+|\chi^\text{Is}_\frac{1}{2}|^2 \big) 
 \nonumber \\ 
Z_{PO^f,PE^s} &= \big(|\chi^{u1_4}_1 |^2 + |\chi^{u1_4}_{-1}|^2 \big)\big( |\chi^\text{Is}_0|^2+|\chi^\text{Is}_\frac{1}{2}|^2 \big) 
 \nonumber \\ 
Z_{PE^f,PE^s} &= \big( \chi^{u1_4}_1 \bar \chi^{u1_4}_{-1} + \chi^{u1_4}_{-1} \bar \chi^{u1_4}_1 \big)|\chi^\text{Is}_\frac{1}{16}|^2 
 \nonumber \\ 
Z_{AO^f,PE^s} &= \big( \chi^{u1_4}_0 \bar \chi^{u1_4}_2 + \chi^{u1_4}_2 \bar \chi^{u1_4}_0 \big)|\chi^\text{Is}_\frac{1}{16}|^2 
 \nonumber \\ 
Z_{AE^f,PO^s} &= \big(|\chi^{u1_4}_0 |^2 + |\chi^{u1_4}_2|^2 \big)|\chi^\text{Is}_\frac{1}{16}|^2 
 \nonumber \\ 
Z_{PO^f,PO^s} &= \big(|\chi^{u1_4}_1 |^2 + |\chi^{u1_4}_{-1}|^2 \big)|\chi^\text{Is}_\frac{1}{16}|^2 
 \nonumber \\ 
Z_{PE^f,PO^s} &= \big( \chi^{u1_4}_1 \bar \chi^{u1_4}_{-1} + \chi^{u1_4}_{-1} \bar \chi^{u1_4}_1 \big)\big( |\chi^\text{Is}_0|^2+|\chi^\text{Is}_\frac{1}{2}|^2 \big) 
 \nonumber \\ 
Z_{AO^f,PO^s} &= \big( \chi^{u1_4}_0 \bar \chi^{u1_4}_2 + \chi^{u1_4}_2 \bar \chi^{u1_4}_0 \big)\big( |\chi^\text{Is}_0|^2+|\chi^\text{Is}_\frac{1}{2}|^2 \big) 
 \nonumber \\ 
Z_{AE^f,AE^s} &= \big(|\chi^{u1_4}_1 |^2 + |\chi^{u1_4}_{-1}|^2 \big)|\chi^\text{Is}_\frac{1}{16}|^2 
 \nonumber \\ 
Z_{PO^f,AE^s} &= \big(|\chi^{u1_4}_0 |^2 + |\chi^{u1_4}_2|^2 \big)|\chi^\text{Is}_\frac{1}{16}|^2 
 \nonumber \\ 
Z_{PE^f,AE^s} &= \big( \chi^{u1_4}_0 \bar \chi^{u1_4}_2 + \chi^{u1_4}_2 \bar \chi^{u1_4}_0 \big)\big(\chi^\text{Is}_0 \bar \chi^\text{Is}_\frac{1}{2} + \chi^\text{Is}_\frac12 \bar \chi^\text{Is}_0\big) 
 \nonumber \\ 
Z_{AO^f,AE^s} &= \big( \chi^{u1_4}_1 \bar \chi^{u1_4}_{-1} + \chi^{u1_4}_{-1} \bar \chi^{u1_4}_1 \big)\big(\chi^\text{Is}_0 \bar \chi^\text{Is}_\frac{1}{2} + \chi^\text{Is}_\frac12 \bar \chi^\text{Is}_0\big) 
 \nonumber \\ 
Z_{AE^f,AO^s} &= \big(|\chi^{u1_4}_1 |^2 + |\chi^{u1_4}_{-1}|^2 \big)\big(\chi^\text{Is}_0 \bar \chi^\text{Is}_\frac{1}{2} + \chi^\text{Is}_\frac12 \bar \chi^\text{Is}_0\big) 
 \nonumber \\ 
Z_{PO^f,AO^s} &= \big(|\chi^{u1_4}_0 |^2 + |\chi^{u1_4}_2|^2 \big)\big(\chi^\text{Is}_0 \bar \chi^\text{Is}_\frac{1}{2} + \chi^\text{Is}_\frac12 \bar \chi^\text{Is}_0\big) 
 \nonumber \\ 
Z_{PE^f,AO^s} &= \big( \chi^{u1_4}_0 \bar \chi^{u1_4}_2 + \chi^{u1_4}_2 \bar \chi^{u1_4}_0 \big)|\chi^\text{Is}_\frac{1}{16}|^2 
 \nonumber \\ 
Z_{AO^f,AO^s} &= \big( \chi^{u1_4}_1 \bar \chi^{u1_4}_{-1} + \chi^{u1_4}_{-1} \bar \chi^{u1_4}_1 \big)|\chi^\text{Is}_\frac{1}{16}|^2. 
\end{align}

\begin{align}
\label{sol10}
 Z_{AE^f,PE^s} &= \big(|\chi^{u1_4}_0 |^2 + |\chi^{u1_4}_2|^2 \big)\big( |\chi^\text{Is}_0|^2+|\chi^\text{Is}_\frac{1}{2}|^2 \big) 
 \nonumber \\ 
Z_{PO^f,PE^s} &= \big( \chi^{u1_4}_1 \bar \chi^{u1_4}_{-1} + \chi^{u1_4}_{-1} \bar \chi^{u1_4}_1 \big)\big( |\chi^\text{Is}_0|^2+|\chi^\text{Is}_\frac{1}{2}|^2 \big) 
 \nonumber \\ 
Z_{PE^f,PE^s} &= \big(|\chi^{u1_4}_1 |^2 + |\chi^{u1_4}_{-1}|^2 \big)|\chi^\text{Is}_\frac{1}{16}|^2 
 \nonumber \\ 
Z_{AO^f,PE^s} &= \big( \chi^{u1_4}_0 \bar \chi^{u1_4}_2 + \chi^{u1_4}_2 \bar \chi^{u1_4}_0 \big)|\chi^\text{Is}_\frac{1}{16}|^2 
 \nonumber \\ 
Z_{AE^f,PO^s} &= \big(|\chi^{u1_4}_0 |^2 + |\chi^{u1_4}_2|^2 \big)|\chi^\text{Is}_\frac{1}{16}|^2 
 \nonumber \\ 
Z_{PO^f,PO^s} &= \big( \chi^{u1_4}_1 \bar \chi^{u1_4}_{-1} + \chi^{u1_4}_{-1} \bar \chi^{u1_4}_1 \big)|\chi^\text{Is}_\frac{1}{16}|^2 
 \nonumber \\ 
Z_{PE^f,PO^s} &= \big(|\chi^{u1_4}_1 |^2 + |\chi^{u1_4}_{-1}|^2 \big)\big( |\chi^\text{Is}_0|^2+|\chi^\text{Is}_\frac{1}{2}|^2 \big) 
 \nonumber \\ 
Z_{AO^f,PO^s} &= \big( \chi^{u1_4}_0 \bar \chi^{u1_4}_2 + \chi^{u1_4}_2 \bar \chi^{u1_4}_0 \big)\big( |\chi^\text{Is}_0|^2+|\chi^\text{Is}_\frac{1}{2}|^2 \big) 
 \nonumber \\ 
Z_{AE^f,AE^s} &= \big( \chi^{u1_4}_1 \bar \chi^{u1_4}_{-1} + \chi^{u1_4}_{-1} \bar \chi^{u1_4}_1 \big)|\chi^\text{Is}_\frac{1}{16}|^2 
 \nonumber \\ 
Z_{PO^f,AE^s} &= \big(|\chi^{u1_4}_0 |^2 + |\chi^{u1_4}_2|^2 \big)|\chi^\text{Is}_\frac{1}{16}|^2 
 \nonumber \\ 
Z_{PE^f,AE^s} &= \big( \chi^{u1_4}_0 \bar \chi^{u1_4}_2 + \chi^{u1_4}_2 \bar \chi^{u1_4}_0 \big)\big(\chi^\text{Is}_0 \bar \chi^\text{Is}_\frac{1}{2} + \chi^\text{Is}_\frac12 \bar \chi^\text{Is}_0\big) 
 \nonumber \\ 
Z_{AO^f,AE^s} &= \big(|\chi^{u1_4}_1 |^2 + |\chi^{u1_4}_{-1}|^2 \big)\big(\chi^\text{Is}_0 \bar \chi^\text{Is}_\frac{1}{2} + \chi^\text{Is}_\frac12 \bar \chi^\text{Is}_0\big) 
 \nonumber \\ 
Z_{AE^f,AO^s} &= \big( \chi^{u1_4}_1 \bar \chi^{u1_4}_{-1} + \chi^{u1_4}_{-1} \bar \chi^{u1_4}_1 \big)\big(\chi^\text{Is}_0 \bar \chi^\text{Is}_\frac{1}{2} + \chi^\text{Is}_\frac12 \bar \chi^\text{Is}_0\big) 
 \nonumber \\ 
Z_{PO^f,AO^s} &= \big(|\chi^{u1_4}_0 |^2 + |\chi^{u1_4}_2|^2 \big)\big(\chi^\text{Is}_0 \bar \chi^\text{Is}_\frac{1}{2} + \chi^\text{Is}_\frac12 \bar \chi^\text{Is}_0\big) 
 \nonumber \\ 
Z_{PE^f,AO^s} &= \big( \chi^{u1_4}_0 \bar \chi^{u1_4}_2 + \chi^{u1_4}_2 \bar \chi^{u1_4}_0 \big)|\chi^\text{Is}_\frac{1}{16}|^2 
 \nonumber \\ 
Z_{AO^f,AO^s} &= \big(|\chi^{u1_4}_1 |^2 + |\chi^{u1_4}_{-1}|^2 \big)|\chi^\text{Is}_\frac{1}{16}|^2. 
\end{align}

\begin{align}
\label{sol11}
 Z_{AE^f,PE^s} &= \big(|\chi^{u1_4}_0 |^2 + |\chi^{u1_4}_2|^2 \big)\big( |\chi^\text{Is}_0|^2+|\chi^\text{Is}_\frac{1}{2}|^2 \big) 
 \nonumber \\ 
Z_{PO^f,PE^s} &= \big( \chi^{u1_4}_1 \bar \chi^{u1_4}_{-1} + \chi^{u1_4}_{-1} \bar \chi^{u1_4}_1 \big)|\chi^\text{Is}_\frac{1}{16}|^2 
 \nonumber \\ 
Z_{PE^f,PE^s} &= \big( \chi^{u1_4}_1 \bar \chi^{u1_4}_{-1} + \chi^{u1_4}_{-1} \bar \chi^{u1_4}_1 \big)|\chi^\text{Is}_\frac{1}{16}|^2 
 \nonumber \\ 
Z_{AO^f,PE^s} &= \big(|\chi^{u1_4}_0 |^2 + |\chi^{u1_4}_2|^2 \big)\big(\chi^\text{Is}_0 \bar \chi^\text{Is}_\frac{1}{2} + \chi^\text{Is}_\frac12 \bar \chi^\text{Is}_0\big) 
 \nonumber \\ 
Z_{AE^f,PO^s} &= \big( \chi^{u1_4}_0 \bar \chi^{u1_4}_2 + \chi^{u1_4}_2 \bar \chi^{u1_4}_0 \big)\big(\chi^\text{Is}_0 \bar \chi^\text{Is}_\frac{1}{2} + \chi^\text{Is}_\frac12 \bar \chi^\text{Is}_0\big) 
 \nonumber \\ 
Z_{PO^f,PO^s} &= \big(|\chi^{u1_4}_1 |^2 + |\chi^{u1_4}_{-1}|^2 \big)|\chi^\text{Is}_\frac{1}{16}|^2 
 \nonumber \\ 
Z_{PE^f,PO^s} &= \big(|\chi^{u1_4}_1 |^2 + |\chi^{u1_4}_{-1}|^2 \big)|\chi^\text{Is}_\frac{1}{16}|^2 
 \nonumber \\ 
Z_{AO^f,PO^s} &= \big( \chi^{u1_4}_0 \bar \chi^{u1_4}_2 + \chi^{u1_4}_2 \bar \chi^{u1_4}_0 \big)\big( |\chi^\text{Is}_0|^2+|\chi^\text{Is}_\frac{1}{2}|^2 \big) 
 \nonumber \\ 
Z_{AE^f,AE^s} &= \big( \chi^{u1_4}_1 \bar \chi^{u1_4}_{-1} + \chi^{u1_4}_{-1} \bar \chi^{u1_4}_1 \big)\big( |\chi^\text{Is}_0|^2+|\chi^\text{Is}_\frac{1}{2}|^2 \big) 
 \nonumber \\ 
Z_{PO^f,AE^s} &= \big(|\chi^{u1_4}_0 |^2 + |\chi^{u1_4}_2|^2 \big)|\chi^\text{Is}_\frac{1}{16}|^2 
 \nonumber \\ 
Z_{PE^f,AE^s} &= \big(|\chi^{u1_4}_0 |^2 + |\chi^{u1_4}_2|^2 \big)|\chi^\text{Is}_\frac{1}{16}|^2 
 \nonumber \\ 
Z_{AO^f,AE^s} &= \big( \chi^{u1_4}_1 \bar \chi^{u1_4}_{-1} + \chi^{u1_4}_{-1} \bar \chi^{u1_4}_1 \big)\big(\chi^\text{Is}_0 \bar \chi^\text{Is}_\frac{1}{2} + \chi^\text{Is}_\frac12 \bar \chi^\text{Is}_0\big) 
 \nonumber \\ 
Z_{AE^f,AO^s} &= \big(|\chi^{u1_4}_1 |^2 + |\chi^{u1_4}_{-1}|^2 \big)\big(\chi^\text{Is}_0 \bar \chi^\text{Is}_\frac{1}{2} + \chi^\text{Is}_\frac12 \bar \chi^\text{Is}_0\big) 
 \nonumber \\ 
Z_{PO^f,AO^s} &= \big( \chi^{u1_4}_0 \bar \chi^{u1_4}_2 + \chi^{u1_4}_2 \bar \chi^{u1_4}_0 \big)|\chi^\text{Is}_\frac{1}{16}|^2 
 \nonumber \\ 
Z_{PE^f,AO^s} &= \big( \chi^{u1_4}_0 \bar \chi^{u1_4}_2 + \chi^{u1_4}_2 \bar \chi^{u1_4}_0 \big)|\chi^\text{Is}_\frac{1}{16}|^2 
 \nonumber \\ 
Z_{AO^f,AO^s} &= \big(|\chi^{u1_4}_1 |^2 + |\chi^{u1_4}_{-1}|^2 \big)\big( |\chi^\text{Is}_0|^2+|\chi^\text{Is}_\frac{1}{2}|^2 \big). 
\end{align}

\begin{align}
\label{sol12}
 Z_{AE^f,PE^s} &= \big(|\chi^{u1_4}_0 |^2 + |\chi^{u1_4}_2|^2 \big)\big( |\chi^\text{Is}_0|^2+|\chi^\text{Is}_\frac{1}{2}|^2 \big) 
 \nonumber \\ 
Z_{PO^f,PE^s} &= \big(|\chi^{u1_4}_1 |^2 + |\chi^{u1_4}_{-1}|^2 \big)|\chi^\text{Is}_\frac{1}{16}|^2 
 \nonumber \\ 
Z_{PE^f,PE^s} &= \big(|\chi^{u1_4}_1 |^2 + |\chi^{u1_4}_{-1}|^2 \big)|\chi^\text{Is}_\frac{1}{16}|^2 
 \nonumber \\ 
Z_{AO^f,PE^s} &= \big(|\chi^{u1_4}_0 |^2 + |\chi^{u1_4}_2|^2 \big)\big(\chi^\text{Is}_0 \bar \chi^\text{Is}_\frac{1}{2} + \chi^\text{Is}_\frac12 \bar \chi^\text{Is}_0\big) 
 \nonumber \\ 
Z_{AE^f,PO^s} &= \big( \chi^{u1_4}_0 \bar \chi^{u1_4}_2 + \chi^{u1_4}_2 \bar \chi^{u1_4}_0 \big)\big(\chi^\text{Is}_0 \bar \chi^\text{Is}_\frac{1}{2} + \chi^\text{Is}_\frac12 \bar \chi^\text{Is}_0\big) 
 \nonumber \\ 
Z_{PO^f,PO^s} &= \big( \chi^{u1_4}_1 \bar \chi^{u1_4}_{-1} + \chi^{u1_4}_{-1} \bar \chi^{u1_4}_1 \big)|\chi^\text{Is}_\frac{1}{16}|^2 
 \nonumber \\ 
Z_{PE^f,PO^s} &= \big( \chi^{u1_4}_1 \bar \chi^{u1_4}_{-1} + \chi^{u1_4}_{-1} \bar \chi^{u1_4}_1 \big)|\chi^\text{Is}_\frac{1}{16}|^2 
 \nonumber \\ 
Z_{AO^f,PO^s} &= \big( \chi^{u1_4}_0 \bar \chi^{u1_4}_2 + \chi^{u1_4}_2 \bar \chi^{u1_4}_0 \big)\big( |\chi^\text{Is}_0|^2+|\chi^\text{Is}_\frac{1}{2}|^2 \big) 
 \nonumber \\ 
Z_{AE^f,AE^s} &= \big(|\chi^{u1_4}_1 |^2 + |\chi^{u1_4}_{-1}|^2 \big)\big( |\chi^\text{Is}_0|^2+|\chi^\text{Is}_\frac{1}{2}|^2 \big) 
 \nonumber \\ 
Z_{PO^f,AE^s} &= \big(|\chi^{u1_4}_0 |^2 + |\chi^{u1_4}_2|^2 \big)|\chi^\text{Is}_\frac{1}{16}|^2 
 \nonumber \\ 
Z_{PE^f,AE^s} &= \big(|\chi^{u1_4}_0 |^2 + |\chi^{u1_4}_2|^2 \big)|\chi^\text{Is}_\frac{1}{16}|^2 
 \\ 
Z_{AO^f,AE^s} &= \big(|\chi^{u1_4}_1 |^2 + |\chi^{u1_4}_{-1}|^2 \big)\big(\chi^\text{Is}_0 \bar \chi^\text{Is}_\frac{1}{2} + \chi^\text{Is}_\frac12 \bar \chi^\text{Is}_0\big) 
 \nonumber \\ 
Z_{AE^f,AO^s} &= \big( \chi^{u1_4}_1 \bar \chi^{u1_4}_{-1} + \chi^{u1_4}_{-1} \bar \chi^{u1_4}_1 \big)\big(\chi^\text{Is}_0 \bar \chi^\text{Is}_\frac{1}{2} + \chi^\text{Is}_\frac12 \bar \chi^\text{Is}_0\big) 
 \nonumber \\ 
Z_{PO^f,AO^s} &= \big( \chi^{u1_4}_0 \bar \chi^{u1_4}_2 + \chi^{u1_4}_2 \bar \chi^{u1_4}_0 \big)|\chi^\text{Is}_\frac{1}{16}|^2 
 \nonumber \\ 
Z_{PE^f,AO^s} &= \big( \chi^{u1_4}_0 \bar \chi^{u1_4}_2 + \chi^{u1_4}_2 \bar \chi^{u1_4}_0 \big)|\chi^\text{Is}_\frac{1}{16}|^2 
 \nonumber \\ 
Z_{AO^f,AO^s} &= \big( \chi^{u1_4}_1 \bar \chi^{u1_4}_{-1} + \chi^{u1_4}_{-1} \bar \chi^{u1_4}_1 \big)\big( |\chi^\text{Is}_0|^2+|\chi^\text{Is}_\frac{1}{2}|^2 \big).
\nonumber 
\end{align}

\begin{align}
\label{sol13}
 Z_{AE^f,PE^s} &= \big(|\chi^{u1_4}_0 |^2 + |\chi^{u1_4}_2|^2 \big)\big( |\chi^\text{Is}_0|^2+|\chi^\text{Is}_\frac{1}{2}|^2 \big) 
 \nonumber \\ 
Z_{PO^f,PE^s} &= \big( \chi^{u1_4}_1 \bar \chi^{u1_4}_{-1} + \chi^{u1_4}_{-1} \bar \chi^{u1_4}_1 \big)|\chi^\text{Is}_\frac{1}{16}|^2 
 \nonumber \\ 
Z_{PE^f,PE^s} &= \big(|\chi^{u1_4}_1 |^2 + |\chi^{u1_4}_{-1}|^2 \big)\big( |\chi^\text{Is}_0|^2+|\chi^\text{Is}_\frac{1}{2}|^2 \big) 
 \nonumber \\ 
Z_{AO^f,PE^s} &= \big( \chi^{u1_4}_0 \bar \chi^{u1_4}_2 + \chi^{u1_4}_2 \bar \chi^{u1_4}_0 \big)|\chi^\text{Is}_\frac{1}{16}|^2 
 \nonumber \\ 
Z_{AE^f,PO^s} &= \big(|\chi^{u1_4}_1 |^2 + |\chi^{u1_4}_{-1}|^2 \big)|\chi^\text{Is}_\frac{1}{16}|^2 
 \nonumber \\ 
Z_{PO^f,PO^s} &= \big( \chi^{u1_4}_0 \bar \chi^{u1_4}_2 + \chi^{u1_4}_2 \bar \chi^{u1_4}_0 \big)\big(\chi^\text{Is}_0 \bar \chi^\text{Is}_\frac{1}{2} + \chi^\text{Is}_\frac12 \bar \chi^\text{Is}_0\big) 
 \nonumber \\ 
Z_{PE^f,PO^s} &= \big(|\chi^{u1_4}_0 |^2 + |\chi^{u1_4}_2|^2 \big)|\chi^\text{Is}_\frac{1}{16}|^2 
 \nonumber \\ 
Z_{AO^f,PO^s} &= \big( \chi^{u1_4}_1 \bar \chi^{u1_4}_{-1} + \chi^{u1_4}_{-1} \bar \chi^{u1_4}_1 \big)\big(\chi^\text{Is}_0 \bar \chi^\text{Is}_\frac{1}{2} + \chi^\text{Is}_\frac12 \bar \chi^\text{Is}_0\big) 
 \nonumber \\ 
Z_{AE^f,AE^s} &= \big(|\chi^{u1_4}_0 |^2 + |\chi^{u1_4}_2|^2 \big)|\chi^\text{Is}_\frac{1}{16}|^2 
 \nonumber \\ 
Z_{PO^f,AE^s} &= \big( \chi^{u1_4}_1 \bar \chi^{u1_4}_{-1} + \chi^{u1_4}_{-1} \bar \chi^{u1_4}_1 \big)\big( |\chi^\text{Is}_0|^2+|\chi^\text{Is}_\frac{1}{2}|^2 \big) 
 \nonumber \\ 
Z_{PE^f,AE^s} &= \big(|\chi^{u1_4}_1 |^2 + |\chi^{u1_4}_{-1}|^2 \big)|\chi^\text{Is}_\frac{1}{16}|^2 
 \nonumber \\ 
Z_{AO^f,AE^s} &= \big( \chi^{u1_4}_0 \bar \chi^{u1_4}_2 + \chi^{u1_4}_2 \bar \chi^{u1_4}_0 \big)\big( |\chi^\text{Is}_0|^2+|\chi^\text{Is}_\frac{1}{2}|^2 \big) 
 \nonumber \\ 
Z_{AE^f,AO^s} &= \big(|\chi^{u1_4}_1 |^2 + |\chi^{u1_4}_{-1}|^2 \big)\big(\chi^\text{Is}_0 \bar \chi^\text{Is}_\frac{1}{2} + \chi^\text{Is}_\frac12 \bar \chi^\text{Is}_0\big) 
 \nonumber \\ 
Z_{PO^f,AO^s} &= \big( \chi^{u1_4}_0 \bar \chi^{u1_4}_2 + \chi^{u1_4}_2 \bar \chi^{u1_4}_0 \big)|\chi^\text{Is}_\frac{1}{16}|^2 
 \nonumber \\ 
Z_{PE^f,AO^s} &= \big(|\chi^{u1_4}_0 |^2 + |\chi^{u1_4}_2|^2 \big)\big(\chi^\text{Is}_0 \bar \chi^\text{Is}_\frac{1}{2} + \chi^\text{Is}_\frac12 \bar \chi^\text{Is}_0\big) 
 \nonumber \\ 
Z_{AO^f,AO^s} &= \big( \chi^{u1_4}_1 \bar \chi^{u1_4}_{-1} + \chi^{u1_4}_{-1} \bar \chi^{u1_4}_1 \big)|\chi^\text{Is}_\frac{1}{16}|^2. 
\end{align}

\begin{align}
\label{sol14}
 Z_{AE^f,PE^s} &= \big(|\chi^{u1_4}_0 |^2 + |\chi^{u1_4}_2|^2 \big)\big( |\chi^\text{Is}_0|^2+|\chi^\text{Is}_\frac{1}{2}|^2 \big) 
 \nonumber \\ 
Z_{PO^f,PE^s} &= \big(|\chi^{u1_4}_1 |^2 + |\chi^{u1_4}_{-1}|^2 \big)|\chi^\text{Is}_\frac{1}{16}|^2 
 \nonumber \\ 
Z_{PE^f,PE^s} &= \big( \chi^{u1_4}_1 \bar \chi^{u1_4}_{-1} + \chi^{u1_4}_{-1} \bar \chi^{u1_4}_1 \big)\big( |\chi^\text{Is}_0|^2+|\chi^\text{Is}_\frac{1}{2}|^2 \big) 
 \nonumber \\ 
Z_{AO^f,PE^s} &= \big( \chi^{u1_4}_0 \bar \chi^{u1_4}_2 + \chi^{u1_4}_2 \bar \chi^{u1_4}_0 \big)|\chi^\text{Is}_\frac{1}{16}|^2 
 \nonumber \\ 
Z_{AE^f,PO^s} &= \big( \chi^{u1_4}_1 \bar \chi^{u1_4}_{-1} + \chi^{u1_4}_{-1} \bar \chi^{u1_4}_1 \big)|\chi^\text{Is}_\frac{1}{16}|^2 
 \nonumber \\ 
Z_{PO^f,PO^s} &= \big( \chi^{u1_4}_0 \bar \chi^{u1_4}_2 + \chi^{u1_4}_2 \bar \chi^{u1_4}_0 \big)\big(\chi^\text{Is}_0 \bar \chi^\text{Is}_\frac{1}{2} + \chi^\text{Is}_\frac12 \bar \chi^\text{Is}_0\big) 
 \nonumber \\ 
Z_{PE^f,PO^s} &= \big(|\chi^{u1_4}_0 |^2 + |\chi^{u1_4}_2|^2 \big)|\chi^\text{Is}_\frac{1}{16}|^2 
 \nonumber \\ 
Z_{AO^f,PO^s} &= \big(|\chi^{u1_4}_1 |^2 + |\chi^{u1_4}_{-1}|^2 \big)\big(\chi^\text{Is}_0 \bar \chi^\text{Is}_\frac{1}{2} + \chi^\text{Is}_\frac12 \bar \chi^\text{Is}_0\big) 
 \nonumber \\ 
Z_{AE^f,AE^s} &= \big(|\chi^{u1_4}_0 |^2 + |\chi^{u1_4}_2|^2 \big)|\chi^\text{Is}_\frac{1}{16}|^2 
 \nonumber \\ 
Z_{PO^f,AE^s} &= \big(|\chi^{u1_4}_1 |^2 + |\chi^{u1_4}_{-1}|^2 \big)\big( |\chi^\text{Is}_0|^2+|\chi^\text{Is}_\frac{1}{2}|^2 \big) 
 \nonumber \\ 
Z_{PE^f,AE^s} &= \big( \chi^{u1_4}_1 \bar \chi^{u1_4}_{-1} + \chi^{u1_4}_{-1} \bar \chi^{u1_4}_1 \big)|\chi^\text{Is}_\frac{1}{16}|^2 
 \nonumber \\ 
Z_{AO^f,AE^s} &= \big( \chi^{u1_4}_0 \bar \chi^{u1_4}_2 + \chi^{u1_4}_2 \bar \chi^{u1_4}_0 \big)\big( |\chi^\text{Is}_0|^2+|\chi^\text{Is}_\frac{1}{2}|^2 \big) 
 \nonumber \\ 
Z_{AE^f,AO^s} &= \big( \chi^{u1_4}_1 \bar \chi^{u1_4}_{-1} + \chi^{u1_4}_{-1} \bar \chi^{u1_4}_1 \big)\big(\chi^\text{Is}_0 \bar \chi^\text{Is}_\frac{1}{2} + \chi^\text{Is}_\frac12 \bar \chi^\text{Is}_0\big) 
 \nonumber \\ 
Z_{PO^f,AO^s} &= \big( \chi^{u1_4}_0 \bar \chi^{u1_4}_2 + \chi^{u1_4}_2 \bar \chi^{u1_4}_0 \big)|\chi^\text{Is}_\frac{1}{16}|^2 
 \nonumber \\ 
Z_{PE^f,AO^s} &= \big(|\chi^{u1_4}_0 |^2 + |\chi^{u1_4}_2|^2 \big)\big(\chi^\text{Is}_0 \bar \chi^\text{Is}_\frac{1}{2} + \chi^\text{Is}_\frac12 \bar \chi^\text{Is}_0\big) 
 \nonumber \\ 
Z_{AO^f,AO^s} &= \big(|\chi^{u1_4}_1 |^2 + |\chi^{u1_4}_{-1}|^2 \big)|\chi^\text{Is}_\frac{1}{16}|^2. 
\end{align}

\begin{align}
\label{sol15}
 Z_{AE^f,PE^s} &= \big(|\chi^{u1_4}_0 |^2 + |\chi^{u1_4}_2|^2 \big)\big( |\chi^\text{Is}_0|^2+|\chi^\text{Is}_\frac{1}{2}|^2 \big) 
 \nonumber \\ 
Z_{PO^f,PE^s} &= \big(|\chi^{u1_4}_1 |^2 + |\chi^{u1_4}_{-1}|^2 \big)\big( |\chi^\text{Is}_0|^2+|\chi^\text{Is}_\frac{1}{2}|^2 \big) 
 \nonumber \\ 
Z_{PE^f,PE^s} &= \big( \chi^{u1_4}_1 \bar \chi^{u1_4}_{-1} + \chi^{u1_4}_{-1} \bar \chi^{u1_4}_1 \big)|\chi^\text{Is}_\frac{1}{16}|^2 
 \nonumber \\ 
Z_{AO^f,PE^s} &= \big( \chi^{u1_4}_0 \bar \chi^{u1_4}_2 + \chi^{u1_4}_2 \bar \chi^{u1_4}_0 \big)|\chi^\text{Is}_\frac{1}{16}|^2 
 \nonumber \\ 
Z_{AE^f,PO^s} &= \big(|\chi^{u1_4}_1 |^2 + |\chi^{u1_4}_{-1}|^2 \big)|\chi^\text{Is}_\frac{1}{16}|^2 
 \nonumber \\ 
Z_{PO^f,PO^s} &= \big(|\chi^{u1_4}_0 |^2 + |\chi^{u1_4}_2|^2 \big)|\chi^\text{Is}_\frac{1}{16}|^2 
 \nonumber \\ 
Z_{PE^f,PO^s} &= \big( \chi^{u1_4}_0 \bar \chi^{u1_4}_2 + \chi^{u1_4}_2 \bar \chi^{u1_4}_0 \big)\big(\chi^\text{Is}_0 \bar \chi^\text{Is}_\frac{1}{2} + \chi^\text{Is}_\frac12 \bar \chi^\text{Is}_0\big) 
 \nonumber \\ 
Z_{AO^f,PO^s} &= \big( \chi^{u1_4}_1 \bar \chi^{u1_4}_{-1} + \chi^{u1_4}_{-1} \bar \chi^{u1_4}_1 \big)\big(\chi^\text{Is}_0 \bar \chi^\text{Is}_\frac{1}{2} + \chi^\text{Is}_\frac12 \bar \chi^\text{Is}_0\big) 
 \nonumber \\ 
Z_{AE^f,AE^s} &= \big(|\chi^{u1_4}_0 |^2 + |\chi^{u1_4}_2|^2 \big)|\chi^\text{Is}_\frac{1}{16}|^2 
 \nonumber \\ 
Z_{PO^f,AE^s} &= \big(|\chi^{u1_4}_1 |^2 + |\chi^{u1_4}_{-1}|^2 \big)|\chi^\text{Is}_\frac{1}{16}|^2 
 \nonumber \\ 
Z_{PE^f,AE^s} &= \big( \chi^{u1_4}_1 \bar \chi^{u1_4}_{-1} + \chi^{u1_4}_{-1} \bar \chi^{u1_4}_1 \big)\big( |\chi^\text{Is}_0|^2+|\chi^\text{Is}_\frac{1}{2}|^2 \big) 
 \nonumber \\ 
Z_{AO^f,AE^s} &= \big( \chi^{u1_4}_0 \bar \chi^{u1_4}_2 + \chi^{u1_4}_2 \bar \chi^{u1_4}_0 \big)\big( |\chi^\text{Is}_0|^2+|\chi^\text{Is}_\frac{1}{2}|^2 \big) 
 \nonumber \\ 
Z_{AE^f,AO^s} &= \big(|\chi^{u1_4}_1 |^2 + |\chi^{u1_4}_{-1}|^2 \big)\big(\chi^\text{Is}_0 \bar \chi^\text{Is}_\frac{1}{2} + \chi^\text{Is}_\frac12 \bar \chi^\text{Is}_0\big) 
 \nonumber \\ 
Z_{PO^f,AO^s} &= \big(|\chi^{u1_4}_0 |^2 + |\chi^{u1_4}_2|^2 \big)\big(\chi^\text{Is}_0 \bar \chi^\text{Is}_\frac{1}{2} + \chi^\text{Is}_\frac12 \bar \chi^\text{Is}_0\big) 
 \nonumber \\ 
Z_{PE^f,AO^s} &= \big( \chi^{u1_4}_0 \bar \chi^{u1_4}_2 + \chi^{u1_4}_2 \bar \chi^{u1_4}_0 \big)|\chi^\text{Is}_\frac{1}{16}|^2 
 \nonumber \\ 
Z_{AO^f,AO^s} &= \big( \chi^{u1_4}_1 \bar \chi^{u1_4}_{-1} + \chi^{u1_4}_{-1} \bar \chi^{u1_4}_1 \big)|\chi^\text{Is}_\frac{1}{16}|^2. 
\end{align}

\begin{align}
\label{sol16}
 Z_{AE^f,PE^s} &= \big(|\chi^{u1_4}_0 |^2 + |\chi^{u1_4}_2|^2 \big)\big( |\chi^\text{Is}_0|^2+|\chi^\text{Is}_\frac{1}{2}|^2 \big) 
 \nonumber \\ 
Z_{PO^f,PE^s} &= \big( \chi^{u1_4}_1 \bar \chi^{u1_4}_{-1} + \chi^{u1_4}_{-1} \bar \chi^{u1_4}_1 \big)\big( |\chi^\text{Is}_0|^2+|\chi^\text{Is}_\frac{1}{2}|^2 \big) 
 \nonumber \\ 
Z_{PE^f,PE^s} &= \big(|\chi^{u1_4}_1 |^2 + |\chi^{u1_4}_{-1}|^2 \big)|\chi^\text{Is}_\frac{1}{16}|^2 
 \nonumber \\ 
Z_{AO^f,PE^s} &= \big( \chi^{u1_4}_0 \bar \chi^{u1_4}_2 + \chi^{u1_4}_2 \bar \chi^{u1_4}_0 \big)|\chi^\text{Is}_\frac{1}{16}|^2 
 \nonumber \\ 
Z_{AE^f,PO^s} &= \big( \chi^{u1_4}_1 \bar \chi^{u1_4}_{-1} + \chi^{u1_4}_{-1} \bar \chi^{u1_4}_1 \big)|\chi^\text{Is}_\frac{1}{16}|^2 
 \nonumber \\ 
Z_{PO^f,PO^s} &= \big(|\chi^{u1_4}_0 |^2 + |\chi^{u1_4}_2|^2 \big)|\chi^\text{Is}_\frac{1}{16}|^2 
 \nonumber \\ 
Z_{PE^f,PO^s} &= \big( \chi^{u1_4}_0 \bar \chi^{u1_4}_2 + \chi^{u1_4}_2 \bar \chi^{u1_4}_0 \big)\big(\chi^\text{Is}_0 \bar \chi^\text{Is}_\frac{1}{2} + \chi^\text{Is}_\frac12 \bar \chi^\text{Is}_0\big) 
 \nonumber \\ 
Z_{AO^f,PO^s} &= \big(|\chi^{u1_4}_1 |^2 + |\chi^{u1_4}_{-1}|^2 \big)\big(\chi^\text{Is}_0 \bar \chi^\text{Is}_\frac{1}{2} + \chi^\text{Is}_\frac12 \bar \chi^\text{Is}_0\big) 
 \nonumber \\ 
Z_{AE^f,AE^s} &= \big(|\chi^{u1_4}_0 |^2 + |\chi^{u1_4}_2|^2 \big)|\chi^\text{Is}_\frac{1}{16}|^2 
 \nonumber \\ 
Z_{PO^f,AE^s} &= \big( \chi^{u1_4}_1 \bar \chi^{u1_4}_{-1} + \chi^{u1_4}_{-1} \bar \chi^{u1_4}_1 \big)|\chi^\text{Is}_\frac{1}{16}|^2 
 \nonumber \\ 
Z_{PE^f,AE^s} &= \big(|\chi^{u1_4}_1 |^2 + |\chi^{u1_4}_{-1}|^2 \big)\big( |\chi^\text{Is}_0|^2+|\chi^\text{Is}_\frac{1}{2}|^2 \big) 
 \nonumber \\ 
Z_{AO^f,AE^s} &= \big( \chi^{u1_4}_0 \bar \chi^{u1_4}_2 + \chi^{u1_4}_2 \bar \chi^{u1_4}_0 \big)\big( |\chi^\text{Is}_0|^2+|\chi^\text{Is}_\frac{1}{2}|^2 \big) 
 \nonumber \\ 
Z_{AE^f,AO^s} &= \big( \chi^{u1_4}_1 \bar \chi^{u1_4}_{-1} + \chi^{u1_4}_{-1} \bar \chi^{u1_4}_1 \big)\big(\chi^\text{Is}_0 \bar \chi^\text{Is}_\frac{1}{2} + \chi^\text{Is}_\frac12 \bar \chi^\text{Is}_0\big) 
 \nonumber \\ 
Z_{PO^f,AO^s} &= \big(|\chi^{u1_4}_0 |^2 + |\chi^{u1_4}_2|^2 \big)\big(\chi^\text{Is}_0 \bar \chi^\text{Is}_\frac{1}{2} + \chi^\text{Is}_\frac12 \bar \chi^\text{Is}_0\big) 
 \nonumber \\ 
Z_{PE^f,AO^s} &= \big( \chi^{u1_4}_0 \bar \chi^{u1_4}_2 + \chi^{u1_4}_2 \bar \chi^{u1_4}_0 \big)|\chi^\text{Is}_\frac{1}{16}|^2 
 \nonumber \\ 
Z_{AO^f,AO^s} &= \big(|\chi^{u1_4}_1 |^2 + |\chi^{u1_4}_{-1}|^2 \big)|\chi^\text{Is}_\frac{1}{16}|^2.
\end{align}

\begin{align}
\label{sol17}
 Z_{AE^f,PE^s} &= \big(|\chi^{u1_4}_0 |^2 + |\chi^{u1_4}_2|^2 \big)\big( |\chi^\text{Is}_0|^2+|\chi^\text{Is}_\frac{1}{2}|^2 \big) 
 \nonumber \\ 
Z_{PO^f,PE^s} &= \big( \chi^{u1_4}_1 \bar \chi^{u1_4}_{-1} + \chi^{u1_4}_{-1} \bar \chi^{u1_4}_1 \big)|\chi^\text{Is}_\frac{1}{16}|^2 
 \nonumber \\ 
Z_{PE^f,PE^s} &= \big( \chi^{u1_4}_1 \bar \chi^{u1_4}_{-1} + \chi^{u1_4}_{-1} \bar \chi^{u1_4}_1 \big)|\chi^\text{Is}_\frac{1}{16}|^2 
 \nonumber \\ 
Z_{AO^f,PE^s} &= \big(|\chi^{u1_4}_0 |^2 + |\chi^{u1_4}_2|^2 \big)\big(\chi^\text{Is}_0 \bar \chi^\text{Is}_\frac{1}{2} + \chi^\text{Is}_\frac12 \bar \chi^\text{Is}_0\big) 
 \nonumber \\ 
Z_{AE^f,PO^s} &= \big( \chi^{u1_4}_1 \bar \chi^{u1_4}_{-1} + \chi^{u1_4}_{-1} \bar \chi^{u1_4}_1 \big)\big( |\chi^\text{Is}_0|^2+|\chi^\text{Is}_\frac{1}{2}|^2 \big) 
 \nonumber \\ 
Z_{PO^f,PO^s} &= \big(|\chi^{u1_4}_0 |^2 + |\chi^{u1_4}_2|^2 \big)|\chi^\text{Is}_\frac{1}{16}|^2 
 \nonumber \\ 
Z_{PE^f,PO^s} &= \big(|\chi^{u1_4}_0 |^2 + |\chi^{u1_4}_2|^2 \big)|\chi^\text{Is}_\frac{1}{16}|^2 
 \nonumber \\ 
Z_{AO^f,PO^s} &= \big( \chi^{u1_4}_1 \bar \chi^{u1_4}_{-1} + \chi^{u1_4}_{-1} \bar \chi^{u1_4}_1 \big)\big(\chi^\text{Is}_0 \bar \chi^\text{Is}_\frac{1}{2} + \chi^\text{Is}_\frac12 \bar \chi^\text{Is}_0\big) 
 \nonumber \\ 
Z_{AE^f,AE^s} &= \big( \chi^{u1_4}_0 \bar \chi^{u1_4}_2 + \chi^{u1_4}_2 \bar \chi^{u1_4}_0 \big)\big(\chi^\text{Is}_0 \bar \chi^\text{Is}_\frac{1}{2} + \chi^\text{Is}_\frac12 \bar \chi^\text{Is}_0\big) 
 \nonumber \\ 
Z_{PO^f,AE^s} &= \big(|\chi^{u1_4}_1 |^2 + |\chi^{u1_4}_{-1}|^2 \big)|\chi^\text{Is}_\frac{1}{16}|^2 
 \nonumber \\ 
Z_{PE^f,AE^s} &= \big(|\chi^{u1_4}_1 |^2 + |\chi^{u1_4}_{-1}|^2 \big)|\chi^\text{Is}_\frac{1}{16}|^2 
 \nonumber \\ 
Z_{AO^f,AE^s} &= \big( \chi^{u1_4}_0 \bar \chi^{u1_4}_2 + \chi^{u1_4}_2 \bar \chi^{u1_4}_0 \big)\big( |\chi^\text{Is}_0|^2+|\chi^\text{Is}_\frac{1}{2}|^2 \big) 
 \nonumber \\ 
Z_{AE^f,AO^s} &= \big(|\chi^{u1_4}_1 |^2 + |\chi^{u1_4}_{-1}|^2 \big)\big(\chi^\text{Is}_0 \bar \chi^\text{Is}_\frac{1}{2} + \chi^\text{Is}_\frac12 \bar \chi^\text{Is}_0\big) 
 \nonumber \\ 
Z_{PO^f,AO^s} &= \big( \chi^{u1_4}_0 \bar \chi^{u1_4}_2 + \chi^{u1_4}_2 \bar \chi^{u1_4}_0 \big)|\chi^\text{Is}_\frac{1}{16}|^2 
 \nonumber \\ 
Z_{PE^f,AO^s} &= \big( \chi^{u1_4}_0 \bar \chi^{u1_4}_2 + \chi^{u1_4}_2 \bar \chi^{u1_4}_0 \big)|\chi^\text{Is}_\frac{1}{16}|^2 
 \nonumber \\ 
Z_{AO^f,AO^s} &= \big(|\chi^{u1_4}_1 |^2 + |\chi^{u1_4}_{-1}|^2 \big)\big( |\chi^\text{Is}_0|^2+|\chi^\text{Is}_\frac{1}{2}|^2 \big). 
\end{align}

\begin{align}
\label{sol18}
 Z_{AE^f,PE^s} &= \big(|\chi^{u1_4}_0 |^2 + |\chi^{u1_4}_2|^2 \big)\big( |\chi^\text{Is}_0|^2+|\chi^\text{Is}_\frac{1}{2}|^2 \big) 
 \nonumber \\ 
Z_{PO^f,PE^s} &= \big(|\chi^{u1_4}_1 |^2 + |\chi^{u1_4}_{-1}|^2 \big)|\chi^\text{Is}_\frac{1}{16}|^2 
 \nonumber \\ 
Z_{PE^f,PE^s} &= \big(|\chi^{u1_4}_1 |^2 + |\chi^{u1_4}_{-1}|^2 \big)|\chi^\text{Is}_\frac{1}{16}|^2 
 \nonumber \\ 
Z_{AO^f,PE^s} &= \big(|\chi^{u1_4}_0 |^2 + |\chi^{u1_4}_2|^2 \big)\big(\chi^\text{Is}_0 \bar \chi^\text{Is}_\frac{1}{2} + \chi^\text{Is}_\frac12 \bar \chi^\text{Is}_0\big) 
 \nonumber \\ 
Z_{AE^f,PO^s} &= \big(|\chi^{u1_4}_1 |^2 + |\chi^{u1_4}_{-1}|^2 \big)\big( |\chi^\text{Is}_0|^2+|\chi^\text{Is}_\frac{1}{2}|^2 \big) 
 \nonumber \\ 
Z_{PO^f,PO^s} &= \big(|\chi^{u1_4}_0 |^2 + |\chi^{u1_4}_2|^2 \big)|\chi^\text{Is}_\frac{1}{16}|^2 
 \nonumber \\ 
Z_{PE^f,PO^s} &= \big(|\chi^{u1_4}_0 |^2 + |\chi^{u1_4}_2|^2 \big)|\chi^\text{Is}_\frac{1}{16}|^2 
 \nonumber \\ 
Z_{AO^f,PO^s} &= \big(|\chi^{u1_4}_1 |^2 + |\chi^{u1_4}_{-1}|^2 \big)\big(\chi^\text{Is}_0 \bar \chi^\text{Is}_\frac{1}{2} + \chi^\text{Is}_\frac12 \bar \chi^\text{Is}_0\big) 
 \nonumber \\ 
Z_{AE^f,AE^s} &= \big( \chi^{u1_4}_0 \bar \chi^{u1_4}_2 + \chi^{u1_4}_2 \bar \chi^{u1_4}_0 \big)\big(\chi^\text{Is}_0 \bar \chi^\text{Is}_\frac{1}{2} + \chi^\text{Is}_\frac12 \bar \chi^\text{Is}_0\big) 
 \nonumber \\ 
Z_{PO^f,AE^s} &= \big( \chi^{u1_4}_1 \bar \chi^{u1_4}_{-1} + \chi^{u1_4}_{-1} \bar \chi^{u1_4}_1 \big)|\chi^\text{Is}_\frac{1}{16}|^2 
 \nonumber \\ 
Z_{PE^f,AE^s} &= \big( \chi^{u1_4}_1 \bar \chi^{u1_4}_{-1} + \chi^{u1_4}_{-1} \bar \chi^{u1_4}_1 \big)|\chi^\text{Is}_\frac{1}{16}|^2 
 \nonumber \\ 
Z_{AO^f,AE^s} &= \big( \chi^{u1_4}_0 \bar \chi^{u1_4}_2 + \chi^{u1_4}_2 \bar \chi^{u1_4}_0 \big)\big( |\chi^\text{Is}_0|^2+|\chi^\text{Is}_\frac{1}{2}|^2 \big) 
 \nonumber \\ 
Z_{AE^f,AO^s} &= \big( \chi^{u1_4}_1 \bar \chi^{u1_4}_{-1} + \chi^{u1_4}_{-1} \bar \chi^{u1_4}_1 \big)\big(\chi^\text{Is}_0 \bar \chi^\text{Is}_\frac{1}{2} + \chi^\text{Is}_\frac12 \bar \chi^\text{Is}_0\big) 
 \nonumber \\ 
Z_{PO^f,AO^s} &= \big( \chi^{u1_4}_0 \bar \chi^{u1_4}_2 + \chi^{u1_4}_2 \bar \chi^{u1_4}_0 \big)|\chi^\text{Is}_\frac{1}{16}|^2 
 \\ 
Z_{PE^f,AO^s} &= \big( \chi^{u1_4}_0 \bar \chi^{u1_4}_2 + \chi^{u1_4}_2 \bar \chi^{u1_4}_0 \big)|\chi^\text{Is}_\frac{1}{16}|^2 
 \nonumber \\ 
Z_{AO^f,AO^s} &= \big( \chi^{u1_4}_1 \bar \chi^{u1_4}_{-1} + \chi^{u1_4}_{-1} \bar \chi^{u1_4}_1 \big)\big( |\chi^\text{Is}_0|^2+|\chi^\text{Is}_\frac{1}{2}|^2 \big). 
\nonumber 
\end{align}

\begin{align}
\label{sol19}
 Z_{AE^f,PE^s} &= \big(|\chi^{u1_4}_0 |^2 + |\chi^{u1_4}_2|^2 \big)\big( |\chi^\text{Is}_0|^2+|\chi^\text{Is}_\frac{1}{2}|^2 \big) 
 \nonumber \\ 
Z_{PO^f,PE^s} &= \big(|\chi^{u1_4}_0 |^2 + |\chi^{u1_4}_2|^2 \big)|\chi^\text{Is}_\frac{1}{16}|^2 
 \nonumber \\ 
Z_{PE^f,PE^s} &= \big( \chi^{u1_4}_0 \bar \chi^{u1_4}_2 + \chi^{u1_4}_2 \bar \chi^{u1_4}_0 \big)\big(\chi^\text{Is}_0 \bar \chi^\text{Is}_\frac{1}{2} + \chi^\text{Is}_\frac12 \bar \chi^\text{Is}_0\big) 
 \nonumber \\ 
Z_{AO^f,PE^s} &= \big( \chi^{u1_4}_0 \bar \chi^{u1_4}_2 + \chi^{u1_4}_2 \bar \chi^{u1_4}_0 \big)|\chi^\text{Is}_\frac{1}{16}|^2 
 \nonumber \\ 
Z_{AE^f,PO^s} &= \big(|\chi^{u1_4}_1 |^2 + |\chi^{u1_4}_{-1}|^2 \big)|\chi^\text{Is}_\frac{1}{16}|^2 
 \nonumber \\ 
Z_{PO^f,PO^s} &= \big(|\chi^{u1_4}_1 |^2 + |\chi^{u1_4}_{-1}|^2 \big)\big( |\chi^\text{Is}_0|^2+|\chi^\text{Is}_\frac{1}{2}|^2 \big) 
 \nonumber \\ 
Z_{PE^f,PO^s} &= \big( \chi^{u1_4}_1 \bar \chi^{u1_4}_{-1} + \chi^{u1_4}_{-1} \bar \chi^{u1_4}_1 \big)|\chi^\text{Is}_\frac{1}{16}|^2 
 \nonumber \\ 
Z_{AO^f,PO^s} &= \big( \chi^{u1_4}_1 \bar \chi^{u1_4}_{-1} + \chi^{u1_4}_{-1} \bar \chi^{u1_4}_1 \big)\big(\chi^\text{Is}_0 \bar \chi^\text{Is}_\frac{1}{2} + \chi^\text{Is}_\frac12 \bar \chi^\text{Is}_0\big) 
 \nonumber \\ 
Z_{AE^f,AE^s} &= \big( \chi^{u1_4}_1 \bar \chi^{u1_4}_{-1} + \chi^{u1_4}_{-1} \bar \chi^{u1_4}_1 \big)|\chi^\text{Is}_\frac{1}{16}|^2 
 \nonumber \\ 
Z_{PO^f,AE^s} &= \big( \chi^{u1_4}_1 \bar \chi^{u1_4}_{-1} + \chi^{u1_4}_{-1} \bar \chi^{u1_4}_1 \big)\big( |\chi^\text{Is}_0|^2+|\chi^\text{Is}_\frac{1}{2}|^2 \big) 
 \nonumber \\ 
Z_{PE^f,AE^s} &= \big(|\chi^{u1_4}_1 |^2 + |\chi^{u1_4}_{-1}|^2 \big)|\chi^\text{Is}_\frac{1}{16}|^2 
 \nonumber \\ 
Z_{AO^f,AE^s} &= \big(|\chi^{u1_4}_1 |^2 + |\chi^{u1_4}_{-1}|^2 \big)\big(\chi^\text{Is}_0 \bar \chi^\text{Is}_\frac{1}{2} + \chi^\text{Is}_\frac12 \bar \chi^\text{Is}_0\big) 
 \nonumber \\ 
Z_{AE^f,AO^s} &= \big( \chi^{u1_4}_0 \bar \chi^{u1_4}_2 + \chi^{u1_4}_2 \bar \chi^{u1_4}_0 \big)\big( |\chi^\text{Is}_0|^2+|\chi^\text{Is}_\frac{1}{2}|^2 \big) 
 \nonumber \\ 
Z_{PO^f,AO^s} &= \big( \chi^{u1_4}_0 \bar \chi^{u1_4}_2 + \chi^{u1_4}_2 \bar \chi^{u1_4}_0 \big)|\chi^\text{Is}_\frac{1}{16}|^2 
 \nonumber \\ 
Z_{PE^f,AO^s} &= \big(|\chi^{u1_4}_0 |^2 + |\chi^{u1_4}_2|^2 \big)\big(\chi^\text{Is}_0 \bar \chi^\text{Is}_\frac{1}{2} + \chi^\text{Is}_\frac12 \bar \chi^\text{Is}_0\big) 
 \nonumber \\ 
Z_{AO^f,AO^s} &= \big(|\chi^{u1_4}_0 |^2 + |\chi^{u1_4}_2|^2 \big)|\chi^\text{Is}_\frac{1}{16}|^2. 
\end{align}

\begin{align}
\label{sol20}
 Z_{AE^f,PE^s} &= \big(|\chi^{u1_4}_0 |^2 + |\chi^{u1_4}_2|^2 \big)\big( |\chi^\text{Is}_0|^2+|\chi^\text{Is}_\frac{1}{2}|^2 \big) 
 \nonumber \\ 
Z_{PO^f,PE^s} &= \big(|\chi^{u1_4}_0 |^2 + |\chi^{u1_4}_2|^2 \big)|\chi^\text{Is}_\frac{1}{16}|^2 
 \nonumber \\ 
Z_{PE^f,PE^s} &= \big( \chi^{u1_4}_0 \bar \chi^{u1_4}_2 + \chi^{u1_4}_2 \bar \chi^{u1_4}_0 \big)\big(\chi^\text{Is}_0 \bar \chi^\text{Is}_\frac{1}{2} + \chi^\text{Is}_\frac12 \bar \chi^\text{Is}_0\big) 
 \nonumber \\ 
Z_{AO^f,PE^s} &= \big( \chi^{u1_4}_0 \bar \chi^{u1_4}_2 + \chi^{u1_4}_2 \bar \chi^{u1_4}_0 \big)|\chi^\text{Is}_\frac{1}{16}|^2 
 \nonumber \\ 
Z_{AE^f,PO^s} &= \big( \chi^{u1_4}_1 \bar \chi^{u1_4}_{-1} + \chi^{u1_4}_{-1} \bar \chi^{u1_4}_1 \big)|\chi^\text{Is}_\frac{1}{16}|^2 
 \nonumber \\ 
Z_{PO^f,PO^s} &= \big( \chi^{u1_4}_1 \bar \chi^{u1_4}_{-1} + \chi^{u1_4}_{-1} \bar \chi^{u1_4}_1 \big)\big( |\chi^\text{Is}_0|^2+|\chi^\text{Is}_\frac{1}{2}|^2 \big) 
 \nonumber \\ 
Z_{PE^f,PO^s} &= \big(|\chi^{u1_4}_1 |^2 + |\chi^{u1_4}_{-1}|^2 \big)|\chi^\text{Is}_\frac{1}{16}|^2 
 \nonumber \\ 
Z_{AO^f,PO^s} &= \big(|\chi^{u1_4}_1 |^2 + |\chi^{u1_4}_{-1}|^2 \big)\big(\chi^\text{Is}_0 \bar \chi^\text{Is}_\frac{1}{2} + \chi^\text{Is}_\frac12 \bar \chi^\text{Is}_0\big) 
 \nonumber \\ 
Z_{AE^f,AE^s} &= \big(|\chi^{u1_4}_1 |^2 + |\chi^{u1_4}_{-1}|^2 \big)|\chi^\text{Is}_\frac{1}{16}|^2 
 \nonumber \\ 
Z_{PO^f,AE^s} &= \big(|\chi^{u1_4}_1 |^2 + |\chi^{u1_4}_{-1}|^2 \big)\big( |\chi^\text{Is}_0|^2+|\chi^\text{Is}_\frac{1}{2}|^2 \big) 
 \nonumber \\ 
Z_{PE^f,AE^s} &= \big( \chi^{u1_4}_1 \bar \chi^{u1_4}_{-1} + \chi^{u1_4}_{-1} \bar \chi^{u1_4}_1 \big)|\chi^\text{Is}_\frac{1}{16}|^2 
 \nonumber \\ 
Z_{AO^f,AE^s} &= \big( \chi^{u1_4}_1 \bar \chi^{u1_4}_{-1} + \chi^{u1_4}_{-1} \bar \chi^{u1_4}_1 \big)\big(\chi^\text{Is}_0 \bar \chi^\text{Is}_\frac{1}{2} + \chi^\text{Is}_\frac12 \bar \chi^\text{Is}_0\big) 
 \nonumber \\ 
Z_{AE^f,AO^s} &= \big( \chi^{u1_4}_0 \bar \chi^{u1_4}_2 + \chi^{u1_4}_2 \bar \chi^{u1_4}_0 \big)\big( |\chi^\text{Is}_0|^2+|\chi^\text{Is}_\frac{1}{2}|^2 \big) 
 \nonumber \\ 
Z_{PO^f,AO^s} &= \big( \chi^{u1_4}_0 \bar \chi^{u1_4}_2 + \chi^{u1_4}_2 \bar \chi^{u1_4}_0 \big)|\chi^\text{Is}_\frac{1}{16}|^2 
 \nonumber \\ 
Z_{PE^f,AO^s} &= \big(|\chi^{u1_4}_0 |^2 + |\chi^{u1_4}_2|^2 \big)\big(\chi^\text{Is}_0 \bar \chi^\text{Is}_\frac{1}{2} + \chi^\text{Is}_\frac12 \bar \chi^\text{Is}_0\big) 
 \nonumber \\ 
Z_{AO^f,AO^s} &= \big(|\chi^{u1_4}_0 |^2 + |\chi^{u1_4}_2|^2 \big)|\chi^\text{Is}_\frac{1}{16}|^2. 
\end{align}

\begin{align}
\label{sol21}
 Z_{AE^f,PE^s} &= \big(|\chi^{u1_4}_0 |^2 + |\chi^{u1_4}_2|^2 \big)\big( |\chi^\text{Is}_0|^2+|\chi^\text{Is}_\frac{1}{2}|^2 \big) 
 \nonumber \\ 
Z_{PO^f,PE^s} &= \big( \chi^{u1_4}_0 \bar \chi^{u1_4}_2 + \chi^{u1_4}_2 \bar \chi^{u1_4}_0 \big)\big(\chi^\text{Is}_0 \bar \chi^\text{Is}_\frac{1}{2} + \chi^\text{Is}_\frac12 \bar \chi^\text{Is}_0\big) 
 \nonumber \\ 
Z_{PE^f,PE^s} &= \big(|\chi^{u1_4}_0 |^2 + |\chi^{u1_4}_2|^2 \big)|\chi^\text{Is}_\frac{1}{16}|^2 
 \nonumber \\ 
Z_{AO^f,PE^s} &= \big( \chi^{u1_4}_0 \bar \chi^{u1_4}_2 + \chi^{u1_4}_2 \bar \chi^{u1_4}_0 \big)|\chi^\text{Is}_\frac{1}{16}|^2 
 \nonumber \\ 
Z_{AE^f,PO^s} &= \big(|\chi^{u1_4}_1 |^2 + |\chi^{u1_4}_{-1}|^2 \big)|\chi^\text{Is}_\frac{1}{16}|^2 
 \nonumber \\ 
Z_{PO^f,PO^s} &= \big( \chi^{u1_4}_1 \bar \chi^{u1_4}_{-1} + \chi^{u1_4}_{-1} \bar \chi^{u1_4}_1 \big)|\chi^\text{Is}_\frac{1}{16}|^2 
 \nonumber \\ 
Z_{PE^f,PO^s} &= \big(|\chi^{u1_4}_1 |^2 + |\chi^{u1_4}_{-1}|^2 \big)\big( |\chi^\text{Is}_0|^2+|\chi^\text{Is}_\frac{1}{2}|^2 \big) 
 \nonumber \\ 
Z_{AO^f,PO^s} &= \big( \chi^{u1_4}_1 \bar \chi^{u1_4}_{-1} + \chi^{u1_4}_{-1} \bar \chi^{u1_4}_1 \big)\big(\chi^\text{Is}_0 \bar \chi^\text{Is}_\frac{1}{2} + \chi^\text{Is}_\frac12 \bar \chi^\text{Is}_0\big) 
 \nonumber \\ 
Z_{AE^f,AE^s} &= \big( \chi^{u1_4}_1 \bar \chi^{u1_4}_{-1} + \chi^{u1_4}_{-1} \bar \chi^{u1_4}_1 \big)|\chi^\text{Is}_\frac{1}{16}|^2 
 \nonumber \\ 
Z_{PO^f,AE^s} &= \big(|\chi^{u1_4}_1 |^2 + |\chi^{u1_4}_{-1}|^2 \big)|\chi^\text{Is}_\frac{1}{16}|^2 
 \nonumber \\ 
Z_{PE^f,AE^s} &= \big( \chi^{u1_4}_1 \bar \chi^{u1_4}_{-1} + \chi^{u1_4}_{-1} \bar \chi^{u1_4}_1 \big)\big( |\chi^\text{Is}_0|^2+|\chi^\text{Is}_\frac{1}{2}|^2 \big) 
 \nonumber \\ 
Z_{AO^f,AE^s} &= \big(|\chi^{u1_4}_1 |^2 + |\chi^{u1_4}_{-1}|^2 \big)\big(\chi^\text{Is}_0 \bar \chi^\text{Is}_\frac{1}{2} + \chi^\text{Is}_\frac12 \bar \chi^\text{Is}_0\big) 
 \nonumber \\ 
Z_{AE^f,AO^s} &= \big( \chi^{u1_4}_0 \bar \chi^{u1_4}_2 + \chi^{u1_4}_2 \bar \chi^{u1_4}_0 \big)\big( |\chi^\text{Is}_0|^2+|\chi^\text{Is}_\frac{1}{2}|^2 \big) 
 \nonumber \\ 
Z_{PO^f,AO^s} &= \big(|\chi^{u1_4}_0 |^2 + |\chi^{u1_4}_2|^2 \big)\big(\chi^\text{Is}_0 \bar \chi^\text{Is}_\frac{1}{2} + \chi^\text{Is}_\frac12 \bar \chi^\text{Is}_0\big) 
 \nonumber \\ 
Z_{PE^f,AO^s} &= \big( \chi^{u1_4}_0 \bar \chi^{u1_4}_2 + \chi^{u1_4}_2 \bar \chi^{u1_4}_0 \big)|\chi^\text{Is}_\frac{1}{16}|^2 
 \nonumber \\ 
Z_{AO^f,AO^s} &= \big(|\chi^{u1_4}_0 |^2 + |\chi^{u1_4}_2|^2 \big)|\chi^\text{Is}_\frac{1}{16}|^2. 
\end{align}

\begin{align}
\label{sol22}
 Z_{AE^f,PE^s} &= \big(|\chi^{u1_4}_0 |^2 + |\chi^{u1_4}_2|^2 \big)\big( |\chi^\text{Is}_0|^2+|\chi^\text{Is}_\frac{1}{2}|^2 \big) 
 \nonumber \\ 
Z_{PO^f,PE^s} &= \big( \chi^{u1_4}_0 \bar \chi^{u1_4}_2 + \chi^{u1_4}_2 \bar \chi^{u1_4}_0 \big)\big(\chi^\text{Is}_0 \bar \chi^\text{Is}_\frac{1}{2} + \chi^\text{Is}_\frac12 \bar \chi^\text{Is}_0\big) 
 \nonumber \\ 
Z_{PE^f,PE^s} &= \big(|\chi^{u1_4}_0 |^2 + |\chi^{u1_4}_2|^2 \big)|\chi^\text{Is}_\frac{1}{16}|^2 
 \nonumber \\ 
Z_{AO^f,PE^s} &= \big( \chi^{u1_4}_0 \bar \chi^{u1_4}_2 + \chi^{u1_4}_2 \bar \chi^{u1_4}_0 \big)|\chi^\text{Is}_\frac{1}{16}|^2 
 \nonumber \\ 
Z_{AE^f,PO^s} &= \big( \chi^{u1_4}_1 \bar \chi^{u1_4}_{-1} + \chi^{u1_4}_{-1} \bar \chi^{u1_4}_1 \big)|\chi^\text{Is}_\frac{1}{16}|^2 
 \nonumber \\ 
Z_{PO^f,PO^s} &= \big(|\chi^{u1_4}_1 |^2 + |\chi^{u1_4}_{-1}|^2 \big)|\chi^\text{Is}_\frac{1}{16}|^2 
 \nonumber \\ 
Z_{PE^f,PO^s} &= \big( \chi^{u1_4}_1 \bar \chi^{u1_4}_{-1} + \chi^{u1_4}_{-1} \bar \chi^{u1_4}_1 \big)\big( |\chi^\text{Is}_0|^2+|\chi^\text{Is}_\frac{1}{2}|^2 \big) 
 \nonumber \\ 
Z_{AO^f,PO^s} &= \big(|\chi^{u1_4}_1 |^2 + |\chi^{u1_4}_{-1}|^2 \big)\big(\chi^\text{Is}_0 \bar \chi^\text{Is}_\frac{1}{2} + \chi^\text{Is}_\frac12 \bar \chi^\text{Is}_0\big) 
 \nonumber \\ 
Z_{AE^f,AE^s} &= \big(|\chi^{u1_4}_1 |^2 + |\chi^{u1_4}_{-1}|^2 \big)|\chi^\text{Is}_\frac{1}{16}|^2 
 \nonumber \\ 
Z_{PO^f,AE^s} &= \big( \chi^{u1_4}_1 \bar \chi^{u1_4}_{-1} + \chi^{u1_4}_{-1} \bar \chi^{u1_4}_1 \big)|\chi^\text{Is}_\frac{1}{16}|^2 
 \nonumber \\ 
Z_{PE^f,AE^s} &= \big(|\chi^{u1_4}_1 |^2 + |\chi^{u1_4}_{-1}|^2 \big)\big( |\chi^\text{Is}_0|^2+|\chi^\text{Is}_\frac{1}{2}|^2 \big) 
 \nonumber \\ 
Z_{AO^f,AE^s} &= \big( \chi^{u1_4}_1 \bar \chi^{u1_4}_{-1} + \chi^{u1_4}_{-1} \bar \chi^{u1_4}_1 \big)\big(\chi^\text{Is}_0 \bar \chi^\text{Is}_\frac{1}{2} + \chi^\text{Is}_\frac12 \bar \chi^\text{Is}_0\big) 
 \nonumber \\ 
Z_{AE^f,AO^s} &= \big( \chi^{u1_4}_0 \bar \chi^{u1_4}_2 + \chi^{u1_4}_2 \bar \chi^{u1_4}_0 \big)\big( |\chi^\text{Is}_0|^2+|\chi^\text{Is}_\frac{1}{2}|^2 \big) 
 \nonumber \\ 
Z_{PO^f,AO^s} &= \big(|\chi^{u1_4}_0 |^2 + |\chi^{u1_4}_2|^2 \big)\big(\chi^\text{Is}_0 \bar \chi^\text{Is}_\frac{1}{2} + \chi^\text{Is}_\frac12 \bar \chi^\text{Is}_0\big) 
 \nonumber \\ 
Z_{PE^f,AO^s} &= \big( \chi^{u1_4}_0 \bar \chi^{u1_4}_2 + \chi^{u1_4}_2 \bar \chi^{u1_4}_0 \big)|\chi^\text{Is}_\frac{1}{16}|^2 
 \nonumber \\ 
Z_{AO^f,AO^s} &= \big(|\chi^{u1_4}_0 |^2 + |\chi^{u1_4}_2|^2 \big)|\chi^\text{Is}_\frac{1}{16}|^2. 
\end{align}

\begin{align}
\label{sol23}
 Z_{AE^f,PE^s} &= \big(|\chi^{u1_4}_0 |^2 + |\chi^{u1_4}_2|^2 \big)\big( |\chi^\text{Is}_0|^2+|\chi^\text{Is}_\frac{1}{2}|^2 \big) 
 \nonumber \\ 
Z_{PO^f,PE^s} &= \big(|\chi^{u1_4}_0 |^2 + |\chi^{u1_4}_2|^2 \big)|\chi^\text{Is}_\frac{1}{16}|^2 
 \nonumber \\ 
Z_{PE^f,PE^s} &= \big(|\chi^{u1_4}_0 |^2 + |\chi^{u1_4}_2|^2 \big)|\chi^\text{Is}_\frac{1}{16}|^2 
 \nonumber \\ 
Z_{AO^f,PE^s} &= \big(|\chi^{u1_4}_0 |^2 + |\chi^{u1_4}_2|^2 \big)\big(\chi^\text{Is}_0 \bar \chi^\text{Is}_\frac{1}{2} + \chi^\text{Is}_\frac12 \bar \chi^\text{Is}_0\big) 
 \nonumber \\ 
Z_{AE^f,PO^s} &= \big( \chi^{u1_4}_1 \bar \chi^{u1_4}_{-1} + \chi^{u1_4}_{-1} \bar \chi^{u1_4}_1 \big)\big( |\chi^\text{Is}_0|^2+|\chi^\text{Is}_\frac{1}{2}|^2 \big) 
 \nonumber \\ 
Z_{PO^f,PO^s} &= \big( \chi^{u1_4}_1 \bar \chi^{u1_4}_{-1} + \chi^{u1_4}_{-1} \bar \chi^{u1_4}_1 \big)|\chi^\text{Is}_\frac{1}{16}|^2 
 \nonumber \\ 
Z_{PE^f,PO^s} &= \big( \chi^{u1_4}_1 \bar \chi^{u1_4}_{-1} + \chi^{u1_4}_{-1} \bar \chi^{u1_4}_1 \big)|\chi^\text{Is}_\frac{1}{16}|^2 
 \nonumber \\ 
Z_{AO^f,PO^s} &= \big( \chi^{u1_4}_1 \bar \chi^{u1_4}_{-1} + \chi^{u1_4}_{-1} \bar \chi^{u1_4}_1 \big)\big(\chi^\text{Is}_0 \bar \chi^\text{Is}_\frac{1}{2} + \chi^\text{Is}_\frac12 \bar \chi^\text{Is}_0\big) 
 \nonumber \\ 
Z_{AE^f,AE^s} &= \big(|\chi^{u1_4}_1 |^2 + |\chi^{u1_4}_{-1}|^2 \big)\big( |\chi^\text{Is}_0|^2+|\chi^\text{Is}_\frac{1}{2}|^2 \big) 
 \nonumber \\ 
Z_{PO^f,AE^s} &= \big(|\chi^{u1_4}_1 |^2 + |\chi^{u1_4}_{-1}|^2 \big)|\chi^\text{Is}_\frac{1}{16}|^2 
 \nonumber \\ 
Z_{PE^f,AE^s} &= \big(|\chi^{u1_4}_1 |^2 + |\chi^{u1_4}_{-1}|^2 \big)|\chi^\text{Is}_\frac{1}{16}|^2 
 \nonumber \\ 
Z_{AO^f,AE^s} &= \big(|\chi^{u1_4}_1 |^2 + |\chi^{u1_4}_{-1}|^2 \big)\big(\chi^\text{Is}_0 \bar \chi^\text{Is}_\frac{1}{2} + \chi^\text{Is}_\frac12 \bar \chi^\text{Is}_0\big) 
 \nonumber \\ 
Z_{AE^f,AO^s} &= \big( \chi^{u1_4}_0 \bar \chi^{u1_4}_2 + \chi^{u1_4}_2 \bar \chi^{u1_4}_0 \big)\big( |\chi^\text{Is}_0|^2+|\chi^\text{Is}_\frac{1}{2}|^2 \big) 
 \nonumber \\ 
Z_{PO^f,AO^s} &= \big( \chi^{u1_4}_0 \bar \chi^{u1_4}_2 + \chi^{u1_4}_2 \bar \chi^{u1_4}_0 \big)|\chi^\text{Is}_\frac{1}{16}|^2 
 \\ 
Z_{PE^f,AO^s} &= \big( \chi^{u1_4}_0 \bar \chi^{u1_4}_2 + \chi^{u1_4}_2 \bar \chi^{u1_4}_0 \big)|\chi^\text{Is}_\frac{1}{16}|^2 
 \nonumber \\ 
Z_{AO^f,AO^s} &= \big( \chi^{u1_4}_0 \bar \chi^{u1_4}_2 + \chi^{u1_4}_2 \bar \chi^{u1_4}_0 \big)\big(\chi^\text{Is}_0 \bar \chi^\text{Is}_\frac{1}{2} + \chi^\text{Is}_\frac12 \bar \chi^\text{Is}_0\big).
\nonumber 
\end{align}

\begin{align}
\label{sol24}
 Z_{AE^f,PE^s} &= \big(|\chi^{u1_4}_0 |^2 + |\chi^{u1_4}_2|^2 \big)\big( |\chi^\text{Is}_0|^2+|\chi^\text{Is}_\frac{1}{2}|^2 \big) 
 \nonumber \\ 
Z_{PO^f,PE^s} &= \big(|\chi^{u1_4}_0 |^2 + |\chi^{u1_4}_2|^2 \big)|\chi^\text{Is}_\frac{1}{16}|^2 
 \nonumber \\ 
Z_{PE^f,PE^s} &= \big(|\chi^{u1_4}_0 |^2 + |\chi^{u1_4}_2|^2 \big)|\chi^\text{Is}_\frac{1}{16}|^2 
 \nonumber \\ 
Z_{AO^f,PE^s} &= \big(|\chi^{u1_4}_0 |^2 + |\chi^{u1_4}_2|^2 \big)\big(\chi^\text{Is}_0 \bar \chi^\text{Is}_\frac{1}{2} + \chi^\text{Is}_\frac12 \bar \chi^\text{Is}_0\big) 
 \nonumber \\ 
Z_{AE^f,PO^s} &= \big(|\chi^{u1_4}_1 |^2 + |\chi^{u1_4}_{-1}|^2 \big)\big( |\chi^\text{Is}_0|^2+|\chi^\text{Is}_\frac{1}{2}|^2 \big) 
 \nonumber \\ 
Z_{PO^f,PO^s} &= \big(|\chi^{u1_4}_1 |^2 + |\chi^{u1_4}_{-1}|^2 \big)|\chi^\text{Is}_\frac{1}{16}|^2 
 \nonumber \\ 
Z_{PE^f,PO^s} &= \big(|\chi^{u1_4}_1 |^2 + |\chi^{u1_4}_{-1}|^2 \big)|\chi^\text{Is}_\frac{1}{16}|^2 
 \nonumber \\ 
Z_{AO^f,PO^s} &= \big(|\chi^{u1_4}_1 |^2 + |\chi^{u1_4}_{-1}|^2 \big)\big(\chi^\text{Is}_0 \bar \chi^\text{Is}_\frac{1}{2} + \chi^\text{Is}_\frac12 \bar \chi^\text{Is}_0\big) 
 \nonumber \\ 
Z_{AE^f,AE^s} &= \big( \chi^{u1_4}_1 \bar \chi^{u1_4}_{-1} + \chi^{u1_4}_{-1} \bar \chi^{u1_4}_1 \big)\big( |\chi^\text{Is}_0|^2+|\chi^\text{Is}_\frac{1}{2}|^2 \big) 
 \nonumber \\ 
Z_{PO^f,AE^s} &= \big( \chi^{u1_4}_1 \bar \chi^{u1_4}_{-1} + \chi^{u1_4}_{-1} \bar \chi^{u1_4}_1 \big)|\chi^\text{Is}_\frac{1}{16}|^2 
 \nonumber \\ 
Z_{PE^f,AE^s} &= \big( \chi^{u1_4}_1 \bar \chi^{u1_4}_{-1} + \chi^{u1_4}_{-1} \bar \chi^{u1_4}_1 \big)|\chi^\text{Is}_\frac{1}{16}|^2 
 \\ 
Z_{AO^f,AE^s} &= \big( \chi^{u1_4}_1 \bar \chi^{u1_4}_{-1} + \chi^{u1_4}_{-1} \bar \chi^{u1_4}_1 \big)\big(\chi^\text{Is}_0 \bar \chi^\text{Is}_\frac{1}{2} + \chi^\text{Is}_\frac12 \bar \chi^\text{Is}_0\big) 
 \nonumber \\ 
Z_{AE^f,AO^s} &= \big( \chi^{u1_4}_0 \bar \chi^{u1_4}_2 + \chi^{u1_4}_2 \bar \chi^{u1_4}_0 \big)\big( |\chi^\text{Is}_0|^2+|\chi^\text{Is}_\frac{1}{2}|^2 \big) 
 \nonumber \\ 
Z_{PO^f,AO^s} &= \big( \chi^{u1_4}_0 \bar \chi^{u1_4}_2 + \chi^{u1_4}_2 \bar \chi^{u1_4}_0 \big)|\chi^\text{Is}_\frac{1}{16}|^2 
 \nonumber \\ 
Z_{PE^f,AO^s} &= \big( \chi^{u1_4}_0 \bar \chi^{u1_4}_2 + \chi^{u1_4}_2 \bar \chi^{u1_4}_0 \big)|\chi^\text{Is}_\frac{1}{16}|^2 
 \nonumber \\ 
Z_{AO^f,AO^s} &= \big( \chi^{u1_4}_0 \bar \chi^{u1_4}_2 + \chi^{u1_4}_2 \bar \chi^{u1_4}_0 \big)\big(\chi^\text{Is}_0 \bar \chi^\text{Is}_\frac{1}{2} + \chi^\text{Is}_\frac12 \bar \chi^\text{Is}_0\big). 
\nonumber 
\end{align}

\begin{align}
\label{sol25}
 Z_{AE^f,PE^s} &= \big(|\chi^{u1_4}_0 |^2 + |\chi^{u1_4}_2|^2 \big)\big( |\chi^\text{Is}_0|^2+|\chi^\text{Is}_\frac{1}{2}|^2 \big) 
 \nonumber \\ 
Z_{PO^f,PE^s} &= \big( \chi^{u1_4}_0 \bar \chi^{u1_4}_2 + \chi^{u1_4}_2 \bar \chi^{u1_4}_0 \big)\big(\chi^\text{Is}_0 \bar \chi^\text{Is}_\frac{1}{2} + \chi^\text{Is}_\frac12 \bar \chi^\text{Is}_0\big) 
 \nonumber \\ 
Z_{PE^f,PE^s} &= \big(|\chi^{u1_4}_1 |^2 + |\chi^{u1_4}_{-1}|^2 \big)\big( |\chi^\text{Is}_0|^2+|\chi^\text{Is}_\frac{1}{2}|^2 \big) 
 \nonumber \\ 
Z_{AO^f,PE^s} &= \big( \chi^{u1_4}_1 \bar \chi^{u1_4}_{-1} + \chi^{u1_4}_{-1} \bar \chi^{u1_4}_1 \big)\big(\chi^\text{Is}_0 \bar \chi^\text{Is}_\frac{1}{2} + \chi^\text{Is}_\frac12 \bar \chi^\text{Is}_0\big) 
 \nonumber \\ 
Z_{AE^f,PO^s} &= \big(|\chi^{u1_4}_1 |^2 + |\chi^{u1_4}_{-1}|^2 \big)|\chi^\text{Is}_\frac{1}{16}|^2 
 \nonumber \\ 
Z_{PO^f,PO^s} &= \big( \chi^{u1_4}_1 \bar \chi^{u1_4}_{-1} + \chi^{u1_4}_{-1} \bar \chi^{u1_4}_1 \big)|\chi^\text{Is}_\frac{1}{16}|^2 
 \nonumber \\ 
Z_{PE^f,PO^s} &= \big(|\chi^{u1_4}_0 |^2 + |\chi^{u1_4}_2|^2 \big)|\chi^\text{Is}_\frac{1}{16}|^2 
 \nonumber \\ 
Z_{AO^f,PO^s} &= \big( \chi^{u1_4}_0 \bar \chi^{u1_4}_2 + \chi^{u1_4}_2 \bar \chi^{u1_4}_0 \big)|\chi^\text{Is}_\frac{1}{16}|^2 
 \nonumber \\ 
Z_{AE^f,AE^s} &= \big(|\chi^{u1_4}_1 |^2 + |\chi^{u1_4}_{-1}|^2 \big)|\chi^\text{Is}_\frac{1}{16}|^2 
  \\ 
Z_{PO^f,AE^s} &= \big( \chi^{u1_4}_1 \bar \chi^{u1_4}_{-1} + \chi^{u1_4}_{-1} \bar \chi^{u1_4}_1 \big)|\chi^\text{Is}_\frac{1}{16}|^2 
 \nonumber \\ 
Z_{PE^f,AE^s} &= \big(|\chi^{u1_4}_0 |^2 + |\chi^{u1_4}_2|^2 \big)|\chi^\text{Is}_\frac{1}{16}|^2 
 \nonumber \\ 
Z_{AO^f,AE^s} &= \big( \chi^{u1_4}_0 \bar \chi^{u1_4}_2 + \chi^{u1_4}_2 \bar \chi^{u1_4}_0 \big)|\chi^\text{Is}_\frac{1}{16}|^2 
 \nonumber \\ 
Z_{AE^f,AO^s} &= \big(|\chi^{u1_4}_0 |^2 + |\chi^{u1_4}_2|^2 \big)\big(\chi^\text{Is}_0 \bar \chi^\text{Is}_\frac{1}{2} + \chi^\text{Is}_\frac12 \bar \chi^\text{Is}_0\big) 
 \nonumber \\ 
Z_{PO^f,AO^s} &= \big( \chi^{u1_4}_0 \bar \chi^{u1_4}_2 + \chi^{u1_4}_2 \bar \chi^{u1_4}_0 \big)\big( |\chi^\text{Is}_0|^2+|\chi^\text{Is}_\frac{1}{2}|^2 \big) 
 \nonumber \\ 
Z_{PE^f,AO^s} &= \big(|\chi^{u1_4}_1 |^2 + |\chi^{u1_4}_{-1}|^2 \big)\big(\chi^\text{Is}_0 \bar \chi^\text{Is}_\frac{1}{2} + \chi^\text{Is}_\frac12 \bar \chi^\text{Is}_0\big) 
 \nonumber \\ 
Z_{AO^f,AO^s} &= \big( \chi^{u1_4}_1 \bar \chi^{u1_4}_{-1} + \chi^{u1_4}_{-1} \bar \chi^{u1_4}_1 \big)\big( |\chi^\text{Is}_0|^2+|\chi^\text{Is}_\frac{1}{2}|^2 \big).
\nonumber 
\end{align}

\begin{align}
\label{sol26}
 Z_{AE^f,PE^s} &= \big(|\chi^{u1_4}_0 |^2 + |\chi^{u1_4}_2|^2 \big)\big( |\chi^\text{Is}_0|^2+|\chi^\text{Is}_\frac{1}{2}|^2 \big) 
 \nonumber \\ 
Z_{PO^f,PE^s} &= \big( \chi^{u1_4}_0 \bar \chi^{u1_4}_2 + \chi^{u1_4}_2 \bar \chi^{u1_4}_0 \big)\big(\chi^\text{Is}_0 \bar \chi^\text{Is}_\frac{1}{2} + \chi^\text{Is}_\frac12 \bar \chi^\text{Is}_0\big) 
 \nonumber \\ 
Z_{PE^f,PE^s} &= \big( \chi^{u1_4}_1 \bar \chi^{u1_4}_{-1} + \chi^{u1_4}_{-1} \bar \chi^{u1_4}_1 \big)\big( |\chi^\text{Is}_0|^2+|\chi^\text{Is}_\frac{1}{2}|^2 \big) 
 \nonumber \\ 
Z_{AO^f,PE^s} &= \big(|\chi^{u1_4}_1 |^2 + |\chi^{u1_4}_{-1}|^2 \big)\big(\chi^\text{Is}_0 \bar \chi^\text{Is}_\frac{1}{2} + \chi^\text{Is}_\frac12 \bar \chi^\text{Is}_0\big) 
 \nonumber \\ 
Z_{AE^f,PO^s} &= \big( \chi^{u1_4}_1 \bar \chi^{u1_4}_{-1} + \chi^{u1_4}_{-1} \bar \chi^{u1_4}_1 \big)|\chi^\text{Is}_\frac{1}{16}|^2 
 \nonumber \\ 
Z_{PO^f,PO^s} &= \big(|\chi^{u1_4}_1 |^2 + |\chi^{u1_4}_{-1}|^2 \big)|\chi^\text{Is}_\frac{1}{16}|^2 
 \nonumber \\ 
Z_{PE^f,PO^s} &= \big(|\chi^{u1_4}_0 |^2 + |\chi^{u1_4}_2|^2 \big)|\chi^\text{Is}_\frac{1}{16}|^2 
 \nonumber \\ 
Z_{AO^f,PO^s} &= \big( \chi^{u1_4}_0 \bar \chi^{u1_4}_2 + \chi^{u1_4}_2 \bar \chi^{u1_4}_0 \big)|\chi^\text{Is}_\frac{1}{16}|^2 
 \nonumber \\ 
Z_{AE^f,AE^s} &= \big( \chi^{u1_4}_1 \bar \chi^{u1_4}_{-1} + \chi^{u1_4}_{-1} \bar \chi^{u1_4}_1 \big)|\chi^\text{Is}_\frac{1}{16}|^2 
 \nonumber \\ 
Z_{PO^f,AE^s} &= \big(|\chi^{u1_4}_1 |^2 + |\chi^{u1_4}_{-1}|^2 \big)|\chi^\text{Is}_\frac{1}{16}|^2 
 \nonumber \\ 
Z_{PE^f,AE^s} &= \big(|\chi^{u1_4}_0 |^2 + |\chi^{u1_4}_2|^2 \big)|\chi^\text{Is}_\frac{1}{16}|^2 
 \nonumber \\ 
Z_{AO^f,AE^s} &= \big( \chi^{u1_4}_0 \bar \chi^{u1_4}_2 + \chi^{u1_4}_2 \bar \chi^{u1_4}_0 \big)|\chi^\text{Is}_\frac{1}{16}|^2 
 \nonumber \\ 
Z_{AE^f,AO^s} &= \big(|\chi^{u1_4}_0 |^2 + |\chi^{u1_4}_2|^2 \big)\big(\chi^\text{Is}_0 \bar \chi^\text{Is}_\frac{1}{2} + \chi^\text{Is}_\frac12 \bar \chi^\text{Is}_0\big) 
 \nonumber \\ 
Z_{PO^f,AO^s} &= \big( \chi^{u1_4}_0 \bar \chi^{u1_4}_2 + \chi^{u1_4}_2 \bar \chi^{u1_4}_0 \big)\big( |\chi^\text{Is}_0|^2+|\chi^\text{Is}_\frac{1}{2}|^2 \big) 
 \nonumber \\ 
Z_{PE^f,AO^s} &= \big( \chi^{u1_4}_1 \bar \chi^{u1_4}_{-1} + \chi^{u1_4}_{-1} \bar \chi^{u1_4}_1 \big)\big(\chi^\text{Is}_0 \bar \chi^\text{Is}_\frac{1}{2} + \chi^\text{Is}_\frac12 \bar \chi^\text{Is}_0\big) 
 \nonumber \\ 
Z_{AO^f,AO^s} &= \big(|\chi^{u1_4}_1 |^2 + |\chi^{u1_4}_{-1}|^2 \big)\big( |\chi^\text{Is}_0|^2+|\chi^\text{Is}_\frac{1}{2}|^2 \big). 
\end{align}

\begin{align}
\label{sol27}
 Z_{AE^f,PE^s} &= \big(|\chi^{u1_4}_0 |^2 + |\chi^{u1_4}_2|^2 \big)\big( |\chi^\text{Is}_0|^2+|\chi^\text{Is}_\frac{1}{2}|^2 \big) 
 \nonumber \\ 
Z_{PO^f,PE^s} &= \big(|\chi^{u1_4}_0 |^2 + |\chi^{u1_4}_2|^2 \big)|\chi^\text{Is}_\frac{1}{16}|^2 
 \nonumber \\ 
Z_{PE^f,PE^s} &= \big( \chi^{u1_4}_1 \bar \chi^{u1_4}_{-1} + \chi^{u1_4}_{-1} \bar \chi^{u1_4}_1 \big)|\chi^\text{Is}_\frac{1}{16}|^2 
 \nonumber \\ 
Z_{AO^f,PE^s} &= \big( \chi^{u1_4}_1 \bar \chi^{u1_4}_{-1} + \chi^{u1_4}_{-1} \bar \chi^{u1_4}_1 \big)\big(\chi^\text{Is}_0 \bar \chi^\text{Is}_\frac{1}{2} + \chi^\text{Is}_\frac12 \bar \chi^\text{Is}_0\big) 
 \nonumber \\ 
Z_{AE^f,PO^s} &= \big(|\chi^{u1_4}_1 |^2 + |\chi^{u1_4}_{-1}|^2 \big)|\chi^\text{Is}_\frac{1}{16}|^2 
 \nonumber \\ 
Z_{PO^f,PO^s} &= \big(|\chi^{u1_4}_1 |^2 + |\chi^{u1_4}_{-1}|^2 \big)\big( |\chi^\text{Is}_0|^2+|\chi^\text{Is}_\frac{1}{2}|^2 \big) 
 \nonumber \\ 
Z_{PE^f,PO^s} &= \big( \chi^{u1_4}_0 \bar \chi^{u1_4}_2 + \chi^{u1_4}_2 \bar \chi^{u1_4}_0 \big)\big(\chi^\text{Is}_0 \bar \chi^\text{Is}_\frac{1}{2} + \chi^\text{Is}_\frac12 \bar \chi^\text{Is}_0\big) 
 \nonumber \\ 
Z_{AO^f,PO^s} &= \big( \chi^{u1_4}_0 \bar \chi^{u1_4}_2 + \chi^{u1_4}_2 \bar \chi^{u1_4}_0 \big)|\chi^\text{Is}_\frac{1}{16}|^2 
 \nonumber \\ 
Z_{AE^f,AE^s} &= \big( \chi^{u1_4}_1 \bar \chi^{u1_4}_{-1} + \chi^{u1_4}_{-1} \bar \chi^{u1_4}_1 \big)\big( |\chi^\text{Is}_0|^2+|\chi^\text{Is}_\frac{1}{2}|^2 \big) 
 \nonumber \\ 
Z_{PO^f,AE^s} &= \big( \chi^{u1_4}_1 \bar \chi^{u1_4}_{-1} + \chi^{u1_4}_{-1} \bar \chi^{u1_4}_1 \big)|\chi^\text{Is}_\frac{1}{16}|^2 
 \nonumber \\ 
Z_{PE^f,AE^s} &= \big(|\chi^{u1_4}_0 |^2 + |\chi^{u1_4}_2|^2 \big)|\chi^\text{Is}_\frac{1}{16}|^2 
 \nonumber \\ 
Z_{AO^f,AE^s} &= \big(|\chi^{u1_4}_0 |^2 + |\chi^{u1_4}_2|^2 \big)\big(\chi^\text{Is}_0 \bar \chi^\text{Is}_\frac{1}{2} + \chi^\text{Is}_\frac12 \bar \chi^\text{Is}_0\big) 
 \nonumber \\ 
Z_{AE^f,AO^s} &= \big( \chi^{u1_4}_0 \bar \chi^{u1_4}_2 + \chi^{u1_4}_2 \bar \chi^{u1_4}_0 \big)|\chi^\text{Is}_\frac{1}{16}|^2 
 \nonumber \\ 
Z_{PO^f,AO^s} &= \big( \chi^{u1_4}_0 \bar \chi^{u1_4}_2 + \chi^{u1_4}_2 \bar \chi^{u1_4}_0 \big)\big( |\chi^\text{Is}_0|^2+|\chi^\text{Is}_\frac{1}{2}|^2 \big) 
 \nonumber \\ 
Z_{PE^f,AO^s} &= \big(|\chi^{u1_4}_1 |^2 + |\chi^{u1_4}_{-1}|^2 \big)\big(\chi^\text{Is}_0 \bar \chi^\text{Is}_\frac{1}{2} + \chi^\text{Is}_\frac12 \bar \chi^\text{Is}_0\big) 
 \nonumber \\ 
Z_{AO^f,AO^s} &= \big(|\chi^{u1_4}_1 |^2 + |\chi^{u1_4}_{-1}|^2 \big)|\chi^\text{Is}_\frac{1}{16}|^2 .
\end{align}

\begin{align}
\label{sol28}
 Z_{AE^f,PE^s} &= \big(|\chi^{u1_4}_0 |^2 + |\chi^{u1_4}_2|^2 \big)\big( |\chi^\text{Is}_0|^2+|\chi^\text{Is}_\frac{1}{2}|^2 \big) 
 \nonumber \\ 
Z_{PO^f,PE^s} &= \big(|\chi^{u1_4}_0 |^2 + |\chi^{u1_4}_2|^2 \big)|\chi^\text{Is}_\frac{1}{16}|^2 
 \nonumber \\ 
Z_{PE^f,PE^s} &= \big(|\chi^{u1_4}_1 |^2 + |\chi^{u1_4}_{-1}|^2 \big)|\chi^\text{Is}_\frac{1}{16}|^2 
 \nonumber \\ 
Z_{AO^f,PE^s} &= \big(|\chi^{u1_4}_1 |^2 + |\chi^{u1_4}_{-1}|^2 \big)\big(\chi^\text{Is}_0 \bar \chi^\text{Is}_\frac{1}{2} + \chi^\text{Is}_\frac12 \bar \chi^\text{Is}_0\big) 
 \nonumber \\ 
Z_{AE^f,PO^s} &= \big( \chi^{u1_4}_1 \bar \chi^{u1_4}_{-1} + \chi^{u1_4}_{-1} \bar \chi^{u1_4}_1 \big)|\chi^\text{Is}_\frac{1}{16}|^2 
 \nonumber \\ 
Z_{PO^f,PO^s} &= \big( \chi^{u1_4}_1 \bar \chi^{u1_4}_{-1} + \chi^{u1_4}_{-1} \bar \chi^{u1_4}_1 \big)\big( |\chi^\text{Is}_0|^2+|\chi^\text{Is}_\frac{1}{2}|^2 \big) 
 \nonumber \\ 
Z_{PE^f,PO^s} &= \big( \chi^{u1_4}_0 \bar \chi^{u1_4}_2 + \chi^{u1_4}_2 \bar \chi^{u1_4}_0 \big)\big(\chi^\text{Is}_0 \bar \chi^\text{Is}_\frac{1}{2} + \chi^\text{Is}_\frac12 \bar \chi^\text{Is}_0\big) 
 \nonumber \\ 
Z_{AO^f,PO^s} &= \big( \chi^{u1_4}_0 \bar \chi^{u1_4}_2 + \chi^{u1_4}_2 \bar \chi^{u1_4}_0 \big)|\chi^\text{Is}_\frac{1}{16}|^2 
 \nonumber \\ 
Z_{AE^f,AE^s} &= \big(|\chi^{u1_4}_1 |^2 + |\chi^{u1_4}_{-1}|^2 \big)\big( |\chi^\text{Is}_0|^2+|\chi^\text{Is}_\frac{1}{2}|^2 \big) 
 \nonumber \\ 
Z_{PO^f,AE^s} &= \big(|\chi^{u1_4}_1 |^2 + |\chi^{u1_4}_{-1}|^2 \big)|\chi^\text{Is}_\frac{1}{16}|^2 
 \nonumber \\ 
Z_{PE^f,AE^s} &= \big(|\chi^{u1_4}_0 |^2 + |\chi^{u1_4}_2|^2 \big)|\chi^\text{Is}_\frac{1}{16}|^2 
 \nonumber \\ 
Z_{AO^f,AE^s} &= \big(|\chi^{u1_4}_0 |^2 + |\chi^{u1_4}_2|^2 \big)\big(\chi^\text{Is}_0 \bar \chi^\text{Is}_\frac{1}{2} + \chi^\text{Is}_\frac12 \bar \chi^\text{Is}_0\big) 
 \nonumber \\ 
Z_{AE^f,AO^s} &= \big( \chi^{u1_4}_0 \bar \chi^{u1_4}_2 + \chi^{u1_4}_2 \bar \chi^{u1_4}_0 \big)|\chi^\text{Is}_\frac{1}{16}|^2 
 \nonumber \\ 
Z_{PO^f,AO^s} &= \big( \chi^{u1_4}_0 \bar \chi^{u1_4}_2 + \chi^{u1_4}_2 \bar \chi^{u1_4}_0 \big)\big( |\chi^\text{Is}_0|^2+|\chi^\text{Is}_\frac{1}{2}|^2 \big) 
 \nonumber \\ 
Z_{PE^f,AO^s} &= \big( \chi^{u1_4}_1 \bar \chi^{u1_4}_{-1} + \chi^{u1_4}_{-1} \bar \chi^{u1_4}_1 \big)\big(\chi^\text{Is}_0 \bar \chi^\text{Is}_\frac{1}{2} + \chi^\text{Is}_\frac12 \bar \chi^\text{Is}_0\big) 
 \nonumber \\ 
Z_{AO^f,AO^s} &= \big( \chi^{u1_4}_1 \bar \chi^{u1_4}_{-1} + \chi^{u1_4}_{-1} \bar \chi^{u1_4}_1 \big)|\chi^\text{Is}_\frac{1}{16}|^2. 
\end{align}

\begin{align}
\label{sol29}
 Z_{AE^f,PE^s} &= \big(|\chi^{u1_4}_0 |^2 + |\chi^{u1_4}_2|^2 \big)\big( |\chi^\text{Is}_0|^2+|\chi^\text{Is}_\frac{1}{2}|^2 \big) 
 \nonumber \\ 
Z_{PO^f,PE^s} &= \big(|\chi^{u1_4}_0 |^2 + |\chi^{u1_4}_2|^2 \big)|\chi^\text{Is}_\frac{1}{16}|^2 
 \nonumber \\ 
Z_{PE^f,PE^s} &= \big( \chi^{u1_4}_1 \bar \chi^{u1_4}_{-1} + \chi^{u1_4}_{-1} \bar \chi^{u1_4}_1 \big)|\chi^\text{Is}_\frac{1}{16}|^2 
 \nonumber \\ 
Z_{AO^f,PE^s} &= \big( \chi^{u1_4}_1 \bar \chi^{u1_4}_{-1} + \chi^{u1_4}_{-1} \bar \chi^{u1_4}_1 \big)\big(\chi^\text{Is}_0 \bar \chi^\text{Is}_\frac{1}{2} + \chi^\text{Is}_\frac12 \bar \chi^\text{Is}_0\big) 
 \nonumber \\ 
Z_{AE^f,PO^s} &= \big( \chi^{u1_4}_1 \bar \chi^{u1_4}_{-1} + \chi^{u1_4}_{-1} \bar \chi^{u1_4}_1 \big)\big( |\chi^\text{Is}_0|^2+|\chi^\text{Is}_\frac{1}{2}|^2 \big) 
 \nonumber \\ 
Z_{PO^f,PO^s} &= \big( \chi^{u1_4}_1 \bar \chi^{u1_4}_{-1} + \chi^{u1_4}_{-1} \bar \chi^{u1_4}_1 \big)|\chi^\text{Is}_\frac{1}{16}|^2 
 \nonumber \\ 
Z_{PE^f,PO^s} &= \big(|\chi^{u1_4}_0 |^2 + |\chi^{u1_4}_2|^2 \big)|\chi^\text{Is}_\frac{1}{16}|^2 
 \nonumber \\ 
Z_{AO^f,PO^s} &= \big(|\chi^{u1_4}_0 |^2 + |\chi^{u1_4}_2|^2 \big)\big(\chi^\text{Is}_0 \bar \chi^\text{Is}_\frac{1}{2} + \chi^\text{Is}_\frac12 \bar \chi^\text{Is}_0\big) 
 \nonumber \\ 
Z_{AE^f,AE^s} &= \big(|\chi^{u1_4}_1 |^2 + |\chi^{u1_4}_{-1}|^2 \big)|\chi^\text{Is}_\frac{1}{16}|^2 
 \nonumber \\ 
Z_{PO^f,AE^s} &= \big(|\chi^{u1_4}_1 |^2 + |\chi^{u1_4}_{-1}|^2 \big)\big( |\chi^\text{Is}_0|^2+|\chi^\text{Is}_\frac{1}{2}|^2 \big) 
 \nonumber \\ 
Z_{PE^f,AE^s} &= \big( \chi^{u1_4}_0 \bar \chi^{u1_4}_2 + \chi^{u1_4}_2 \bar \chi^{u1_4}_0 \big)\big(\chi^\text{Is}_0 \bar \chi^\text{Is}_\frac{1}{2} + \chi^\text{Is}_\frac12 \bar \chi^\text{Is}_0\big) 
 \nonumber \\ 
Z_{AO^f,AE^s} &= \big( \chi^{u1_4}_0 \bar \chi^{u1_4}_2 + \chi^{u1_4}_2 \bar \chi^{u1_4}_0 \big)|\chi^\text{Is}_\frac{1}{16}|^2 
 \nonumber \\ 
Z_{AE^f,AO^s} &= \big( \chi^{u1_4}_0 \bar \chi^{u1_4}_2 + \chi^{u1_4}_2 \bar \chi^{u1_4}_0 \big)|\chi^\text{Is}_\frac{1}{16}|^2 
 \nonumber \\ 
Z_{PO^f,AO^s} &= \big( \chi^{u1_4}_0 \bar \chi^{u1_4}_2 + \chi^{u1_4}_2 \bar \chi^{u1_4}_0 \big)\big( |\chi^\text{Is}_0|^2+|\chi^\text{Is}_\frac{1}{2}|^2 \big) 
 \nonumber \\ 
Z_{PE^f,AO^s} &= \big(|\chi^{u1_4}_1 |^2 + |\chi^{u1_4}_{-1}|^2 \big)\big(\chi^\text{Is}_0 \bar \chi^\text{Is}_\frac{1}{2} + \chi^\text{Is}_\frac12 \bar \chi^\text{Is}_0\big) 
 \nonumber \\ 
Z_{AO^f,AO^s} &= \big(|\chi^{u1_4}_1 |^2 + |\chi^{u1_4}_{-1}|^2 \big)|\chi^\text{Is}_\frac{1}{16}|^2 .
\end{align}

\begin{align}
\label{sol30}
 Z_{AE^f,PE^s} &= \big(|\chi^{u1_4}_0 |^2 + |\chi^{u1_4}_2|^2 \big)\big( |\chi^\text{Is}_0|^2+|\chi^\text{Is}_\frac{1}{2}|^2 \big) 
 \nonumber \\ 
Z_{PO^f,PE^s} &= \big(|\chi^{u1_4}_0 |^2 + |\chi^{u1_4}_2|^2 \big)|\chi^\text{Is}_\frac{1}{16}|^2 
 \nonumber \\ 
Z_{PE^f,PE^s} &= \big(|\chi^{u1_4}_1 |^2 + |\chi^{u1_4}_{-1}|^2 \big)|\chi^\text{Is}_\frac{1}{16}|^2 
 \nonumber \\ 
Z_{AO^f,PE^s} &= \big(|\chi^{u1_4}_1 |^2 + |\chi^{u1_4}_{-1}|^2 \big)\big(\chi^\text{Is}_0 \bar \chi^\text{Is}_\frac{1}{2} + \chi^\text{Is}_\frac12 \bar \chi^\text{Is}_0\big) 
 \nonumber \\ 
Z_{AE^f,PO^s} &= \big(|\chi^{u1_4}_1 |^2 + |\chi^{u1_4}_{-1}|^2 \big)\big( |\chi^\text{Is}_0|^2+|\chi^\text{Is}_\frac{1}{2}|^2 \big) 
 \nonumber \\ 
Z_{PO^f,PO^s} &= \big(|\chi^{u1_4}_1 |^2 + |\chi^{u1_4}_{-1}|^2 \big)|\chi^\text{Is}_\frac{1}{16}|^2 
 \nonumber \\ 
Z_{PE^f,PO^s} &= \big(|\chi^{u1_4}_0 |^2 + |\chi^{u1_4}_2|^2 \big)|\chi^\text{Is}_\frac{1}{16}|^2 
 \nonumber \\ 
Z_{AO^f,PO^s} &= \big(|\chi^{u1_4}_0 |^2 + |\chi^{u1_4}_2|^2 \big)\big(\chi^\text{Is}_0 \bar \chi^\text{Is}_\frac{1}{2} + \chi^\text{Is}_\frac12 \bar \chi^\text{Is}_0\big) 
 \nonumber \\ 
Z_{AE^f,AE^s} &= \big( \chi^{u1_4}_1 \bar \chi^{u1_4}_{-1} + \chi^{u1_4}_{-1} \bar \chi^{u1_4}_1 \big)|\chi^\text{Is}_\frac{1}{16}|^2 
 \nonumber \\ 
Z_{PO^f,AE^s} &= \big( \chi^{u1_4}_1 \bar \chi^{u1_4}_{-1} + \chi^{u1_4}_{-1} \bar \chi^{u1_4}_1 \big)\big( |\chi^\text{Is}_0|^2+|\chi^\text{Is}_\frac{1}{2}|^2 \big) 
 \nonumber \\ 
Z_{PE^f,AE^s} &= \big( \chi^{u1_4}_0 \bar \chi^{u1_4}_2 + \chi^{u1_4}_2 \bar \chi^{u1_4}_0 \big)\big(\chi^\text{Is}_0 \bar \chi^\text{Is}_\frac{1}{2} + \chi^\text{Is}_\frac12 \bar \chi^\text{Is}_0\big) 
 \nonumber \\ 
Z_{AO^f,AE^s} &= \big( \chi^{u1_4}_0 \bar \chi^{u1_4}_2 + \chi^{u1_4}_2 \bar \chi^{u1_4}_0 \big)|\chi^\text{Is}_\frac{1}{16}|^2 
 \nonumber \\ 
Z_{AE^f,AO^s} &= \big( \chi^{u1_4}_0 \bar \chi^{u1_4}_2 + \chi^{u1_4}_2 \bar \chi^{u1_4}_0 \big)|\chi^\text{Is}_\frac{1}{16}|^2 
 \nonumber \\ 
Z_{PO^f,AO^s} &= \big( \chi^{u1_4}_0 \bar \chi^{u1_4}_2 + \chi^{u1_4}_2 \bar \chi^{u1_4}_0 \big)\big( |\chi^\text{Is}_0|^2+|\chi^\text{Is}_\frac{1}{2}|^2 \big) 
 \nonumber \\ 
Z_{PE^f,AO^s} &= \big( \chi^{u1_4}_1 \bar \chi^{u1_4}_{-1} + \chi^{u1_4}_{-1} \bar \chi^{u1_4}_1 \big)\big(\chi^\text{Is}_0 \bar \chi^\text{Is}_\frac{1}{2} + \chi^\text{Is}_\frac12 \bar \chi^\text{Is}_0\big) 
 \nonumber \\ 
Z_{AO^f,AO^s} &= \big( \chi^{u1_4}_1 \bar \chi^{u1_4}_{-1} + \chi^{u1_4}_{-1} \bar \chi^{u1_4}_1 \big)|\chi^\text{Is}_\frac{1}{16}|^2. 
\end{align}

\begin{align}
\label{sol31}
 Z_{AE^f,PE^s} &= \big(|\chi^{u1_4}_0 |^2 + |\chi^{u1_4}_2|^2 \big)\big( |\chi^\text{Is}_0|^2+|\chi^\text{Is}_\frac{1}{2}|^2 \big) 
 \nonumber \\ 
Z_{PO^f,PE^s} &= \big(|\chi^{u1_4}_1 |^2 + |\chi^{u1_4}_{-1}|^2 \big)\big( |\chi^\text{Is}_0|^2+|\chi^\text{Is}_\frac{1}{2}|^2 \big) 
 \nonumber \\ 
Z_{PE^f,PE^s} &= \big( \chi^{u1_4}_0 \bar \chi^{u1_4}_2 + \chi^{u1_4}_2 \bar \chi^{u1_4}_0 \big)\big(\chi^\text{Is}_0 \bar \chi^\text{Is}_\frac{1}{2} + \chi^\text{Is}_\frac12 \bar \chi^\text{Is}_0\big) 
 \nonumber \\ 
Z_{AO^f,PE^s} &= \big( \chi^{u1_4}_1 \bar \chi^{u1_4}_{-1} + \chi^{u1_4}_{-1} \bar \chi^{u1_4}_1 \big)\big(\chi^\text{Is}_0 \bar \chi^\text{Is}_\frac{1}{2} + \chi^\text{Is}_\frac12 \bar \chi^\text{Is}_0\big) 
 \nonumber \\ 
Z_{AE^f,PO^s} &= \big(|\chi^{u1_4}_1 |^2 + |\chi^{u1_4}_{-1}|^2 \big)|\chi^\text{Is}_\frac{1}{16}|^2 
 \nonumber \\ 
Z_{PO^f,PO^s} &= \big(|\chi^{u1_4}_0 |^2 + |\chi^{u1_4}_2|^2 \big)|\chi^\text{Is}_\frac{1}{16}|^2 
 \\ 
Z_{PE^f,PO^s} &= \big( \chi^{u1_4}_1 \bar \chi^{u1_4}_{-1} + \chi^{u1_4}_{-1} \bar \chi^{u1_4}_1 \big)|\chi^\text{Is}_\frac{1}{16}|^2 
 \nonumber \\ 
Z_{AO^f,PO^s} &= \big( \chi^{u1_4}_0 \bar \chi^{u1_4}_2 + \chi^{u1_4}_2 \bar \chi^{u1_4}_0 \big)|\chi^\text{Is}_\frac{1}{16}|^2 
 \nonumber \\ 
Z_{AE^f,AE^s} &= \big(|\chi^{u1_4}_1 |^2 + |\chi^{u1_4}_{-1}|^2 \big)|\chi^\text{Is}_\frac{1}{16}|^2 
 \nonumber \\ 
Z_{PO^f,AE^s} &= \big(|\chi^{u1_4}_0 |^2 + |\chi^{u1_4}_2|^2 \big)|\chi^\text{Is}_\frac{1}{16}|^2 
 \nonumber \\ 
Z_{PE^f,AE^s} &= \big( \chi^{u1_4}_1 \bar \chi^{u1_4}_{-1} + \chi^{u1_4}_{-1} \bar \chi^{u1_4}_1 \big)|\chi^\text{Is}_\frac{1}{16}|^2 
 \nonumber \\ 
Z_{AO^f,AE^s} &= \big( \chi^{u1_4}_0 \bar \chi^{u1_4}_2 + \chi^{u1_4}_2 \bar \chi^{u1_4}_0 \big)|\chi^\text{Is}_\frac{1}{16}|^2 
 \nonumber \\ 
Z_{AE^f,AO^s} &= \big(|\chi^{u1_4}_0 |^2 + |\chi^{u1_4}_2|^2 \big)\big(\chi^\text{Is}_0 \bar \chi^\text{Is}_\frac{1}{2} + \chi^\text{Is}_\frac12 \bar \chi^\text{Is}_0\big) 
 \nonumber \\ 
Z_{PO^f,AO^s} &= \big(|\chi^{u1_4}_1 |^2 + |\chi^{u1_4}_{-1}|^2 \big)\big(\chi^\text{Is}_0 \bar \chi^\text{Is}_\frac{1}{2} + \chi^\text{Is}_\frac12 \bar \chi^\text{Is}_0\big) 
 \nonumber \\ 
Z_{PE^f,AO^s} &= \big( \chi^{u1_4}_0 \bar \chi^{u1_4}_2 + \chi^{u1_4}_2 \bar \chi^{u1_4}_0 \big)\big( |\chi^\text{Is}_0|^2+|\chi^\text{Is}_\frac{1}{2}|^2 \big) 
 \nonumber \\ 
Z_{AO^f,AO^s} &= \big( \chi^{u1_4}_1 \bar \chi^{u1_4}_{-1} + \chi^{u1_4}_{-1} \bar \chi^{u1_4}_1 \big)\big( |\chi^\text{Is}_0|^2+|\chi^\text{Is}_\frac{1}{2}|^2 \big). 
\nonumber 
\end{align}

\begin{align}
\label{sol32}
 Z_{AE^f,PE^s} &= \big(|\chi^{u1_4}_0 |^2 + |\chi^{u1_4}_2|^2 \big)\big( |\chi^\text{Is}_0|^2+|\chi^\text{Is}_\frac{1}{2}|^2 \big) 
 \nonumber \\ 
Z_{PO^f,PE^s} &= \big( \chi^{u1_4}_1 \bar \chi^{u1_4}_{-1} + \chi^{u1_4}_{-1} \bar \chi^{u1_4}_1 \big)\big( |\chi^\text{Is}_0|^2+|\chi^\text{Is}_\frac{1}{2}|^2 \big) 
 \nonumber \\ 
Z_{PE^f,PE^s} &= \big( \chi^{u1_4}_0 \bar \chi^{u1_4}_2 + \chi^{u1_4}_2 \bar \chi^{u1_4}_0 \big)\big(\chi^\text{Is}_0 \bar \chi^\text{Is}_\frac{1}{2} + \chi^\text{Is}_\frac12 \bar \chi^\text{Is}_0\big) 
 \nonumber \\ 
Z_{AO^f,PE^s} &= \big(|\chi^{u1_4}_1 |^2 + |\chi^{u1_4}_{-1}|^2 \big)\big(\chi^\text{Is}_0 \bar \chi^\text{Is}_\frac{1}{2} + \chi^\text{Is}_\frac12 \bar \chi^\text{Is}_0\big) 
 \nonumber \\ 
Z_{AE^f,PO^s} &= \big( \chi^{u1_4}_1 \bar \chi^{u1_4}_{-1} + \chi^{u1_4}_{-1} \bar \chi^{u1_4}_1 \big)|\chi^\text{Is}_\frac{1}{16}|^2 
 \nonumber \\ 
Z_{PO^f,PO^s} &= \big(|\chi^{u1_4}_0 |^2 + |\chi^{u1_4}_2|^2 \big)|\chi^\text{Is}_\frac{1}{16}|^2 
 \nonumber \\ 
Z_{PE^f,PO^s} &= \big(|\chi^{u1_4}_1 |^2 + |\chi^{u1_4}_{-1}|^2 \big)|\chi^\text{Is}_\frac{1}{16}|^2 
 \nonumber \\ 
Z_{AO^f,PO^s} &= \big( \chi^{u1_4}_0 \bar \chi^{u1_4}_2 + \chi^{u1_4}_2 \bar \chi^{u1_4}_0 \big)|\chi^\text{Is}_\frac{1}{16}|^2 
 \nonumber \\ 
Z_{AE^f,AE^s} &= \big( \chi^{u1_4}_1 \bar \chi^{u1_4}_{-1} + \chi^{u1_4}_{-1} \bar \chi^{u1_4}_1 \big)|\chi^\text{Is}_\frac{1}{16}|^2 
 \nonumber \\ 
Z_{PO^f,AE^s} &= \big(|\chi^{u1_4}_0 |^2 + |\chi^{u1_4}_2|^2 \big)|\chi^\text{Is}_\frac{1}{16}|^2 
 \nonumber \\ 
Z_{PE^f,AE^s} &= \big(|\chi^{u1_4}_1 |^2 + |\chi^{u1_4}_{-1}|^2 \big)|\chi^\text{Is}_\frac{1}{16}|^2 
 \nonumber \\ 
Z_{AO^f,AE^s} &= \big( \chi^{u1_4}_0 \bar \chi^{u1_4}_2 + \chi^{u1_4}_2 \bar \chi^{u1_4}_0 \big)|\chi^\text{Is}_\frac{1}{16}|^2 
 \nonumber \\ 
Z_{AE^f,AO^s} &= \big(|\chi^{u1_4}_0 |^2 + |\chi^{u1_4}_2|^2 \big)\big(\chi^\text{Is}_0 \bar \chi^\text{Is}_\frac{1}{2} + \chi^\text{Is}_\frac12 \bar \chi^\text{Is}_0\big) 
 \nonumber \\ 
Z_{PO^f,AO^s} &= \big( \chi^{u1_4}_1 \bar \chi^{u1_4}_{-1} + \chi^{u1_4}_{-1} \bar \chi^{u1_4}_1 \big)\big(\chi^\text{Is}_0 \bar \chi^\text{Is}_\frac{1}{2} + \chi^\text{Is}_\frac12 \bar \chi^\text{Is}_0\big) 
 \nonumber \\ 
Z_{PE^f,AO^s} &= \big( \chi^{u1_4}_0 \bar \chi^{u1_4}_2 + \chi^{u1_4}_2 \bar \chi^{u1_4}_0 \big)\big( |\chi^\text{Is}_0|^2+|\chi^\text{Is}_\frac{1}{2}|^2 \big) 
 \nonumber \\ 
Z_{AO^f,AO^s} &= \big(|\chi^{u1_4}_1 |^2 + |\chi^{u1_4}_{-1}|^2 \big)\big( |\chi^\text{Is}_0|^2+|\chi^\text{Is}_\frac{1}{2}|^2 \big). 
\end{align}

\begin{align}
\label{sol33}
 Z_{AE^f,PE^s} &= \big(|\chi^{u1_4}_0 |^2 + |\chi^{u1_4}_2|^2 \big)\big( |\chi^\text{Is}_0|^2+|\chi^\text{Is}_\frac{1}{2}|^2 \big) 
 \nonumber \\ 
Z_{PO^f,PE^s} &= \big( \chi^{u1_4}_1 \bar \chi^{u1_4}_{-1} + \chi^{u1_4}_{-1} \bar \chi^{u1_4}_1 \big)|\chi^\text{Is}_\frac{1}{16}|^2 
 \nonumber \\ 
Z_{PE^f,PE^s} &= \big(|\chi^{u1_4}_0 |^2 + |\chi^{u1_4}_2|^2 \big)|\chi^\text{Is}_\frac{1}{16}|^2 
 \nonumber \\ 
Z_{AO^f,PE^s} &= \big( \chi^{u1_4}_1 \bar \chi^{u1_4}_{-1} + \chi^{u1_4}_{-1} \bar \chi^{u1_4}_1 \big)\big(\chi^\text{Is}_0 \bar \chi^\text{Is}_\frac{1}{2} + \chi^\text{Is}_\frac12 \bar \chi^\text{Is}_0\big) 
 \nonumber \\ 
Z_{AE^f,PO^s} &= \big(|\chi^{u1_4}_1 |^2 + |\chi^{u1_4}_{-1}|^2 \big)|\chi^\text{Is}_\frac{1}{16}|^2 
 \nonumber \\ 
Z_{PO^f,PO^s} &= \big( \chi^{u1_4}_0 \bar \chi^{u1_4}_2 + \chi^{u1_4}_2 \bar \chi^{u1_4}_0 \big)\big(\chi^\text{Is}_0 \bar \chi^\text{Is}_\frac{1}{2} + \chi^\text{Is}_\frac12 \bar \chi^\text{Is}_0\big) 
 \nonumber \\ 
Z_{PE^f,PO^s} &= \big(|\chi^{u1_4}_1 |^2 + |\chi^{u1_4}_{-1}|^2 \big)\big( |\chi^\text{Is}_0|^2+|\chi^\text{Is}_\frac{1}{2}|^2 \big) 
 \nonumber \\ 
Z_{AO^f,PO^s} &= \big( \chi^{u1_4}_0 \bar \chi^{u1_4}_2 + \chi^{u1_4}_2 \bar \chi^{u1_4}_0 \big)|\chi^\text{Is}_\frac{1}{16}|^2 
 \nonumber \\ 
Z_{AE^f,AE^s} &= \big( \chi^{u1_4}_1 \bar \chi^{u1_4}_{-1} + \chi^{u1_4}_{-1} \bar \chi^{u1_4}_1 \big)\big( |\chi^\text{Is}_0|^2+|\chi^\text{Is}_\frac{1}{2}|^2 \big) 
 \nonumber \\ 
Z_{PO^f,AE^s} &= \big(|\chi^{u1_4}_0 |^2 + |\chi^{u1_4}_2|^2 \big)|\chi^\text{Is}_\frac{1}{16}|^2 
 \nonumber \\ 
Z_{PE^f,AE^s} &= \big( \chi^{u1_4}_1 \bar \chi^{u1_4}_{-1} + \chi^{u1_4}_{-1} \bar \chi^{u1_4}_1 \big)|\chi^\text{Is}_\frac{1}{16}|^2 
 \nonumber \\ 
Z_{AO^f,AE^s} &= \big(|\chi^{u1_4}_0 |^2 + |\chi^{u1_4}_2|^2 \big)\big(\chi^\text{Is}_0 \bar \chi^\text{Is}_\frac{1}{2} + \chi^\text{Is}_\frac12 \bar \chi^\text{Is}_0\big) 
 \nonumber \\ 
Z_{AE^f,AO^s} &= \big( \chi^{u1_4}_0 \bar \chi^{u1_4}_2 + \chi^{u1_4}_2 \bar \chi^{u1_4}_0 \big)|\chi^\text{Is}_\frac{1}{16}|^2 
 \nonumber \\ 
Z_{PO^f,AO^s} &= \big(|\chi^{u1_4}_1 |^2 + |\chi^{u1_4}_{-1}|^2 \big)\big(\chi^\text{Is}_0 \bar \chi^\text{Is}_\frac{1}{2} + \chi^\text{Is}_\frac12 \bar \chi^\text{Is}_0\big) 
 \nonumber \\ 
Z_{PE^f,AO^s} &= \big( \chi^{u1_4}_0 \bar \chi^{u1_4}_2 + \chi^{u1_4}_2 \bar \chi^{u1_4}_0 \big)\big( |\chi^\text{Is}_0|^2+|\chi^\text{Is}_\frac{1}{2}|^2 \big) 
 \nonumber \\ 
Z_{AO^f,AO^s} &= \big(|\chi^{u1_4}_1 |^2 + |\chi^{u1_4}_{-1}|^2 \big)|\chi^\text{Is}_\frac{1}{16}|^2. 
\end{align}

\begin{align}
\label{sol34}
 Z_{AE^f,PE^s} &= \big(|\chi^{u1_4}_0 |^2 + |\chi^{u1_4}_2|^2 \big)\big( |\chi^\text{Is}_0|^2+|\chi^\text{Is}_\frac{1}{2}|^2 \big) 
 \nonumber \\ 
Z_{PO^f,PE^s} &= \big(|\chi^{u1_4}_1 |^2 + |\chi^{u1_4}_{-1}|^2 \big)|\chi^\text{Is}_\frac{1}{16}|^2 
 \nonumber \\ 
Z_{PE^f,PE^s} &= \big(|\chi^{u1_4}_0 |^2 + |\chi^{u1_4}_2|^2 \big)|\chi^\text{Is}_\frac{1}{16}|^2 
 \nonumber \\ 
Z_{AO^f,PE^s} &= \big(|\chi^{u1_4}_1 |^2 + |\chi^{u1_4}_{-1}|^2 \big)\big(\chi^\text{Is}_0 \bar \chi^\text{Is}_\frac{1}{2} + \chi^\text{Is}_\frac12 \bar \chi^\text{Is}_0\big) 
 \nonumber \\ 
Z_{AE^f,PO^s} &= \big( \chi^{u1_4}_1 \bar \chi^{u1_4}_{-1} + \chi^{u1_4}_{-1} \bar \chi^{u1_4}_1 \big)|\chi^\text{Is}_\frac{1}{16}|^2 
 \nonumber \\ 
Z_{PO^f,PO^s} &= \big( \chi^{u1_4}_0 \bar \chi^{u1_4}_2 + \chi^{u1_4}_2 \bar \chi^{u1_4}_0 \big)\big(\chi^\text{Is}_0 \bar \chi^\text{Is}_\frac{1}{2} + \chi^\text{Is}_\frac12 \bar \chi^\text{Is}_0\big) 
 \nonumber \\ 
Z_{PE^f,PO^s} &= \big( \chi^{u1_4}_1 \bar \chi^{u1_4}_{-1} + \chi^{u1_4}_{-1} \bar \chi^{u1_4}_1 \big)\big( |\chi^\text{Is}_0|^2+|\chi^\text{Is}_\frac{1}{2}|^2 \big) 
 \nonumber \\ 
Z_{AO^f,PO^s} &= \big( \chi^{u1_4}_0 \bar \chi^{u1_4}_2 + \chi^{u1_4}_2 \bar \chi^{u1_4}_0 \big)|\chi^\text{Is}_\frac{1}{16}|^2 
 \nonumber \\ 
Z_{AE^f,AE^s} &= \big(|\chi^{u1_4}_1 |^2 + |\chi^{u1_4}_{-1}|^2 \big)\big( |\chi^\text{Is}_0|^2+|\chi^\text{Is}_\frac{1}{2}|^2 \big) 
 \nonumber \\ 
Z_{PO^f,AE^s} &= \big(|\chi^{u1_4}_0 |^2 + |\chi^{u1_4}_2|^2 \big)|\chi^\text{Is}_\frac{1}{16}|^2 
 \nonumber \\ 
Z_{PE^f,AE^s} &= \big(|\chi^{u1_4}_1 |^2 + |\chi^{u1_4}_{-1}|^2 \big)|\chi^\text{Is}_\frac{1}{16}|^2 
 \nonumber \\ 
Z_{AO^f,AE^s} &= \big(|\chi^{u1_4}_0 |^2 + |\chi^{u1_4}_2|^2 \big)\big(\chi^\text{Is}_0 \bar \chi^\text{Is}_\frac{1}{2} + \chi^\text{Is}_\frac12 \bar \chi^\text{Is}_0\big) 
 \nonumber \\ 
Z_{AE^f,AO^s} &= \big( \chi^{u1_4}_0 \bar \chi^{u1_4}_2 + \chi^{u1_4}_2 \bar \chi^{u1_4}_0 \big)|\chi^\text{Is}_\frac{1}{16}|^2 
 \nonumber \\ 
Z_{PO^f,AO^s} &= \big( \chi^{u1_4}_1 \bar \chi^{u1_4}_{-1} + \chi^{u1_4}_{-1} \bar \chi^{u1_4}_1 \big)\big(\chi^\text{Is}_0 \bar \chi^\text{Is}_\frac{1}{2} + \chi^\text{Is}_\frac12 \bar \chi^\text{Is}_0\big) 
 \nonumber \\ 
Z_{PE^f,AO^s} &= \big( \chi^{u1_4}_0 \bar \chi^{u1_4}_2 + \chi^{u1_4}_2 \bar \chi^{u1_4}_0 \big)\big( |\chi^\text{Is}_0|^2+|\chi^\text{Is}_\frac{1}{2}|^2 \big) 
 \nonumber \\ 
Z_{AO^f,AO^s} &= \big( \chi^{u1_4}_1 \bar \chi^{u1_4}_{-1} + \chi^{u1_4}_{-1} \bar \chi^{u1_4}_1 \big)|\chi^\text{Is}_\frac{1}{16}|^2. 
\end{align}

\begin{align}
\label{sol35}
 Z_{AE^f,PE^s} &= \big(|\chi^{u1_4}_0 |^2 + |\chi^{u1_4}_2|^2 \big)\big( |\chi^\text{Is}_0|^2+|\chi^\text{Is}_\frac{1}{2}|^2 \big) 
 \nonumber \\ 
Z_{PO^f,PE^s} &= \big( \chi^{u1_4}_1 \bar \chi^{u1_4}_{-1} + \chi^{u1_4}_{-1} \bar \chi^{u1_4}_1 \big)|\chi^\text{Is}_\frac{1}{16}|^2 
 \nonumber \\ 
Z_{PE^f,PE^s} &= \big(|\chi^{u1_4}_0 |^2 + |\chi^{u1_4}_2|^2 \big)|\chi^\text{Is}_\frac{1}{16}|^2 
 \nonumber \\ 
Z_{AO^f,PE^s} &= \big( \chi^{u1_4}_1 \bar \chi^{u1_4}_{-1} + \chi^{u1_4}_{-1} \bar \chi^{u1_4}_1 \big)\big(\chi^\text{Is}_0 \bar \chi^\text{Is}_\frac{1}{2} + \chi^\text{Is}_\frac12 \bar \chi^\text{Is}_0\big) 
 \nonumber \\ 
Z_{AE^f,PO^s} &= \big( \chi^{u1_4}_1 \bar \chi^{u1_4}_{-1} + \chi^{u1_4}_{-1} \bar \chi^{u1_4}_1 \big)\big( |\chi^\text{Is}_0|^2+|\chi^\text{Is}_\frac{1}{2}|^2 \big) 
 \nonumber \\ 
Z_{PO^f,PO^s} &= \big(|\chi^{u1_4}_0 |^2 + |\chi^{u1_4}_2|^2 \big)|\chi^\text{Is}_\frac{1}{16}|^2 
 \nonumber \\ 
Z_{PE^f,PO^s} &= \big( \chi^{u1_4}_1 \bar \chi^{u1_4}_{-1} + \chi^{u1_4}_{-1} \bar \chi^{u1_4}_1 \big)|\chi^\text{Is}_\frac{1}{16}|^2 
 \nonumber \\ 
Z_{AO^f,PO^s} &= \big(|\chi^{u1_4}_0 |^2 + |\chi^{u1_4}_2|^2 \big)\big(\chi^\text{Is}_0 \bar \chi^\text{Is}_\frac{1}{2} + \chi^\text{Is}_\frac12 \bar \chi^\text{Is}_0\big) 
 \nonumber \\ 
Z_{AE^f,AE^s} &= \big(|\chi^{u1_4}_1 |^2 + |\chi^{u1_4}_{-1}|^2 \big)|\chi^\text{Is}_\frac{1}{16}|^2 
 \nonumber \\ 
Z_{PO^f,AE^s} &= \big( \chi^{u1_4}_0 \bar \chi^{u1_4}_2 + \chi^{u1_4}_2 \bar \chi^{u1_4}_0 \big)\big(\chi^\text{Is}_0 \bar \chi^\text{Is}_\frac{1}{2} + \chi^\text{Is}_\frac12 \bar \chi^\text{Is}_0\big) 
 \nonumber \\ 
Z_{PE^f,AE^s} &= \big(|\chi^{u1_4}_1 |^2 + |\chi^{u1_4}_{-1}|^2 \big)\big( |\chi^\text{Is}_0|^2+|\chi^\text{Is}_\frac{1}{2}|^2 \big) 
 \nonumber \\ 
Z_{AO^f,AE^s} &= \big( \chi^{u1_4}_0 \bar \chi^{u1_4}_2 + \chi^{u1_4}_2 \bar \chi^{u1_4}_0 \big)|\chi^\text{Is}_\frac{1}{16}|^2 
 \nonumber \\ 
Z_{AE^f,AO^s} &= \big( \chi^{u1_4}_0 \bar \chi^{u1_4}_2 + \chi^{u1_4}_2 \bar \chi^{u1_4}_0 \big)|\chi^\text{Is}_\frac{1}{16}|^2 
 \nonumber \\ 
Z_{PO^f,AO^s} &= \big(|\chi^{u1_4}_1 |^2 + |\chi^{u1_4}_{-1}|^2 \big)\big(\chi^\text{Is}_0 \bar \chi^\text{Is}_\frac{1}{2} + \chi^\text{Is}_\frac12 \bar \chi^\text{Is}_0\big) 
 \nonumber \\ 
Z_{PE^f,AO^s} &= \big( \chi^{u1_4}_0 \bar \chi^{u1_4}_2 + \chi^{u1_4}_2 \bar \chi^{u1_4}_0 \big)\big( |\chi^\text{Is}_0|^2+|\chi^\text{Is}_\frac{1}{2}|^2 \big) 
 \nonumber \\ 
Z_{AO^f,AO^s} &= \big(|\chi^{u1_4}_1 |^2 + |\chi^{u1_4}_{-1}|^2 \big)|\chi^\text{Is}_\frac{1}{16}|^2. 
\end{align}

\begin{align}
\label{sol36}
 Z_{AE^f,PE^s} &= \big(|\chi^{u1_4}_0 |^2 + |\chi^{u1_4}_2|^2 \big)\big( |\chi^\text{Is}_0|^2+|\chi^\text{Is}_\frac{1}{2}|^2 \big) 
 \nonumber \\ 
Z_{PO^f,PE^s} &= \big(|\chi^{u1_4}_1 |^2 + |\chi^{u1_4}_{-1}|^2 \big)|\chi^\text{Is}_\frac{1}{16}|^2 
 \nonumber \\ 
Z_{PE^f,PE^s} &= \big(|\chi^{u1_4}_0 |^2 + |\chi^{u1_4}_2|^2 \big)|\chi^\text{Is}_\frac{1}{16}|^2 
 \nonumber \\ 
Z_{AO^f,PE^s} &= \big(|\chi^{u1_4}_1 |^2 + |\chi^{u1_4}_{-1}|^2 \big)\big(\chi^\text{Is}_0 \bar \chi^\text{Is}_\frac{1}{2} + \chi^\text{Is}_\frac12 \bar \chi^\text{Is}_0\big) 
 \nonumber \\ 
Z_{AE^f,PO^s} &= \big(|\chi^{u1_4}_1 |^2 + |\chi^{u1_4}_{-1}|^2 \big)\big( |\chi^\text{Is}_0|^2+|\chi^\text{Is}_\frac{1}{2}|^2 \big) 
 \nonumber \\ 
Z_{PO^f,PO^s} &= \big(|\chi^{u1_4}_0 |^2 + |\chi^{u1_4}_2|^2 \big)|\chi^\text{Is}_\frac{1}{16}|^2 
 \nonumber \\ 
Z_{PE^f,PO^s} &= \big(|\chi^{u1_4}_1 |^2 + |\chi^{u1_4}_{-1}|^2 \big)|\chi^\text{Is}_\frac{1}{16}|^2 
 \nonumber \\ 
Z_{AO^f,PO^s} &= \big(|\chi^{u1_4}_0 |^2 + |\chi^{u1_4}_2|^2 \big)\big(\chi^\text{Is}_0 \bar \chi^\text{Is}_\frac{1}{2} + \chi^\text{Is}_\frac12 \bar \chi^\text{Is}_0\big) 
 \nonumber \\ 
Z_{AE^f,AE^s} &= \big( \chi^{u1_4}_1 \bar \chi^{u1_4}_{-1} + \chi^{u1_4}_{-1} \bar \chi^{u1_4}_1 \big)|\chi^\text{Is}_\frac{1}{16}|^2 
 \nonumber \\ 
Z_{PO^f,AE^s} &= \big( \chi^{u1_4}_0 \bar \chi^{u1_4}_2 + \chi^{u1_4}_2 \bar \chi^{u1_4}_0 \big)\big(\chi^\text{Is}_0 \bar \chi^\text{Is}_\frac{1}{2} + \chi^\text{Is}_\frac12 \bar \chi^\text{Is}_0\big) 
 \nonumber \\ 
Z_{PE^f,AE^s} &= \big( \chi^{u1_4}_1 \bar \chi^{u1_4}_{-1} + \chi^{u1_4}_{-1} \bar \chi^{u1_4}_1 \big)\big( |\chi^\text{Is}_0|^2+|\chi^\text{Is}_\frac{1}{2}|^2 \big) 
 \nonumber \\ 
Z_{AO^f,AE^s} &= \big( \chi^{u1_4}_0 \bar \chi^{u1_4}_2 + \chi^{u1_4}_2 \bar \chi^{u1_4}_0 \big)|\chi^\text{Is}_\frac{1}{16}|^2 
 \nonumber \\ 
Z_{AE^f,AO^s} &= \big( \chi^{u1_4}_0 \bar \chi^{u1_4}_2 + \chi^{u1_4}_2 \bar \chi^{u1_4}_0 \big)|\chi^\text{Is}_\frac{1}{16}|^2 
 \nonumber \\ 
Z_{PO^f,AO^s} &= \big( \chi^{u1_4}_1 \bar \chi^{u1_4}_{-1} + \chi^{u1_4}_{-1} \bar \chi^{u1_4}_1 \big)\big(\chi^\text{Is}_0 \bar \chi^\text{Is}_\frac{1}{2} + \chi^\text{Is}_\frac12 \bar \chi^\text{Is}_0\big) 
 \nonumber \\ 
Z_{PE^f,AO^s} &= \big( \chi^{u1_4}_0 \bar \chi^{u1_4}_2 + \chi^{u1_4}_2 \bar \chi^{u1_4}_0 \big)\big( |\chi^\text{Is}_0|^2+|\chi^\text{Is}_\frac{1}{2}|^2 \big) 
 \nonumber \\ 
Z_{AO^f,AO^s} &= \big( \chi^{u1_4}_1 \bar \chi^{u1_4}_{-1} + \chi^{u1_4}_{-1} \bar \chi^{u1_4}_1 \big)|\chi^\text{Is}_\frac{1}{16}|^2.
\end{align}

\bibliography{../../bib/wencross,../../bib/all,../../bib/allnew,../../bib/publst,local}

\end{document}